\documentclass{article}
\pdfoutput=1
\usepackage{jheppub}
\usepackage{datetime}
\usepackage{float}
\usepackage[utf8]{inputenc}

\def\le{\left}
\def\ri{\right}

\def\){\right)}
\def\({\left( }
\def\]{\right] }
\def\[{\left[ }

\def\NO{\nonumber}

\newcommand{\be}{\begin{equation}}
\newcommand{\ee}{\end{equation}}

\def\bea{\begin{eqnarray}}
\def\eea{\end{eqnarray}}

\def\bal#1\eal{\begin{align}#1\end{align}}

\def\bald{\begin{aligned}}
\def\eald{\end{aligned}}

\def\bsub{\begin{subequations}}
\def\esub{\end{subequations}}

\def\beqx{\begin{displaymath}}
\def\eeqx{\end{displaymath}}

\newcommand{\bmat}{\left(\begin{array}}
\newcommand{\emat}{\end{array}\right)}

\def\a{\alpha}
\def\b{\beta}
\def\c{\chi}
\def\d{\delta}
\def\e{\epsilon}

\def\g{\gamma}
\def\h{\eta}
\def\j{\psi}
\def\k{\kappa}

\def\m{\mu}
\def\n{\nu}
\def\o{\omega}
    
\def\p{\pi}

\def\r{\rho}
\def\s{\sigma}
\def\t{\tau}
\def\x{\xi}
\def\z{\zeta}
\def\D{\Delta}
\def\F{\Phi}

\def\L{\Lambda}

    \def\vth{\vartheta}

\def\vf{\varphi}

\def\ca{{\cal A}}

\def\cd{{\cal D}}

\def\cf{{\cal F}}
\def\cg{{\cal G}}
\def\ch{{\cal H}}
\def\ci{{\cal I}}
\def\cj{{\cal J}}

\def\cl{{\cal L}}
\def\cm{{\cal M}}
\def\cn{{\cal N}}
\def\co{{\cal O}}

\def\cq{{\cal Q}}
\def\car{{\cal R}}
\def\cs{{\cal S}}
\def\ct{{\cal T}}
\def\cu{{\cal U}}

\def\cx{{\cal X}}
\def\cy{{\cal Y}}

\def\bo{{\raise-.3ex\hbox{\large$\Box$}}}               % D'Alembertian
\def\pa{\partial}                                       % curly d
\def\face{{\raise.2ex\hbox{$\displaystyle \bigodot$}\mskip-2.2mu \llap {$\ddot
        \smile$}}}                                   % happy face
\def\>{\rangle}                                      %right angle
\def\<{\langle}                                      %left angle

\def\sbtx#1{{}_{\rm #1}}                           % text sub"
\def\wt#1{\widetilde{#1}}                            % big tilde
\def\Hat#1{\widehat{#1}}                             % big hat
\def\leftrightarrowfill{$\mathsurround=0pt \mathord\leftarrow \mkern-6mu
        \cleaders\hbox{$\mkern-2mu \mathord- \mkern-2mu$}\hfill
        \mkern-6mu \mathord\rightarrow$}        % <--> double differential
\def\dvec#1{\vbox{\ialign{##\crcr
        \leftrightarrowfill\crcr\noalign{\kern-1pt\nointerlineskip}
        $\hfil\displaystyle{#1}\hfil$\crcr}}}           % <--> accent
\def\-{\hphantom{-}}

\preprint{KIAS-P18074}

\begin{document}

\title{5D Rotating Black Holes and the nAdS$_2$/nCFT$_1$ Correspondence}

    \author[a]{Alejandra Castro,}  \author[b]{Finn Larsen} \author[c]{and Ioannis Papadimitriou}
  
    \affiliation[a]{Institute for Theoretical Physics Amsterdam and Delta Institute for Theoretical Physics, University of Amsterdam, Science Park 904, 1098 XH Amsterdam, The Netherlands}
    \affiliation[b]{Department of Physics and Leinweber Center for Theoretical Physics, University of Michigan, Ann Arbor, MI 48109-1120, USA}
 \affiliation[c]{School of Physics, Korea Institute for Advanced Study, 85 Hoegi-ro, Dongdaemun-gu, Seoul 02455, Korea}
 
\noindent\today \\

\abstract{We study rotating black holes in five dimensions using the nAdS$_2$/nCFT$_1$ correspondence. A consistent truncation of pure Einstein gravity (with a cosmological constant) in five dimensions to two dimensions gives a generalization of the Jackiw-Teitelboim theory that has two scalar fields: a dilaton and a squashing parameter that breaks spherical symmetry. The interplay between these two scalar fields is non trivial and leads to interesting new features.  We study the holographic description of this theory  and apply the results to the thermodynamics of the rotating black hole from a two dimensional point of view. This setup challenges notions of universality that have been advanced based on simpler models: we find that the mass gap of Kerr-AdS$_5$ corresponds to an undetermined effective coupling in the nAdS$_2$/nCFT$_1$ theory which depends on ultraviolet data.
}

\maketitle

%%%%%%%%%%%%%%%%%%%%%%%%%%%%%%%%%%%%%%%%%%%%%%%%%%%%%%%%%%%%
\section{Introduction}

Holographic dualities and specifically the AdS/CFT correspondence have proven invaluable to the quantum description of black holes. One might have thought that the simplest model of this type would be AdS$_2$/CFT$_1$ since this amounts to gravity in just two spacetime dimensions, typically identified as the radial and temporal directions with the angular variables suppressed. However, any such description faces several complications: pure gravity in two dimensions is over-constrained by its symmetries so it is mandatory to include matter, at least the equivalent of one scalar field. Moreover, the symmetries of AdS$_2$ preclude excitations above the ground state so non trivial dynamics requires a deformation away from the ideal AdS$_2$ limit \cite{Strominger:1998yg,Maldacena:1998uz}. It is only in the last few years that a detailed proposal addressing these obstacles was made in the
form of the duality known as the nAdS$_2$/nCFT$_1$ correspondence \cite{Almheiri:2014cka,Maldacena:2016upp}.\footnote{See \cite{Sarosi2018} for an overview on these recent  developments, and further references.}  

The linchpin of nAdS$_2$/nCFT$_1$ is the non linear realization of symmetry. The conformal symmetry of AdS$_2$ is spontaneously broken and also broken by an anomaly. This symmetry breaking pattern is realized by the IR behavior of quantum systems like the SYK model \cite{SachdevYe,Kitaev,Maldacena2016a,Polchinski2016,GrossRosenhaus2017c} and its avatars so such systems have been the subject of intense study in the last few years. On the gravity side of the correspondence, the preponderance of studies have focused on dilaton gravity, ie. 2D gravity coupled to a single scalar field, with additional minimally coupled matter serving as probes of the theory \cite{Jensen2016,Engelsoy:2016xyb,Almheiri:2016fws,GrumillerSalzerVassilevich2017,Cvetic:2016eiv,ForsteGolla2017,GrumillerMcNeesSalzerEtAl2017,CadoniCiuluTuveri2018,GonzalezGrumillerSalzer2018,GaikwadJoshiMandalEtAl2018,NayakShuklaSoniEtAl2018,KolekarNarayan2018}. However, many interesting black holes involve more elaborate matter content and we expect that such models can realize other symmetry breaking patterns. 

In this paper, we develop a model that is clearly motivated by a ``real" black hole: we study a rotating black hole from the two dimensional viewpoint. Specifically, we consider the Kerr-AdS$_5$ black hole with its two rotation parameters equal. In this setting we develop nAdS$_2$/nCFT$_1$ holography and discuss connections to the Kerr/CFT correspondence \cite{Guica:2008mu}. The starting point for our study, is a {\it consistent reduction of 5D Einstein gravity to 2D} with the option of a cosmological constant in the 5D theory. The resulting 2D geometry corresponds to a base generated by (comoving) time and the radial direction away from the horizon. The main novelty we encounter is the importance of {\it two} scalar fields in the 2D theory. One of them is similar to the dilaton studied in other models and interpreted geometrically as the radius of the radial sphere that grows as we move away from the black hole horizon. The other represents the concurrent ``squashing" of the spatial sphere due to the rotation of the black hole. The interplay between these two scalar fields is non trivial and interesting. In particular, it challenges notions of universality that have been advanced based on simpler models. 

We stress that our truncation is {\it consistent}: the reduction ansatz maps any solution of the 2D theory to an exact solution of the 5D progenitor. For example, we readily find numerous time dependent solutions to the 2D theory and they correspond to black holes with time-dependent ``hair" that are exact solutions to 5D general relativity. The classical expectation is that such hair must be trivial because the no hair theorem ensures that hairy solutions are diffeomorphic to black holes with no hair. However, it may happen that the requisite diffeomorphisms are ``large" in the sense that they act non trivially on boundary conditions. Then these modes become non trivial in the quantum theory. This physical mechanism plays a central role in AdS$_3$ holography \cite{Brown:1986nw,BanadosHenneauxTeitelboimEtAl1993,MaldacenaStrominger1998,Strominger1998} and in the Kerr/CFT correspondence, so it has been studied in great detail \cite{Compere:2012jk}. Large diffeomorphisms are also essential for the nAdS$_2$/nCFT$_1$ correspondence because they are responsible for the Goldstone modes that form the core of the dual boundary theory. We will study diffeomorphism symmetry in detail.

The nAdS$_2$/nCFT$_1$ correspondence applies to the {\it near} horizon region of a black hole that is {\it nearly} extremal. From the 5D point of view the starting point is the conventional Near Horizon Extremal Kerr (NHEK) limit that forms the basis for the Kerr/CFT correspondence \cite{Bardeen:1999px}. The region where this limit applies strictly is interpreted as a trivial IR fixed point of the dual theory. It is the extension of the geometry away from this region that adds dynamics to the theory. In the dual theory the extension corresponds to deformation of the IR theory by irrelevant operators. We find that the operator dual to the mode ${\cal Y}$ that describes the size of the spatial sphere has conformal dimension $\Delta_{\cal Y}=2$. This is the canonical value of the scalar in dilaton gravity so some aspects of our model will coincide with results that are familiar from that context. For example, important aspects of the effective boundary theory are encoded in a Schwarzian action. 

However, our model features two scalars, and they have specific non-minimal couplings to gravity and to each other.  The ``squashing" mode ${\cal X}$ is more irrelevant than the dilaton $\Delta_{\cal X}>\Delta_{\cal Y}$, with $\Delta_{\cal X}=3$ in the case of vanishing cosmological constant. However, these modes generally couple and must be considered together. The only situation where they decouple is for vanishing cosmological constant where it is consistent to keep the squashing mode constant; but such fine-tuning of the effective IR theory is not natural and, indeed, this situation does not correspond to asymptotically flat space. Thus, the generic situation is that the two modes are coupled, with the dilaton dominant and acting as a source of the squashing mode. This non trivial renormalization group flow is a good illustration of effective quantum field theory in holography. Our incorporation of AdS$_5$ boundary conditions ensures that the discussion of such flows makes sense, because the theory is defined in the UV.

It is only marginal operators that have dimensionless coupling constants so the irrelevant operators that appear prominently in nAdS$_2$/nCFT$_1$ are characterized by intrinsic scales. In effective field theory such scales set the cut-off for reliability of the effective description. On the gravity side the scales necessitate some technicalities but those are addressed by conformal perturbation theory adapted to the holographic setting and the needed machinery has been developed elsewhere \cite{KalkkinenMartelliMueck2001,MartelliMueck2003,Papadimitriou:2004ap,Papadimitriou:2010as,ElvangHadjiantonis2016}. The qualitative significance is that the coefficients of these operators introduce symmetry breaking scales into the theory. Interestingly, since the more irrelevant squashing operator dual to ${\cal X}$ is driven by the less irrelevant dilaton operator dual to the mode ${\cal Y}$, in the IR theory there is in fact just one scale in the theory we study. It enters as the overall dimensionful coefficient of the Schwarzian boundary action and can be interpreted physically as the mass-gap of the theory. 

The application to black holes is a central motivation for this work so we discuss black hole thermodynamics in detail. The thermodynamic variables of Kerr-AdS$_5$ depend on the AdS$_5$ radius rather elaborately and the dependence remains non trivial in the near extreme limit.  A microscopic understanding of the black hole entropy would involve accounting for this function. However, in the effective field theory description of the corresponding 2D black hole, the scale of all variables is set by the mass gap which is introduced as an arbitrary IR parameter and offers no intrinsic normalization. Therefore, the function of AdS$_5$ that describes the black hole entropy and other physical variables is not determined by the effective theory. The Kerr-AdS$_5$ black hole differs in this crucial aspect from Reissner-Nordstr\"{o}m-AdS$_5$ and related simple examples considered in the literature hitherto \cite{Almheiri:2016fws,NayakShuklaSoniEtAl2018}. 

This paper is organized as follows. 
In section \ref{sec:5d} we review the thermodynamics of  Kerr-AdS$_5$ black holes.
In section \ref{sec:truncation} we discuss the consistent truncation to 2D of 5D Einstein gravity with a cosmological constant. 
Section \ref{sec:2Deff} discusses the reduction from 5D to 2D in the context of the Kerr-AdS$_5$ black hole and also introduces the near extreme/near horizon limit from 5D and 2D points of view. In section \ref{dictA} we analyze the dynamics of the 2D theory systematically using the Hamilton-Jacobi method. 
These results are used in section \ref{dictB} for the holographic renormalization of the theory, including the discussion of residual symmetries, Ward identities, and the effective Schwarzian action. 
In section \ref{sec:2Dthermo} we discuss the black hole thermodynamics from the 2D point of view. 
Finally, in section \ref{sec:discussion} we conclude with a brief discussion that summarizes our main results and indicate future research directions. 
Several appendices pursue research directions that are not within the main thrust of the paper.

%%%%%%%%%%%%%%%%%%%%%%%%%%%%%%%%%%%%%%%%%%%%%%%%%%%%%%%%%%%%

\section{Black Hole Thermodynamics: 5D Perspective}\label{sec:5d}

In this section we introduce the geometry of the Kerr black hole in AdS$_5$ and we review its thermodynamics. 

We focus on the rotating black holes with ``equal angular momenta". These backgrounds break $SO(4)$ rotational symmetry but preserve
$SO(3)$ through a round $S^2\subset S^3$. We generally assume a geometry that is asymptotically AdS$_5$ 
but the asymptotically flat Myers-Perry black holes are special cases that have particular interest. 

%%%%%%%%%%%%%%%%%%%%%%%%%%%%%%%%
\subsection{5D Black Hole Geometry} \label{sec:bhgeom}
We consider five dimensional Einstein gravity with a negative cosmological constant. It has action:
\begin{equation}
\label{eq:action5d}
I_{\rm 5D} ={1\over 2\kappa_5^2} \int d^5x \sqrt{-g^{(5)}} \left({\cal R}^{(5)}+{12\over \ell_5^2}\right)~,
\end{equation}
where $\ell_5$ is identified as the radius of the vacuum AdS$_5$ background. 

The ``equal angular momentum" family of solutions depends on two parameters $(m,a)$, in addition to the AdS$_5$ scale $\ell_5$. 
It has metric
\begin{equation}
\label{eq:a1}
ds^2_5=g_{\mu\nu}^{(5)}dx^\mu dx^\nu= -{1\over \Xi}\Delta(r)e^{U_2-U_1} dt^2 + {r^2 dr^2\over (r^2+a^2)\Delta(r)}  +  e^{-U_1} d\Omega_2^2 + e^{-U_2}\left(\sigma^3+A \right)^2~, 
\end{equation}
where
\begin{eqnarray}
\label{eq:a2}
 e^{-U_2}&=& {r^2+a^2\over 4 \Xi} +{m a^2\over 2\Xi^2 (r^2+a^2)}~,\cr
 e^{-U_1}&=& {r^2+a^2\over 4 \Xi} ~, \cr
 A&= & A_t dt ={a\over 2\Xi}\left({r^2+a^2\over \ell_5^2}-{2m\over r^2+a^2}\right)e^{U_2}dt~,  
\end{eqnarray}
with
\begin{eqnarray}
\label{eq:a3}
\Xi & = &  1-{a^2\over \ell_5^2}~,\cr 
\Delta(r) & = &  1 + {r^2\over\ell^2_5} - {2mr^2\over (r^2+a^2)^2}~.
\end{eqnarray}
Our notation for the angular forms is
\begin{eqnarray}
\sigma^1&=&-\sin\psi d\theta+\cos\psi \sin\theta d\phi~,\cr
\sigma^2&=&\cos\psi d\theta+\sin\psi \sin\theta d\phi~,\cr
\sigma^3&=& d\psi+ \cos\theta d\phi~,
\label{eqn:sigmadefs}
\end{eqnarray}
so the solutions exhibit a manifest sphere $S^2$: 
\begin{equation}
d\Omega^2_2= d\theta^2+ \sin^2\theta d\phi^2 =  (\sigma^1)^2+(\sigma^2)^2~.
\label{eqn:omegadef}
\end{equation}
The isometry of this sphere can be identified as an $SU(2)_R$ subgroup of the 5D rotation group 
$SO(4)\approx SU(2)_L \times SU(2)_R$ that is preserved by the black hole background. 

The parameters $(m, a)$ employed in the explicit formulae above are loosely interpreted as a ``mass parameter" $m$ and an ``angular momentum parameter" $a$. Importantly, these parameters should not be confused with the physical mass $M$ and angular momentum $J$ of the black hole. A careful 
analysis of the asymptotic behavior far from the black hole identify the physical parameters \cite{Papadimitriou:2005ii}: 
\begin{eqnarray}
M  & = & M_C + {2\pi^2 m ( 3 + {a^2\over\ell^2_5})\over\kappa^2_5  ( 1- {a^2\over\ell^2_5})^3}~,\cr
J & = & {8\pi^2 m a \over\kappa^2_5 ( 1- {a^2\over\ell^2_5})^3}~.
\label{eqn:MJeqn}
\end{eqnarray}
 In the case of equal angular momenta, the Casimir energy is
$$
M_C = {3\pi^2\ell^2_5\over 4\kappa^2_5}~.
$$
Since $M_C$ is independent of the black hole parameters, it will not be important for most of our considerations. 

%%%%%%%%%%%%%%%%%%%%%%%%%%%%%%%%
\subsection{Black Hole Thermodynamics} 
The event horizon of the black hole is located at the coordinate $r_+$ that is the largest value where $\Delta(r)$ vanishes. Since it is unilluminating to solve
$\Delta(r_+)=0$ for $r_+^2$ we solve it for $m$ as
$$
m  =  {(r^2_++a^2)^2 ( 1 + {r^2_+\over\ell^2_5})\over 2r^2_+}~,
$$
and henceforth parameterize the physical variables $M,J$ by the two parameters $r_+,a$. In this parameterization the entropy is
\begin{eqnarray}\label{eq:sbh}
S  =  {4\pi^3 (r^2_+ + a^2)^2\over\kappa^2_5r_+ ( 1 - {a^2\over\ell^2_5})^2}~,
\end{eqnarray}
and the thermodynamic potentials dual to $M,J$ are the temperature 
\begin{eqnarray}
T & = &  {r^2_+ - a^2 + {2r^4_+\over \ell^2_5}\over 2\pi r_+(r^2_++a^2)}~,
\label{eqn:TOmega}
\end{eqnarray}
and the rotational velocity
\begin{eqnarray}
\Omega & = & {a( 1 + {r_+^2\over\ell^2_5})\over r^2_+ + a^2}~.
\label{eqn:Omega}
\end{eqnarray}
The expressions are such that the first law of thermodynamics is satisfied, as it 
should be\footnote{This fact is worth stressing for AdS-Kerr black holes since some influential works use erroneous expressions 
for $M$ and/or $J$ that do not satisfy the 1st law. For (correct) discussion and references see \cite{Gibbons:2004ai,Papadimitriou:2005ii}.}
\begin{equation}
T dS = dM - \Omega dJ~.
\label{eqn:1stlaw}
\end{equation}
For some considerations the entropy is not the appropriate thermodynamic potential and it is better to use 
the Gibbs free energy
\begin{equation}
G(T,\Omega) = M - TS-\Omega J = M_C + {\pi^2(r^2_++a^2)^2 ( 1 - {r^2_+\over\ell^2_5})\over \kappa^2_5 r^2_+(1-{a^2\over\ell^2_5})^2 }~, 
\label{eqn:gibbs}
\end{equation} 
where we combined the formulae given above. The Gibbs free energy appears naturally in Euclidean quantum gravity where
the (appropriately renormalized) on-shell action is $I_5 = \beta G$.

%%%%%%%%%%%%%%%%%%%%%%%%%%%%%%%%
\subsection{The Near Extreme Limit} \label{sec:nIR5Dthermo}

The Kerr-AdS$_5$ black holes with given angular momentum $J$ all have masses satisfying
\begin{equation}
M\geq M_{\rm ext}~,
\label{eqn:mextineq}
\end{equation}
with equality defining the extremal limit. The extremal mass $M_{\rm ext}$ depends on the angular momentum $J$ and 
the AdS$_5$ scale $\ell_5$. To find it explicitly we first express the dimensionless variables 
$M\kappa^2_5/\ell^2_5$ and $J\kappa^2_5/\ell^3_5$ formed from \eqref{eqn:MJeqn} 
in terms of dimensionless parameters $x=a/r_+$, $y=a/\ell_5$, and then take the limit where the temperature (\ref{eqn:TOmega}) vanishes 
by imposing the relation $y^2 = {1\over 2}x^2 (x^2-1)$. This procedure gives the extremal mass
\begin{equation}
M_{\rm ext}=  M_C + {4\pi^2\ell^2_5\over\kappa^2_5}   {(x^2-1)(3 + {1\over 2}x^2(x^2-1))\over (2-x^2)^3 } ~,
\label{eqn:mext}
\end{equation}
where $x$, with $1\leq x^2\leq 2$, parameterizes the angular momentum through
\begin{equation}
J=  {8\pi^2 \ell^3_5\over\kappa^2_5}  {x(x^2-1)^{3/2}\over \sqrt{2}(2-x^2)^3 } ~.
\label{eqn:jext}
\end{equation}

The extremal mass given implicitly by (\ref{eqn:mext}-\ref{eqn:jext}) is complicated for general $J \kappa^2_5/\ell^3_5$. 
It simplifies in the ``small" black hole regime $J\ll{\ell^3_5\over\kappa^2_5}$ where
\begin{equation}
M_{\rm ext} (x^2\sim 1) = M_C + \left( {27\pi^2\over 32\kappa^2_5} J^2 \right)^{1/3}~.
\label{eqn:mextsmall}
\end{equation}
The small black hole limit corresponds to black holes in asymptotically flat space so it is unsurprising that the excitation energy 
represented by the extremal black hole is independent of the AdS$_5$ radius $\ell_5$. However, it is interesting that the Casimir 
energy $M_C$ dominates the black hole mass in this limit. 

In the opposite extreme, for ``large" black holes with $J\gg{\ell^3_5\over\kappa^2_5}$ we find
\begin{equation}
M_{\rm ext} (x^2\sim 2)= {1\over 2\sqrt{2}\ell_5} J ~.
\label{eqn:mextlarge}
\end{equation}
The Casimir energy is negligible in this limit. It is intriguing that the extremal mass is proportional to $J$ since that suggests a relatively simple microscopic origin of these black holes. This feature is reminiscent of the Kerr/CFT correspondence for asymptotically flat black 
holes \cite{Guica:2008mu,Compere:2012jk} but the setting here is novel because it involves a highly curved AdS$_5$.

A {\it nearly} extreme black hole has small temperature $T\ll M$ and corresponds to low energy excitations above the extremal state, while keeping the angular momentum $J$ fixed. 
This regime is central to this work because it can be described by effective field theory and by the nAdS$_2$/CFT$_1$ correspondence. 
Near extremality, the mass and temperature are related by 
\begin{equation}
M-M_{\rm ext}= {1\over M_{\rm gap}} T^2~,
\label{eqn:mgapdef}
\end{equation}
where $M_{\rm gap}$ is the ``mass gap".  At the scale $M-M_{\rm ext}\sim M_{\rm gap}$ a typical thermal excitation carries the entire 
available energy of the system.  A thermodynamic description is therefore only justified 
for $M-M_{\rm ext}\gg M_{\rm gap}$ \cite{Maldacena:1998uz, Maldacena:2016upp,Almheiri:2016fws}. 
The mass gap $M_{\rm gap}$ is fundamental for the nAdS$_2$/nCFT$_1$ correspondence because it is a dimensionful parameter that 
breaks scaling symmetry explicitly, albeit by a small amount. We interpret this important scale physically as the smallest possible excitation energy of the black hole. 

The definition \eqref{eqn:mgapdef} of the mass gap is equivalent to an entropy near extremality that is linear in the temperature
$$
S = S_{\rm ext} + {2\over M_{\rm gap}} T~,
$$
due to the first law of thermodynamics (\ref{eqn:1stlaw}). The equivalence is naturally established in terms of the heat capacity
\begin{equation}
C_J = T\left({dS\over dM}\right)_J~,
\label{eqn:gencj}
\end{equation}
and it gives the mass gap
\begin{equation}
M_{\rm gap} = {2T\over C_J (T\to 0)}~.
\label{eqn:mgcj}
\end{equation}
In explicit computations it is straightforward and conceptually transparent to first compute the heat capacity \eqref{eqn:gencj} by parametric 
differentiation of the entropy formula for any temperature, and then determine the mass gap by taking the limit \eqref{eqn:mgcj}. For 
example, we can employ the dimensionless parameters $x,y$ introduced after \eqref{eqn:mextineq} in intermediate computations,
and only then impose vanishing temperature by $y^2 = {1\over 2}x^2 (x^2-1)$. This procedure gives 
\begin{equation}
M_{\rm gap} = {2T\over C_J (T\to 0)}= {\kappa_5^2 \over 2\pi^4\ell^4_5}   {(2 - x^2)^2 (2 x^2-1) \over (3 - x^2) (x^2-1)^2 }~,
\label{eqn:CJT=0}
\end{equation}
where, as before, the parameter $1\leq x^2\leq 2$ is equivalent to the angular momentum through \eqref{eqn:jext}.

The mass gap $M_{\rm gap}$ (\ref{eqn:CJT=0}) is generally a complicated function of the angular momentum, similar in complexity to the extremal mass $M_{\rm ext}$ \eqref{eqn:mext}. A thorough microscopic understanding of near extreme Kerr-AdS$_5$ black holes must ultimately account for both of these functions. 

The mass gap simplifies in the small black hole regime $J\ll{\ell^3_5\over\kappa^2_5}$ where
\begin{equation}
M_{\rm gap} (x^2\sim 1)= {1\over 4 \pi^4}  \left( {J\over 16\pi^2} \right)^{-{4\over 3}}\kappa_5^{-{2\over 3}}~.
\label{eqn:mgapext}
\end{equation}
As noted previously, a small black hole effectively experiences asymptotically flat space so it is expected that the mass gap for a 
small black hole is independent of the AdS$_5$ radius $\ell_5$. Given this feature, the power law $M_{\rm gap}\sim J^{-{4\over 3}}$
is determined by dimensional analysis. The formula for the mass gap in the limit of large black holes $J\gg{\ell^3_5\over\kappa^2_5}$ 
is 
\begin{equation}
M_{\rm gap} (x^2\sim 2) = {3\over 2\pi^4\ell^2_5} \left( {J\over 8\sqrt{2}\pi^2} \right)^{-{2\over 3}}\kappa_5^{{2\over 3}}~.
\label{eqn:mgapultra}
\end{equation}
The dependencies expressed in this formula suggest that the apparent simplicity of the extremal mass \eqref{eqn:mextlarge} does 
not extend to the dominant excitations of the ground state.

%%%%%%%%%%%%%%%%%%%%%%%%%%%%%%%%%%%%%%%%%%%%%%%%%%%%%%%%%%%%
%%%%%%%%%%%%%%%%%%%%%%%%%%%%%%%%%%%%%%%%%%%%%%%%%%%%%%%%%%%%

\section{Consistent Truncation From 5D to 2D}\label{sec:truncation}

In this section we present the consistent truncation of 5D Einstein gravity with a negative cosmological constant (\ref{eq:action5d}) to 2D. The resulting theory in two spacetime dimensions is the setting for our holographic analysis presented in the following sections. However, the dimensional reduction is also interesting in its own right. Similar reductions have been discussed before in \cite{Gouteraux:2011qh}. 

The reduction from 5D to 2D is effectuated by the simple {\it ansatz}: 
\be\label{eq:b1}
ds^2_5=g_{\mu\nu}^{(5)}dx^\mu dx^\nu= ds_{(2)}^2 +  e^{-U_1} d\Omega_2^2 + e^{-U_2}\le(\sigma^3+A \ri)^2~.
\ee
Here $ds_{(2)}^2$ describes a general 2D geometry. The scalar fields $U_{1,2}$ and the one-form $A$ are functions on this 2D base, but independent of the angular variables. Our notation for angles was introduced in (\ref{eqn:sigmadefs}-\ref{eqn:omegadef}).

Given the background \eqref{eq:b1}, it is straightforward to perform a dimensional reduction of the 5D action \eqref{eq:action5d} down to 2D. The resulting effective action is
\begin{eqnarray}
\label{eq:b3}
I_{\rm 2D} ={\pi^2\over\kappa_5^2} \int d^2x \sqrt{-g^{(2)}} \,e^{-U_1-{1\over 2}U_2} &\Big({\cal R}^{(2)} - {1\over 4} e^{-U_2} F_{ab} F^{ab}+ {1\over 2}\partial_a U_1\partial^a U_1 + \partial_a U_2\partial^a U_1 \cr
&-{1\over 2 }e^{-U_2+2U_1} +{2 e^{U_1}} +{12\over \ell_5^2} \Big)~.
\end{eqnarray}
The indices $(a,b)$ run over the two dimensional directions, and all geometrical quantities are defined with respect to the 2D metric $ds_{(2)}^2=g_{ab}^{(2)} dx^adx^b$. The field strength is given as usual by $F_{ab}=\partial_a A_b-\partial_bA_a$, with $A$ the one-form defined 
by the reduction {\it ansatz} \eqref{eq:b1}.  Since the rest of our discussion will mostly focus on two dimensions, we will henceforth drop the index ``(2)''. 

It is important to emphasize that the effective action $I_{\rm 2D}$ is a {\it consistent} truncation of $I_{\rm 5D}$. Any field configuration that solves the equations of motion derived from the 2D
action \eqref{eq:b3} is also a solution to the five dimensional theory. We proved this claim in the most straightforward way possible: we worked out all components of the 5D Einstein equations for the {\it ansatz} (\ref{eq:b1}) and showed that, using the 2D equations of motion, they were all satisfied. The details are rather messy, but they are manageable using Mathematica.  

As we will see, it is not difficult  to find time-dependent solutions to the 2D theory and all such solutions will automatically have constant Ricci curvature in 5D, approaching Ricci flat geometries as $\ell_5\to\infty$. Another example that will play an important role is the existence of solutions with constant scalars and pure AdS$_2$ geometry. It is interesting that in our construction the AdS$_2$ geometry is not supported 
by flux from the higher dimensional view, but by pure geometry. 

The most important example of all is the 5D Kerr-AdS black with one rotational parameter. It was introduced as a 5D geometry in (\ref{eq:a1}). From the 2D perspective
it has metric 
\be\label{eq:aa2}
ds^2= -{1\over \Xi}\Delta(r)e^{U_2-U_1}dt^2 + {r^2 dr^2\over (r^2+a^2)\Delta(r)}  ~,
\ee
where $\Xi,\Delta$ were introduced in (\ref{eq:a3}). The variables $U_1, U_2$ are the same as the scalars fields that, along with the one-form gauge field $A$, support the 
solutions. These variables were introduced in \eqref{eq:a2}, as notation defining the 5D geometry, but from the 2D perspective they are matter fields.

%%%%%%%%%%%%%%%%%%%%%%%%%%%%%%%%%%%%%%%%%%%%%%%%%%%%%%%%%%%%
%%%%%%%%%%%%%%%%%%%%%%%%%%%%%%%%%%%%%%%%%%%%%%%%%%%%%%%%%%%%
\section{2D Equations of Motion and Solutions}\label{sec:2Deff}

In this section we initiate our study of the effective action \eqref{eq:b3}. We make our notation more convenient and present the equations of motion. We find a static solution that describes the IR of the dual theory, study perturbations around it, and compare those results with the dimensional reduction of the 5D black hole to 2D. 

%%%%%%%%%%%%%%%%%%%%%%%%%%%%%%%%
\subsection{Field Redefinitions} 

Our metric {\it ansatz} \eqref{eq:b1}  and action \eqref{eq:b3} were presented in variables mimicking dimensional reduction in other contexts, for easy comparison. 
However, it is awkward that the scalars $e^{-U_i}$ carry units of length squared, and from the 2D perspective it is suboptimal that the couplings in the action \eqref{eq:b3} 
have off diagonal kinetic terms. To address these issues, we recast our metric  \eqref{eq:b1} as
\be
g_{\mu\nu}^{(5)}dx^\mu dx^\nu= e^{2V}ds_{(2)}^2 + R^2 e^{-2\psi+\chi} d\Omega_2^2 +R^2 e^{-2\chi}\le(\sigma^3+A \ri)^2~.
\label{eqn:mansatz}
\ee    
We introduced a scale $R$ that makes all scalars dimensionless and we redefined the scalar fields $U_{1,2}$ as
\be\label{eq:dilatondef}
e^{\chi}=R \,e^{U_2/2}~,\qquad e^{2\psi}=R^{3} e^{U_1 + U_2/2}~.
\ee
We also performed a Weyl rescaling of the 2D metric by a conformal factor 
\be\label{Weyl-extremal}
e^{2V}= e^{\psi +\chi}~,
\ee
that was chosen such that the kinetic term of the field $\psi$ is absent in the action. The new variables realized by the {\it ansatz} \eqref{eqn:mansatz} give the 2D action
\be
\label{2Daction-extremal}
I_{\rm 2D}={1\over 2\kappa_2^2} \int d^2x\sqrt{-g}\,e^{-2\psi}\Big({\cal R}-\frac{R^2}{4}e^{-3\chi-\psi} F^2-\frac{3}{2}(\nabla\chi)^2+{1\over 2R^2}\le(4e^{3\psi}-e^{5\psi-3\chi}\ri)
+{12\over\ell_5^2}e^{\psi+\chi}\Big)~,
\ee
where ${1\over\kappa_2^2}={16\pi^2R^3\over\kappa_5^2}$. 
This effective action is equivalent to \eqref{eq:b3} and it will be our main focus for the remainder of this paper. It is a generalization of the Jackiw-Teitelboim theory considered e.g. in \cite{Maldacena2016a}. Different generalizations of the Jackiw-Teitelboim model were obtained recently via Kaluza-Klein reduction from a higher dimensional theory in \cite{Cvetic:2016eiv,Das:2017pif,Taylor:2017dly,NayakShuklaSoniEtAl2018,Li:2018omr,GaikwadJoshiMandalEtAl2018,KolekarNarayan2018}. In comparisons with work on 2D dilaton gravity 
it may be useful to identify $\psi$ as ``the'' dilaton field. The field $\chi$ then represents the ``additional" field that parameterizes the deformation of
$S^3$ that is needed to accommodate rotation in 5D.

%%%%%%%%%%%%%%%%%%%%%%%%%%%%%%%%
\subsection{2D Bulk Equations of Motion}
\label{eoms}

The equations of motion for the 2D metric $g_{ab}$, the scalars $\psi$, $\chi$, and the 2D gauge field $A_a$ read 
\begin{align}
\label{eoms-extremal}
&e^{2\j}(\nabla_a\nabla_b-g_{ab}\square)e^{-2\j}+ g_{ab}\le(
{1\over 4R^2}\le(4e^{3\psi}-e^{5\psi-3\chi}\ri)
+\frac{R^2}{8}e^{-3\chi-\psi} F^2+{6\over\ell_5^2}e^{\psi+\chi}\ri)\cr &\qquad \qquad+\frac32\Big(\nabla_a\c\nabla_b\c-\frac12 g_{ab}(\nabla\c)^2\Big)=0~,\NO\\
&{\cal R}+{3\over 4}e^{-3\chi+5\psi}\le({1\over R^2}-{R^2\over 2}F^2e^{-6\psi}\ri)-{1\over R^2}e^{3\j}+\frac{6}{\ell_5^2}e^{\j+\c}-\frac32(\nabla\c)^2=0~,\NO\\
&e^{2\j}\nabla_a(e^{-2\j}\nabla^a\c)+\frac{R^2}{4}e^{-3\chi-\psi} F^2+{1\over 2R^2} e^{5\psi-3\chi}+{4\over\ell_5^2}e^{\psi+\chi}=0~,\NO\\
&\nabla_a\Big(e^{-3\j-3\c}F^{ab}\Big)=0~.
\end{align}
These equations of motion are generally rather complicated and we will proceed in stages. 

The simplest first step is to note that Maxwell's equations in 2D can be integrated in covariant form 
\be\label{eq:f2}
F_{ab}= {Q} e^{3\j+3\c} \epsilon_{ab}~,\qquad F^{2}= -2{Q^2} e^{6\j+6\c}~. 
\ee
Here $\epsilon_{ab}$ is the volume form and the charge $Q$ is an integration constant that is proportional to the angular momentum of the 5D black hole with a constant of proportionality we determine later.\footnote{Our conventions are $\epsilon_{t\rho}=\sqrt{-g}$. } 

The next step is to fix diffeomorphism invariance. We use Fefferman-Graham coordinates: 
\be\label{FG-gauge}
ds^2=d\r^2+\g_{tt}(\r,t)dt^2~.
\ee
The solution for the gauge field \eqref{eq:f2} and the coordinate system \eqref{FG-gauge} simplify the equations of motion \eqref{eoms-extremal} to 
\bal
\label{eoms-extremal-gf}
&\Big(\pa_\r^2-K\pa_\r-\square_t+\frac32\dot\c^2-\frac32\g^{tt}(\pa_t\c)^2\Big)e^{-2\j}=0~,\NO\\
%%%%%%%%%%%%%%%%%%%%%%%%%%%%%%%%%%%%%%%%%%%%%%%%%%%%%%%%%%%%%%%%%%%%%%%%%%%%%%%%%%%%%%%%%%%%%%%%%%%%%%%%%%
&\Big(\pa_\r\pa_t-K\pa_t+\frac32\dot\c\pa_t\c\Big)e^{-2\j}=0~,\NO\\
%%%%%%%%%%%%%%%%%%%%%%%%%%%%%%%%%%%%%%%%%%%%%%%%%%%%%%%%%%%%%%%%%%%%%%%%%%%%%%%%%%%%%%%%%%%%%%%%%%%%%%%%%%
&\Big(\pa_\r^2+K\pa_\r+\square_t +{1\over 2R^2}e^{-3\chi+5\psi}(1+{R^4Q^2}e^{6\chi})-{2\over R^2}e^{3\psi}-\frac{12}{\ell_5^2}e^{\j+\c}\Big)e^{-2\j}=0~,\NO\\
%%%%%%%%%%%%%%%%%%%%%%%%%%%%%%%%%%%%%%%%%%%%%%%%%%%%%%%%%%%%%%%%%%%%%%%%%%%%%%%%%%%%%%%%%%%%%%%%%%%%%%%%%%
&\Big(\pa_\r^2-{3\over 8 R^2}e^{-3\chi+5\psi}(1+{R^4Q^2} e^{6\chi})+\frac{1}{2R^2}e^{3\j}-\frac{3}{\ell_5^2}e^{\j+\c}+\frac34\dot\c^2+\frac34\g^{tt}(\pa_t\c)^2\Big)\sqrt{-\g}=0~,\NO\\
%%%%%%%%%%%%%%%%%%%%%%%%%%%%%%%%%%%%%%%%%%%%%%%%%%%%%%%%%%%%%%%%%%%%%%%%%%%%%%%%%%%%%%%%%%%%%%%%%%%%%%%%%%
&\ddot\c+K\dot\c+\square_t\c-2\dot\j\dot\c-2\g^{tt}\pa_t\j\pa_t\c +{1\over 2R^2}e^{-3\chi+5\psi}(1-{R^4 Q^2}e^{6\chi})+\frac{4}{\ell_5^2}e^{\j+\c}=0~.
\eal
The dot denotes the {\it radial} derivative $\dot \chi\equiv \partial_\rho \chi$. The
metric variable enters implicitly through $\sqrt{-\g} = \sqrt{-\gamma_{tt}}$ and
\begin{eqnarray}
\label{K+L}
K &\equiv& \partial_\rho \log \sqrt{-\gamma} ~,\cr
\square_t &\equiv & {1\over \sqrt{-\gamma}}\partial_t \le( \sqrt{-\gamma}\, \gamma^{tt} \partial_t\ri)~.
\end{eqnarray}
Therefore \eqref{eoms-extremal-gf} is a system of differential equations for just three functions $\psi, \chi, \gamma_{tt}$. However, these are coupled nonlinear equations so generally it is difficult to find exact solutions. In some 2D gravity models the analogous equations can be integrated
entirely, yielding the full classical phase space even far from any fixed points. That is the situation for the Jackiw-Teitelboim model and some of its generalizations \cite{Grumiller:2002nm,Almheiri:2014cka,Engelsoy:2016xyb,Almheiri:2016fws, Cvetic:2016eiv,GrumillerMcNeesSalzerEtAl2017,GonzalezGrumillerSalzer2018}. The present case is more complicated and we cannot fully integrate the equations. However, there are several classes of exact solutions that are worth highlighting:
\begin{enumerate}
\item {\it Attractor Solutions}: solutions with {\it constant} scalar fields. These describe the very near horizon region of 5D Kerr-AdS. 
\item {\it Dilaton Gravity}: take $\ell_5^{-1}= 0$ and $\chi$ the constant that minimizes its potential. From a 5D perspective this theory arises naturally 
from an asymptotically Taub-NUT geometry, where the four dimensional base allows for a Reissner-Nordstr\"{o}m black hole. 
The resulting 2D model resembles the models considered in, e.g., \cite{Strominger:1998yg,HartmanStrominger2009,CastroGrumillerLarsenEtAl2008,Almheiri:2016fws,Cvetic:2016eiv}. In appendix \ref{UV-analysis-flat} we discuss aspects of this truncation. 
\item {\it Kerr-AdS}: the static solution \eqref{eq:aa2} of the 2D theory with two non-trivial scalars and a gauge field. 
Some special cases are Schwarzschild-AdS$_5$ and the limit $\ell_5^{-1}= 0$ that gives asymptotically flat space (and so Myers-Perry black holes).
\item {\it Neutral Solutions}: setting the charge $Q=0$ gives 2D gravity coupled to two scalars $U_{1,2}$. We can find the general time dependent solutions for these scalars in AdS$_2$ geometry. One special case is global AdS$_5$. See appendix \ref{app:neutral} for an example.

\end{enumerate}
This list is clearly not exhaustive, but these represent some significant examples. 

Interestingly, the last equation in \eqref{eoms-extremal-gf} shows that if $\chi$ is constant it {\it must} be that either $\psi$ is also constant or $\ell_5\to \infty$. 
Importantly, this is not an artifact of our parameterization of the fields: we need two scalar fields to describe a running dilaton background if $\ell_5$ is finite. The resulting interplay between the two scalars is an interesting feature of our study that we have not seen discussed in other recent examples.  

In the remainder of this section we focus on the attractor solutions and the perturbations around them. This setting allows us to study the near horizon region of Kerr-AdS$_5$ black holes from the 2D point of view. 

%%%%%%%%%%%%%%%%%%%%%%%%%%%%%%%%%%%%%%%%%%%%%%%%%%%%%%%%%%%%

\subsection{The IR Fixed Point}\label{sec:nIRfix}
We define the IR fixed point as solutions to our equations with {\it constant scalars}. 
This corresponds to the attractor fixed point of the black hole background and, as we will see shortly, the metric at the fixed point is locally AdS$_2$.  

The equations that determine the fixed value of the scalars as functions of the parameters $(Q,R,\ell_5)$ are 
\be\label{eqn:psi0eqn}
e^{-2\psi_0}=e^{-3\chi_0}-{R^4 Q^2\over 2 }e^{3\chi_0}~, 
\ee
and
\be
1-{R^4 Q^2}e^{6\chi_0} +{2R^2\over \ell_5^2}e^{-2\chi_0}\le(2-{R^4 Q^2}e^{6\chi_0}\ri)^2=0~.
\label{eqn:chi0eqn}
\ee
We introduced the subscript ``0'' on the fields $\chi_0$ and $\psi_0$ as a reference to their values at the attractor point. At the IR fixed point the scalars are thus constant on the 
2D spacetime, by definition, but the equations of motion then allow for non trivial metric and gauge field  
\begin{eqnarray}
\label{eqn:leading}
\sqrt{-\gamma_0}&=& \alpha(t)e^{\rho/\ell_2}+\beta(t)e^{-\rho/\ell_2}~,\cr
A_t^0 &=& \mu(t)- Q \ell_2 e^{3\chi_0+3\psi_0}\le(\alpha(t)e^{\rho/\ell_2}-\beta(t)e^{-\rho/\ell_2}\ri)~,
%\label{eqn:leadingorder}
\end{eqnarray}
where we imposed the radial gauge 
\be
\label{radialgauge}
A_\r=0~,
\ee
on the gauge field. Importantly, the integration ``constants" $\alpha(t)$, $\beta(t)$, and $\mu(t)$ are arbitrary functions of the temporal variable $t$. 

With this field configuration the equations of motion show that the background geometry is (at least) locally AdS$_2$, with the AdS$_2$ radius 
given by
\begin{equation}
\label{eq:l2}
\ell_2^{-2}={1\over R^2}e^{3\psi_0}\le(1+12 q\ri)~,
\end{equation}
where
\begin{equation}
\label{eq:q}
\qquad q\equiv{1\over 8}e^{2\psi_0}({R^4 Q^2} e^{3\chi_0}-e^{-3\chi_0})~. 
\end{equation}
It follows from \eqref{eqn:chi0eqn} that the dimensionless variable $q$ is related to the AdS$_5$ radius as
\be\label{AdS5-radius}
\ell_5^{-2}=\frac{qe^{2\j_0-\c_0}}{R^2}~,
\ee
such that $q\to 0$ in the limit $\ell_5^{-1}\to 0$ where the 5D geometry changes from asymptotically AdS$_5$ to asymptotically flat space.

%%%%%%%%%%%%%%%%%%%%%%%%%%%%%%%%%%%%%%%%%%%%%%%%%%%%%%%%%%%%
\subsection{Perturbations Around the IR Fixed Point}
\label{sec:IRperturbations}
We now begin the study of small perturbations away from the IR fixed point. To parameterize the deviation of the fields away from their constant values at the IR fixed point 
we define
\begin{eqnarray}
\label{eqn:cycxdef}
 \cy &\equiv & e^{-2\j}-e^{-2\j_0}~,\cr
  \cx &\equiv & \c-\c_0~,\cr
 \sqrt{-\g_1} & \equiv & \sqrt{-\g}-\sqrt{-\g_0}~.
\end{eqnarray}
Although both $\cy$ and $\cx$ are assumed small they need not be of the same order since their fluctuations can be driven by independent couplings. We will revisit this point below. 

Expanding the field equations \eqref{eoms-extremal-gf} around the IR fixed point we find 
\bal
\label{eoms-near-extremal-gf}
&\Big(\pa_\r^2-K_0\pa_\r-\square_t^0\Big)\cy=0~,\NO\\
%%%%%%%%%%%%%%%%%%%%%%%%%%%%%%%%%%%%%%%%%%%%%%%%%%%%%%%%%%%%%%%%%%%%%%%%%%%%%%%%%%%%%%%%%%%%%%%%%%%%%%%%%%
&\Big(\pa_\r\pa_t-K_0\pa_t\Big)\cy=0~,\NO\\
%%%%%%%%%%%%%%%%%%%%%%%%%%%%%%%%%%%%%%%%%%%%%%%%%%%%%%%%%%%%%%%%%%%%%%%%%%%%%%%%%%%%%%%%%%%%%%%%%%%%%%%%%%
&\Big(\pa_\r^2+K_0\pa_\r+\square_t^0-2\ell_2^{-2}\Big)\cy=0~,\NO\\
%%%%%%%%%%%%%%%%%%%%%%%%%%%%%%%%%%%%%%%%%%%%%%%%%%%%%%%%%%%%%%%%%%%%%%%%%%%%%%%%%%%%%%%%%%%%%%%%%%%%%%%%%%
&\Big(\pa_\r^2-\ell_2^{-2}\Big)\sqrt{-\g_1}+{R^{-2}}\Big(3e^{5\psi_0}(1+8q)\cy-12qe^{3\psi_0}\cx\Big)\sqrt{-\g_0}=0~,\NO\\
%%%%%%%%%%%%%%%%%%%%%%%%%%%%%%%%%%%%%%%%%%%%%%%%%%%%%%%%%%%%%%%%%%%%%%%%%%%%%%%%%%%%%%%%%%%%%%%%%%%%%%%%%%
&\Big(\pa_\r^2+K_0\pa_\r+\square_t^0-\(\frac{6+32q}{1+12q}\)\ell_2^{-2}\Big)\cx +8{q\over R^2}\, e^{5\psi_0}\cy=0~,
\eal   
to linear order in $\cy$, $\cx$ and $\sqrt{-\g_1}$. The extrinsic curvature $K_0$ and the d'Alembertian $\square_t^0$ were defined in 
\eqref{K+L}, except for the index ``0" indicating that here they are evaluated in the IR geometry with metric $\gamma_0$. 

We begin the analysis of the system of equations \eqref{eoms-near-extremal-gf} by reading off the AdS$_2$ mass of the scalar fields. These values 
determine the conformal dimensions of the dual scalar operators at the IR fixed point. 

The third equation in \eqref{eoms-near-extremal-gf} implies that the scalar operator dual to the dilaton $\psi$, now represented by the perturbation $\cy$, has conformal 
dimension $\D_\cy=2$ for any 
value of the AdS$_5$ radius $\ell_5$. Our nomenclature that this is ``the" dilaton is based on the fact that this is also the value in simple linear 
dilaton gravity. 

The last equation in \eqref{eoms-near-extremal-gf} similarly determines the conformal dimension of the scalar operator dual to $\cx$ as
\be\label{chi-dim}
\D_\cx=\frac12\(1+5\sqrt{\frac{1+\frac{28}{5}q}{1+12q}}\) ~.
\ee   
The value of $\D_\cx$ decreases monotonically as $q$ varies from the asymptotically flat space $q=0$ to strongly coupled AdS$_5$ $q=\infty$. It satisfies
\be
2<\frac16(3+\sqrt{105})\leq \D_\cx\leq 3~.
\ee
It follows that $\D_\cx > \D_\cy$ for any value of the AdS$_5$ radius $\ell_5$ and so the near IR dynamics is generically dominated by the dilaton fluctuation $\cy$. 

Motivated by this observation we will solve the remainder of the linear equations \eqref{eoms-near-extremal-gf} with boundary conditions corresponding to a non-zero source for the dilaton $\cy$ but no independent source for the fluctuation $\cx$.\footnote{It is in principle 
straightforward to turn on an independent source for $\cx$, but as we will see in subsection \ref{sec:2D5DBH} it is not important for our application to the black hole background. However, we do turn on such a source later on in subsection \ref{sec:effIR}, where it is necessary for developing the holographic dictionary. Moreover, the full homogeneous solution for the fluctuation $\cx$ leads to a dynamical two-point function in the dual theory, which would be interesting to explore.} Since the last equation in \eqref{eoms-near-extremal-gf} has a term proportional to $\cy$, the operator dual to $\cx$ nevertheless will be subject to a source, but only indirectly through the source of $\cy$.  

It is interesting to note that the linearized equations \eqref{eoms-near-extremal-gf} are qualitatively similar to those in e.g. eq.~(3.33) of \cite{KolekarNarayan2018}. In particular, in both cases there is a dilaton field that satisfies a decoupled equation and is dual to a dimension 2 scalar operator. Moreover, in both cases there is a second scalar field that is sourced by the dilaton for generic values of the parameters of the theory. However, in our case the operator dual to this second scalar is always more irrelevant in the IR than the dilaton, i.e. $\D_\cx>\D_{\cy}$, and hence there is a well defined effective IR theory that is dominated by the dilaton dynamics. This is not always the case in \cite{KolekarNarayan2018}, where the second scalar can even be massless for certain values of the parameters of the theory.

We start by solving for $\cy$. Adding the first and third equations in \eqref{eoms-near-extremal-gf} we find the constraint
\be\label{Y-decoupled}
\big(\pa_\r^2-\ell_2^{-2}\big)\cy=0~,
\ee
with the solution
\bal
\label{eq:s1}
\cy=&\;\n(t)e^{\r/\ell_2}+\vth(t)e^{-\r/\ell_2}~.
\eal
We must require that $|\n(t)|\ll e^{-2\j_0}$ since only then there is a non trivial spatial region satisfying $\;|\n(t)|e^{\r/\ell_2}\ll e^{-2\j_0}$ and that is the condition that perturbation theory 
is valid.

The second equation in \eqref{eoms-near-extremal-gf} can be recast as the constraint
\be
\pa_\r\(\frac{\pa_t\cy}{\sqrt{-\g_{0}}}\)=0~.
\label{eqn:spatconstraint}
\ee
The leading order metric $\sqrt{-\gamma_0}$ was given in \eqref{eqn:leading} where it was parameterized in terms of two coefficients $\a(t), \b(t)$.
The constraint \eqref{eqn:spatconstraint} now relates these two functions to their analogues $\n(t), \vth(t)$ in the dilaton profile $\cy$. We find 
\begin{eqnarray}
\label{eq:s2}
	\b(t) &=&-\frac{\ell_2^2}{4}\frac{\a}{\partial_t\n}\pa_t\(\frac{1}{\n}\( c_0+\frac{(\partial_t\n)^2}{\a^2}\)\)~,\cr
	\vth(t)&=&-\frac{\ell_2^2}{4\n}\( c_0+\frac{(\partial_t\n)^2}{\a^2}\)~,
\end{eqnarray}
with the integration constant  $c_0$ spacetime independent. These constraints express the damped $(e^{-\r/\ell_2})$ terms in the 
background metric $\sqrt{-\gamma_0}$ and in the dilaton fluctuation $\cy$ in terms of the arbitrary (finite) boundary source $\alpha(t)$ for the 
metric and the arbitrary (infinitesimal) source $\nu(t)$ for the irrelevant operator dual to the dilaton. 

The inhomogeneous solution for $\cx$ can be determined by comparing the last and third equations in \eqref{eoms-near-extremal-gf}. We find
\be\label{eq:x1}
\cx\sbtx{inhom}=\frac{2q}{1+2q} e^{2\psi_0}\cy~.
\ee
This inhomogeneous solution is a novel feature of our model. In the presence of a non-trivial AdS$_5$ cosmological 
constant $q\neq0$ so turning on an irrelevant deformation for the dilaton $\cy$ requires a non-trivial profile for the matter field $\cx$. This non-minimal coupling is a radical departure from the other recent examples of AdS$_2$ holography, where additional matter fields are minimally coupled or ignored altogether. We stress that the solution  in \eqref{eq:x1} does not have an independent source for $\cx$. 
This would arise from the homogeneous solutions to the last equation in \eqref{eoms-near-extremal-gf}. 

We can now finally use the fourth equation in \eqref{eoms-near-extremal-gf} to determine the metric perturbation. Inserting the inhomogeneous solution \eqref{eq:x1} 
for $\cx$ we find
\be
\Big(\pa_\r^2-\ell_2^{-2}\Big)\sqrt{-\g_1}+\frac{3}{R^{2}} e^{5\psi_0}\frac{(1+10q+8q^2)}{(1+2q)}  \sqrt{-\g_0}\cy=0~.
\ee
The homogeneous equation for $\sqrt{-\g_1}$ in this case is identical to the zero order solution for $\sqrt{-\g_0}$ and, without loss of generality, can be absorbed in the arbitrary functions $\a(t)$ and $\b(t)$ parameterizing the zero order solution. We are therefore only interested in the inhomogeneous solution for $\sqrt{-\g_1}$. Inserting the 
explicit solutions \eqref{eqn:leading} and \eqref{eq:s1} for $\sqrt{-\g_0}$ and $\cy$ it is straightforward to integrate and find the inhomogeneous solution\footnote{
The final term in the square bracket is a rewrite of $-4(\alpha\vth+\beta\nu)$ using the constraints \eqref{eq:s2}.}
\be%\boxed{
\label{eq:s3} 
	\sqrt{-\g_1}=-\frac{(1+10q+8q^2)}{(1+2q)(1+12q)}e^{2\psi_0}\le[\sqrt{-\g_0}\;\cy+{2\ell^2_2}\partial_t\le(\frac{\partial_t\n}{\a}\ri)\ri]~.%}
\ee

In summary, we have solved our linearized system of equations of motion \eqref{eoms-near-extremal-gf} assuming only that there is no source term for $\cx$. 
The solutions for the fields $\cy$, $\cx$, and $\sqrt{-\g_1}$ are given by equations \eqref{eq:s1}, \eqref{eq:x1}, and \eqref{eq:s3}. Recalling the expression 
\eqref{eqn:leading} for the leading order metric $\sqrt{-\g_0}$ and the constraint \eqref{eq:s2} on the time dependent coefficients, all three fields have been determined in terms of the two sources $\alpha(t)$, $\nu(t)$. 

%%%%%%%%%%%%%%%%%%%%%%%%%%%%%%%%%%%%%%%%%%%%%%%%%%%%%%%%%%%%
%%%%%%%%%%%%%%%%%%%%%%%%%%%%%%%%%%%%%%%%%%%%%%%%%%%%%%%%%%%%

\subsection{2D Black Holes from AdS$_5$ Black Holes}
\label{sec:2D5DBH}

In this subsection we identify the near-horizon geometry of near extreme Kerr-AdS$_5$ black holes starting from the complete 5D solution reviewed in subsection \ref{sec:bhgeom}. This will illuminate our perturbative expansion around the IR fixed point and motivate the boundary 
conditions we imposed on the fluctuations $\cy$ and $\cx$ in subsection \ref{sec:nIRfix}. 

The Hawking temperature $T$ vanishes at extremality. At $T=0$ the expression \eqref{eqn:TOmega} for the temperature gives
\be
{a_0^2   } = {\ell_5^2\over 2} x^2(x^2-1)~,
\label{eqn:l5nearh}
\ee
where $x=a_0/r_0$ is defined in terms of $r_0$, the radial coordinate at the extremal horizon, and $a_0$, the extremal value of the rotational parameter. The dimensionless variable $x$ introduced here 
is not identical to $x=a/r_+$ defined in subsection \ref{sec:nIR5Dthermo} but, to the precision we work, we will not need to distinguish them.

Near extremality is a small departure of $(r_+,a)$ from $(r_0,a_0)$, such that we increase slightly the temperature of the black hole (and its mass) while keeping the angular momentum $J$ and $\ell_5$ fixed. We parameterize this departure as 
\be\label{eq:extlimit}
r_+ = r_0 + \varepsilon \lambda ~, \quad a=a_0 + O(\lambda^2)~,
\ee
with $\lambda\varepsilon\ll r_0$ and $\varepsilon$ dimensionless. The deviation of $a$ away from extremality is determined by requiring that $J$ is fixed in the near extremal limit; its precise form is not important for the purpose of this section.\footnote{In the near extremal limit, the matter fields $(\psi,\chi)$ respond linearly with $\lambda$ and this will suffice for later applications.} The entire near-horizon region has $r-r_0\sim \lambda$ and we describe it using a radial coordinate $\rho$ introduced as
\be\label{eq:extlimit1}
r = r_0 + {\lambda\over 2}( e^{\rho/\ell_2}+\varepsilon^2e^{-\rho/\ell_2})~.
\ee
The coordinate $\rho$ is adapted to the scale $\ell_2$ of the near-horizon region. This scale will shortly be identified as the radius of an AdS$_2$ factor with $(t,\rho)$ coordinates. 

The near horizon geometry is isolated by expanding the 5D geometry \eqref{eq:a1} to the leading significant order in 
$\lambda\varepsilon/r_0$. Expanding first the function $\Delta$ defined in \eqref{eq:a3}, we readily find the general form of the near-horizon 
metric  
\begin{align}
g_{\mu\nu}^{(5)}dx^\mu dx^\nu&= e^{2V}ds_{(2)}^2 + R^2 e^{-2\psi+\chi} d\Omega_2^2 +R^2 e^{-2\chi}\le(\sigma^3+A \ri)^2 \cr
&\to e^{2V_0}\le(-\gamma_{tt}^{\rm bh} dt^2 +d\rho^2\ri) + R^2 e^{-2\psi_0+\chi_0} d\Omega_2^2 +R^2 e^{-2\chi_0}\le(\sigma^3+A_t^{\rm bh}dt\ri)^2 +O(\lambda)~,
\label{eqn:metexpand}
\end{align}
with the identification
\be\label{eq:mapl2}
{1\over \ell^{2}_2}= \sqrt{8R^5 x^9\over a_0^9}  (2-x^2)^2(2x^2-1)~. 
\ee
Straightforward expansion to the leading order also determines the attractor values of the scalars
\begin{align}
e^{2\psi_0} &= {2^{5/2} R^3\over \ell_5^3}{(2-x^2)^2\over (x^2-1)^{3/2}} ~,\cr
e^{2\chi_0} &= {2R^2\over \ell_5^2} {(2-x^2)^2\over (x^2-1)}~.
\label{eqn:nearhscalars}
\end{align}

The expansion of the remaining functions in (\ref{eq:a2}-\ref{eq:a3}) shows that the formal near-horizon 
limit $\lambda\varepsilon/r_0\to 0$ does not exist except if we first redefine coordinates
\begin{eqnarray}
\label{eq:extlimit2}
t&\to &\frac{\lambda_0}{\lambda} t~,\cr
\psi &\to &\psi + \frac{\Omega_0}{\lambda}t~,
\end{eqnarray}
where
\begin{eqnarray}\label{eq:l1}
\lambda_0 &=&{\ell_5^2\over 2\ell_2 }{x^4-1\over 2x^2-1 }~,\cr
\Omega_0 &=& {\lambda_0\over a_0}x^2(2-x^2)~,
\end{eqnarray} 
and only then take the limit with the {\it new} $(t,\psi)$ coordinates fixed. This limiting procedure determines the rescaled metric factor 
$\gamma_{tt}^{\rm bh}$ and  gauge field $A_t^{\rm bh}$ introduced in \eqref{eqn:metexpand}. They are
\begin{eqnarray}
\gamma_{tt}^{\rm bh} &=&- (e^{\rho/\ell_2}-\varepsilon^2 e^{-\rho/\ell_2})^2 ~,\cr
A_t^{\rm bh} &=& {x^3(2-x^2)\over a_0^2 (1+x^2)} \lambda_0 (e^{\rho/\ell_2}+\varepsilon^2 e^{-\rho/\ell_2})~.  
\label{eqn:gammatt}
\end{eqnarray}
The result for $\gamma_{tt}^{\rm bh}$ shows that the 2D geometry is AdS$_2$ with radius $\ell_2$, as promised. The intermingling of the near horizon limit with a coordinate transformation \eqref{eq:extlimit2} is characteristic of rotating black holes and well-known from the near horizon Kerr geometry \cite{Bardeen:1999px}.

The near-horizon limit of the 5D Kerr black hole implemented above has constant scalar fields so it must correspond to a 2D geometry at its IR fixed 
point. Those were discussed in subsection \ref{sec:nIRfix}. The near horizon Kerr metric \eqref{eqn:gammatt} and the IR fixed point metric \eqref{eqn:leading} indeed have the same form with the identification.  
\begin{eqnarray}
\alpha(t) &=& 1 ~, \cr
\beta(t) &=& -\varepsilon^2  ~.
\label{eqn:alphabeta}
\end{eqnarray}
Comparing the expressions \eqref{eqn:leading} and \eqref{eqn:gammatt} for the near-horizon gauge field we identify the 2D charge 
\begin{eqnarray}
Q & = & - {a_0^3\over R^5x^2(2-x^2)^3}\cr
&=& -{\kappa_2^2\over  R^2} J~.
\label{eqn:Qconsistency}
\end{eqnarray}
The absolute value of this expression for the charge also follows by equating the horizon values of the scalars for the 5D Kerr in \eqref{eqn:nearhscalars} with the corresponding 2D values (\ref{eqn:psi0eqn}-\ref{eqn:chi0eqn}). 
That this computation agrees with \eqref{eqn:Qconsistency} gives one more consistency check on our algebra. The ``smallness" of the black hole relative to the AdS$_5$ scale was parameterized near the IR fixed point geometry by $q$ in \eqref{eq:q} and by the parameter $x$ in the 5D thermodynamics. They are related as
\begin{equation}
q = {x^2  - 1 \over 4( 2-x^2)}~,
\end{equation}
with $q=0$ and $x=1$ both corresponding to the limit $\ell_5^{-1}\to 0$ where the 5D black hole is asymptotically flat.  

The nAdS$_2$/nCFT$_1$ correspondence would have no dynamics were it not for the small explicit breaking of conformal symmetry captured by the expansion away from the IR fixed point. Starting from the 5D black hole given in \eqref{eq:a2}, and expanding it according 
to the near extremal limit (\ref{eq:extlimit}-\ref{eq:extlimit1}) and \eqref{eq:extlimit2} we find
\begin{align}
e^{2\chi}= e^{2\chi_0}\Big(1 + {x(x^2-1)\over a_0(1+x^2) } \lambda \le(e^{\rho/\ell_2}+\varepsilon^2 e^{-\rho/\ell_2}\ri) + O(\lambda^2)\Big)~,\cr
e^{2\psi}= e^{2\psi_0}\Big(1 - {x(3-x^2)\over 2a_0(1+x^2)} \lambda \le(e^{\rho/\ell_2}+\varepsilon^2 e^{-\rho/\ell_2}\ri) + O(\lambda^2)\Big)~.
\label{eqn:5dkerrchipsi}
\end{align}
We can compare these expressions with our perturbative expansion around the IR fixed point carried out in subsection \ref{sec:nIRfix}. 
The symmetry breaking was introduced in \eqref{eq:s1} as the leading perturbation to the dilaton $\cy = e^{-2\psi}-e^{-2\psi_0}$. 
Comparison with \eqref{eqn:5dkerrchipsi} identifies the dilaton source
\be
\label{eqn:Kerrnutheta}
\nu(t)=  {x(3-x^2)\over 2a_0(1+x^2)}e^{-2\psi_0} \lambda ~,
\ee
for the 5D Kerr black hole. The subleading term in the dilaton perturbation \eqref{eqn:5dkerrchipsi} transcribes to $\vartheta(t) = \varepsilon^2 \nu(t)$ 
so the integration constant $c_0$ defined in \eqref{eq:s2} becomes
\be\label{c0}
c_0=-{4\nu^2\over \ell_2^2} \varepsilon^2~.
\ee
It is then a consistency check that the perturbative formula for $\beta(t)$ given in \eqref{eq:s2} is satisfied with $\beta(t) = -\varepsilon^2$, 
as we found in \eqref{eqn:alphabeta} by expansion of the Kerr-AdS solution. 

The perturbative expansion of Kerr-AdS$_5$ in \eqref{eqn:5dkerrchipsi} shows that generally the ``additional" $\chi$ field is sourced at the same order 
as the dilaton field $\psi$. The perturbation of $\chi$ away from its IR fixed point value vanishes in the flat space limit of the Kerr-AdS$_5$ solution where $x=1$ 
but not in general. However, the perturbation of $\chi$ reported in \eqref{eqn:5dkerrchipsi} for Kerr-AdS$_5$ coincides precisely with the inhomogeneous solution \eqref{eq:x1} computed by the perturbative expansion. In our perturbative analysis in subsection \ref{sec:nIRfix} we imposed boundary conditions that removed the homogeneous solution. We see here that this is the appropriate choice, at least for the Kerr-AdS$_5$ black hole. 

This result nicely illustrates a general feature of effective quantum field theory. Since $\Delta_\cx>\Delta_\cy$ we expect that
the dilaton fluctuation $\cy$ is driving the departure from the IR fixed point. Importantly, this does not mean that other perturbations, 
such as $\cx$, are altogether negligible. Rather, effective field theory predicts that their dual operators do not have independent coefficients, their strengths are determined by the dominant operators. That is precisely what we find here.

%%%%%%%%%%%%%%%%%%%%%%%%%%%%%%%%%%%%%%%%%%%%%%%%%%%%%%%%%%%%
%%%%%%%%%%%%%%%%%%%%%%%%%%%%%%%%%%%%%%%%%%%%%%%%%%%%%%%%%%%%

%%%%%%%%%%%%%%%%%%%%%%%%%%%%%%%%%%%%%%%%%%%%%%%%%%%%%%%%%%%%

\section{Hamilton-Jacobi Formalism}
\label{dictA}

In this section we provide an alternative route to the perturbative solutions near the IR fixed point presented in subsection \ref{sec:IRperturbations} and, in the process, determine the local covariant boundary terms that are needed to holographically renormalize the 2D theory and to construct the related holographic dictionary. 

This alternative route involves a radial Hamiltonian formulation of the bulk dynamics and the associated Hamilton-Jacobi equation. The solution of the radial Hamilton-Jacobi equation determines the ``effective superpotential''
that not only generates first order equations of motion that integrate to the solutions previously found using 
Lagrangian methods, it is also the covariant and local functional of dynamical fields that will serve as holographic counterterms in 
the next section. 

%%%%%%%%%%%%%%%%%%%%%%%%%%%%%%%%%%%%%%%%%%%%%%%%%%%%%%%%%%%%
\subsection{Radial Hamiltonian Dynamics}

The first order flow equations that govern the perturbative solutions near the IR fixed point can be derived systematically by formulating the 2D theory \eqref{2Daction-extremal} in radial Hamiltonian language, which we now briefly review. 

In order to formulate the dynamics of the 2D theory in radial Hamiltonian language we add to the 
2D bulk action \eqref{2Daction-extremal} the Gibbons-Hawking term  
\be
\label{GH}
I\sbtx{GH}=\frac{1}{2\k^2_2}\int_{\pa\cm} dt\sqrt{-\g}\;e^{-2\j}2K~,
\ee
and decompose the 2D metric in the ADM form
\be\label{ADM-metric}
ds^2= N^2 d\r^2 +\g_{tt}(dt+N^t d\rho)^2~,
\ee
in terms of the radial lapse and shift functions, respectively $N$ and $N^t$, as well as the induced metric $\g_{tt}$ on the one dimensional slices of constant radial coordinate $\r$. 

Inserting the metric decomposition \eqref{ADM-metric} in the action \eqref{2Daction-extremal} we find that the total {\em regularized} action, i.e. evaluated with a radial cutoff $\r_c$, takes the form \cite{Cabo-Bizet:2017xdr}
\be\label{reg-action}
I\sbtx{reg}=I\sbtx{2D}+I\sbtx{GH}=\frac{1}{2\k^2_2}\int\limits_{\r=\r_h}\hskip-0.15cm dt\sqrt{-\g}\;e^{-2\j}2K+\int\limits_{\r_h}^{\r_c} d\r\;L~,
\ee
where the radial Lagrangian $L$ is given by 
\bal\label{lagrangian}
L=&\;\frac{1}{2\k^2_2}\int dt\sqrt{-\g}N\Big(-\frac{4}{N}K(\dot\j-N^t\pa_t\j)-\frac{3}{2N^2}(\dot\c-N^t\pa_t\c)^2-\frac32\g^{tt}(\pa_t\c)^2
-\frac{R^2}{2N^2}e^{-\j-3\c}F_{\r t}F_{\r}{}^t\NO\\
&\hskip1.5in+\frac{2}{R^2}e^{3\j}-\frac{1}{2R^2}e^{5\j-3\c}+\frac{12}{\ell^2_5}e^{\j+\c}-2\square_t\Big)e^{-2\j}~,
\eal
and $K=\g^{tt}K_{tt}$ refers to the trace of the extrinsic curvature $K_{tt}$, given by
\be\label{eq:K}
K_{tt}=\frac{1}{2N}\big(\dot\g_{tt}-2D_tN_t\big)~.
\ee
As in the Lagrangian equations of motion \eqref{eoms-extremal-gf}, a dot denotes a derivative with respect to the radial coordinate $\r$ and $D_t$ stands for the covariant derivative with respect to the induced metric $\g_{tt}$. The extrinsic curvature and the covariant Laplacian on the 
radial slice reduce to \eqref{K+L} in the Fefferman-Graham gauge $N=1$, $N_t=0$ that was used in subsection \ref{eoms}. We stress that, 
in writing \eqref{reg-action}, we have explicitly included the possible contributions from the presence of a horizon located at $\r=\r_h$, which will be important when we evaluate the on-shell action later on. 

It is interesting that the radial Lagrangian for 2D gravity \eqref{lagrangian} is qualitatively different from its higher dimensional analogues in that it contains no quadratic terms in the ``velocities'' $\dot\j$ or $K$, but rather a mixed term of the form $K\dot\j$. This is a special property of 2D theories that leads to mixing between the canonical structure of the 1D metric $\g_{tt}$ and of the dilaton $\j$. 

From the radial Lagrangian \eqref{lagrangian} we obtain the canonical momenta 
\bal
\label{momenta}
\p^{tt}&=\frac{\d L}{\d\dot\g_{tt}}=-\frac{1}{2\k^2_2}\sqrt{-\g}e^{-2\j}
\frac{2}{N}\g^{tt}(\dot\j-N^t\pa_t\j)~,\NO\\
\p^{t}&=\frac{\d L}{\d\dot A_{t}}=-\frac{1}{2\k^2_2}\sqrt{-\g}e^{-3\j-3\c}
\frac{R^2}{N}\g^{tt}F_{\r t}~,\NO\\
\p_\j&=\frac{\d L}{\d\dot\j}=-\frac{1}{\k^2_2}\sqrt{-\g}e^{-2\j}2K~,\NO\\
\p_\c&=\frac{\d L}{\d\dot\c}=-\frac{3}{2\k_2^2}\sqrt{-\g}\;e^{-2\j}\frac1N(\dot\c-N^t\pa_t\c)~.
\eal
The canonical momenta conjugate to $N$, $N_t$ and $A_\r$ vanish identically so these fields are non dynamical Lagrange multipliers. The Legendre transform of the Lagrangian \eqref{lagrangian} determines the Hamiltonian 
\be\label{Hamiltonian}
H=\int dt\left(\dot\g_{tt}\p^{tt}+\dot A_t\p^t+\dot\j\p_\j+\dot\c\p_\c\right)-L
=\int  dt\left(N\ch+N_t\ch^t+A_\r\cf\right)~,
\ee
where 
\bal
\label{constraints}	
\ch=&\;-\frac{\k^2_2}{\sqrt{-\g}}e^{2\j}\left(\g_{tt}\p^{tt}\p_\j+\frac{1}{R^2}e^{\j+3\c}\p^t\p_t+\frac13\p_\c^2\right)\NO\\
&\;-\frac{\sqrt{-\g}}{\k^2_2}\(\frac{1}{R^2}e^{3\j}-\frac{1}{4R^2}e^{5\j-3\c}+\frac{6}{\ell_5^2}e^{\j+\c}-\frac34\g^{tt}(\pa_t\c)^2-\square_t\)e^{-2\j}~,\NO\\
\ch^t=&\;-2D_t\p^{tt}+\p_\j\pa^t\j+\p_\c\pa^t\c~,\NO\\
\cf=&\;-D_t\p^t~.
\eal
Hamilton's equations for the Lagrange multipliers $N$, $N_t$ and $A_\r$ are the first class constraints
\be\label{constraints0}
\ch=\ch^t=\cf=0~,
\ee
which reflect the diffeomorphism invariance and $U(1)$ gauge symmetry of the bulk theory. As a result, the Hamiltonian \eqref{Hamiltonian} vanishes identically on the constraint surface, for any choice of the auxiliary fields $N$, $N_t$ and $A_\r$. In the subsequent analysis we will work in the Fefferman-Graham gauge \eqref{FG-gauge}, which corresponds to setting $N=1$, $N_t=0$, and $A_\r=0$. In this gauge, the expressions \eqref{momenta} for the canonical momenta can be inverted to obtain 
\bal
\label{H-eqs}
\dot\g_{tt}&=\frac{\d H}{\d\p^{tt}}=-\frac{\k^2_2}{\sqrt{-\g}}e^{2\j}\p_\j\g_{tt}~,\NO\\
%%%%%%%%%%%%%%%%%%%%%%%%%%%%%%%%%%%%%%%%%%%%%%%%%%%%%%%%%%%%%%%%%%%%%%%%%%%%%%%%%%%%
\dot A_t&=\frac{\d H}{\d\p^{t}}=-\frac{2\k_2^2}{\sqrt{-\g}}\frac{1}{R^2}e^{3\j+3\c}\p_t~,\NO\\
%%%%%%%%%%%%%%%%%%%%%%%%%%%%%%%%%%%%%%%%%%%%%%%%%%%%%%%%%%%%%%%%%%%%%%%%%%%%%%%%%%%%
\dot\j&=\frac{\d H}{\d\p_\j}=-\frac{\k_2^2}{\sqrt{-\g}}e^{2\j}\g_{tt}\p^{tt}~,\NO\\
%%%%%%%%%%%%%%%%%%%%%%%%%%%%%%%%%%%%%%%%%%%%%%%%%%%%%%%%%%%%%%%%%%%%%%%%%%%%%%%%%%%%
\dot\c&=\frac{\d H}{\d\p_\c}=-\frac{2\k_2^2}{3\sqrt{-\g}}e^{2\j}\p_\c~.
%%%%%%%%%%%%%%%%%%%%%%%%%%%%%%%%%%%%%%%%%%%%%%%%%%%%%%%%%%%%%%%%%%%%%%%%
\eal
These equations are half of all of Hamilton's equations. The other half are equations involving the radial derivative of the canonical momenta, derived by varying the Hamiltonian \eqref{Hamiltonian} with respect to the canonical coordinates. Together, all of Hamilton's equations are equivalent to the second order equations of motion \eqref{eoms-extremal-gf} obtained from the Lagrangian. We do not write Hamilton equations involving the radial derivative of the canonical momenta explicitly here
because they are represented differently in Hamilton-Jacobi theory which we develop in the following.  
 
 %%%%%%%%%%%%%%%%%%%%%%%%%%%%%%%%%%%%%%%%%%%%%%%%%%%%%%%%%%%%%%%%%%%%%%%%%
\subsection{Hamilton-Jacobi Formalism}
In the radial Hamiltonian language Hamilton's principal function $\cs[\g_{tt},\j,\c, A_t]$ is a functional of the canonical fields $\g_{tt},\j,\c,  A_t$ and their time derivatives, all evaluated at some fixed radial coordinate $\rho$ which we generally identify with the cutoff $\rho_c$. A defining property of this functional is that all canonical momenta can be expressed as gradients of the functional  $\cs$ with respect to their conjugate fields
\be\label{HJ-momenta-S}
\p^{tt}=\frac{\d\cs}{\d\g_{tt}}~,\quad \p^{t}=\frac{\d\cs}{\d A_{t}}~,\quad \p_\j=\frac{\d\cs}{\d\j}~,\quad \p_\c=\frac{\d\cs}{\d\c}~.
\ee
Bulk diffeomorphism invariance guarantees that $\cs$ depends on the cutoff $\r_c$ only through the canonical fields $\g_{tt},\j,\c,  A_t$. Together with the defining relations \eqref{HJ-momenta-S}, this implies that
\bal\label{reg-action-Sb}
\left.\cs\right|_{\r_c}  =& \int\limits_{\r_h}^{\r_c} d\r\;\int dt\left(\dot\g_{tt}\p^{tt}+\dot A_t\p^t+\dot\j\p_\j+\dot\c\p_\c\right)+ \left.\cs\right|_{\r_h}~,
\eal
where the reference point $\rho_h$ is introduced in order to fix the additive constant that is not specified by \eqref{HJ-momenta-S}. It will ultimately be identified with the position of a possible horizon. 

Hamilton's principal function is closely related to the on-shell value of the regularized action $I_{\rm reg}$. To see this we 
express the radial Lagrangian $L$ in terms of the Hamiltonian $H$ through the Legendre transform \eqref{Hamiltonian} 
and then impose the on-shell constraint $H=0$. Integrating the resulting expression for the Lagrangian with respect to $\rho$ gives the integral
on the right hand side of \eqref{reg-action-Sb}. However, the integral of the Lagrangian also gives the last term in the regularized action 
\eqref{reg-action} and so we find \cite{Cabo-Bizet:2017xdr} 
\bal\label{reg-action-Sa}
I\sbtx{reg} &= \left.\cs\right|_{\r_c}  + \;\frac{1}{2\k^2_2}\int\limits_{\r=\r_h}\hskip-0.15cm dt\sqrt{-\g}\;e^{-2\j}2K - \left.\cs\right|_{\r_h}~.
\eal
Thus the regularized on-shell action \eqref{reg-action} is almost identical to Hamilton's principal function; they differ at most by 
the surface terms at a possible horizon. Powerful methods of analytical mechanics that determine the functional $\cs$ therefore allow us
to find the regularized action. 

Inserting the canonical momenta in the form \eqref{HJ-momenta-S} into Hamilton's equations \eqref{H-eqs} we can express 
the radial derivatives of the canonical variables as a gradient flow generated by the principal function $\cs$
\bal
\label{flow-eqs-S}
\dot\g_{tt}&=-\frac{\k^2_2}{\sqrt{-\g}}e^{2\j}\g_{tt}\frac{\d\cs}{\d\j}~,\NO\\
%%%%%%%%%%%%%%%%%%%%%%%%%%%%%%%%%%%%%%%%%%%%%%%%%%%%%%%%%%%%%%%%%%%%%%%%%%%%%%%%%%%%
\dot A_t&=-\frac{2\k_2^2}{\sqrt{-\g}}\frac{1}{R^2}e^{3\j+3\c}\g_{tt}\frac{\d\cs}{\d A_{t}}~,\NO\\
%%%%%%%%%%%%%%%%%%%%%%%%%%%%%%%%%%%%%%%%%%%%%%%%%%%%%%%%%%%%%%%%%%%%%%%%%%%%%%%%%%%%
\dot\j&=-\frac{\k_2^2}{\sqrt{-\g}}e^{2\j}\g_{tt}\frac{\d\cs}{\d\g_{tt}}~,\NO\\
%%%%%%%%%%%%%%%%%%%%%%%%%%%%%%%%%%%%%%%%%%%%%%%%%%%%%%%%%%%%%%%%%%%%%%%%%%%%%%%%%%%%
\dot\c&=-\frac{2\k_2^2}{3\sqrt{-\g}}e^{2\j}\frac{\d\cs}{\d\c}~.
\eal
These first order equations are reminiscent of those satisfied by BPS solutions in supergravity. This analogy motivates 
 reference to $\cs$ as the ``effective superpotential''.

The Hamilton-Jacobi equations satisfied by Hamilton's principal function $\cs[\g_{tt},\j,\c, A_t]$ are obtained by inserting the expressions \eqref{HJ-momenta-S} for the canonical momenta into the first class constraints \eqref{constraints0}. In particular, the Hamiltonian constraint $\ch=0$ gives
\bal\label{HJ-S}
&-\frac{\k^2_2}{\sqrt{-\g}}e^{2\j}\left(\g_{tt}\frac{\d\cs}{\d\g_{tt}}\frac{\d\cs}{\d\j}+\frac{1}{R^2}e^{\j+3\c}\g_{tt}\(\frac{\d\cs}{\d A_{t}}\)^2+\frac13\(\frac{\d\cs}{\d\c}\)^2\right)~\NO\\
&\;-\frac{\sqrt{-\g}}{\k^2_2}\(\frac{1}{R^2}e^{3\j}-\frac{1}{4R^2}e^{5\j-3\c}+\frac{6}{\ell_5^2}e^{\j+\c}-\frac34\g^{tt}(\pa_t\c)^2-\square_t\)e^{-2\j}=0~.
\eal
It is a standard result of Hamilton-Jacobi theory that a complete integral of the Hamilton-Jacobi equation \eqref{HJ-S}, together with the general solution of the corresponding first order equations \eqref{flow-eqs-S}, are equivalent to the general solution of the second order equations of motion.

%%%%%%%%%%%%%%%%%%%%%%%%%%%%%%%%%%%%%%%%%%%%%%%%%%%%%%%%%%%%
%%%%%%%%%%%%%%%%%%%%%%%%%%%%
\subsection{General Solution to the Hamilton-Jacobi Equations}

The dependence of Hamilton's principal function $\cs$ on the gauge field can be determined once and for all due to the fact that the 
gauge field can be integrated out in two dimensions, as we saw in \eqref{eq:f2}. In the Hamiltonian formalism this can be seen from 
(\ref{Hamiltonian}-\ref{constraints}), which imply that 
\be\label{eq:pit1}
\dot{\p}^t =-\frac{\d H}{\d A_t}=0~.
\ee
Hence, the canonical momentum $\p^t$ is conserved and so the 2D gauge field is entirely captured by one quantum number, a ``charge". The expression \eqref{momenta} for the canonical momentum in terms of the 2D field strength, combined with our convention for the 2D electric 
charge $Q$ introduced in \eqref{eq:f2}, determine that
\be\label{Maxwell-momentum}
\p^t=-\frac{QR^2}{2\k^2_2}~.
\ee

A conserved quantity conjugate to a cyclic variable appears in the Hamilton-Jacobi formalism as a separation constant when 
separating variables in the Hamilton-Jacobi equation. Specifically, the normalization of momenta in  \eqref{HJ-momenta-S} 
shows that we can write Hamilton's principal function as   
\be\label{S-sol}
\cs[\g_{tt},\j,\c, A_t]=\cu[\g_{tt},\j,\c]+\int dt\;\Big(-\frac{QR^2}{2\k^2_2}\Big)A_t~,
\ee
where $\cu[\g_{tt},\j,\c]$ is a functional that is independent of the gauge field. The solution of the Hamilton-Jacobi equations therefore 
simplifies to computing the reduced principal function $\cu[\g_{tt},\j,\c]$, aka. the reduced effective superpotential. 

The system we consider is too complicated to solve completely in general. However, a recursive technique for solving the Hamilton-Jacobi equations asymptotically was developed in \cite{Papadimitriou:2011qb}, as a generalization of the dilatation operator method \cite{Papadimitriou:2004ap}. It relies on a covariant expansion in eigenfunctions of the functional operator
\be
\d_\g=\int dt\; 2\g_{tt}\frac{\d}{\d\g_{tt}}~,
\ee
namely,
\be\label{U-exp}
\cu=\cu_{(0)}+\cu_{(2)}+\cdots~,
\ee 
where the terms $\cu_{(2n)}$ satisfy $\d_\g \cu_{(2n)}=(d-2n)\cu_{(2n)}$. This is a covariant asymptotic expansion in the sense that $\cu_{(2n')}$ is asymptotically subleading relative to $\cu_{(2n)}$ for $n'>n$. In two dimensions this expansion coincides with an expansion in time derivatives, but this is not the case in general. 

In order to obtain the asymptotic solutions of the equations of motion \eqref{eoms-extremal-gf} and evaluate the renormalized on-shell action it is sufficient to determine only the first two terms in the covariant expansion \eqref{U-exp}. 
Covariance on the radial slice, imposed by the momentum constraint in \eqref{constraints}, and locality imply that $\cu_{(0)}$ and $\cu_{(2)}$ can be parameterized in general as
\bal\label{HJ-U}
\cu_{(0)}=&\;\frac{1}{\k_2^2}\int dt\sqrt{-\g}\;W(\j,\c)~,\NO\\
%%%%%%%%%%%%%%%%%%%%%%%%%%%%%%%%%%%%%%%%%%%%%%%%%%%%%%%%%%%%%%%%%%%%%%%%%%%%%%%%%%%%%%%%%%%%%%%
\rule{0cm}{0.8cm}\cu_{(2)}=&\;\frac{1}{\k_2^2}\int dt\sqrt{-\g}\Big(Z_1(\j,\c)\g^{tt}(\pa_t\j)^2+Z_2(\j,\c)\g^{tt}\pa_t\j\pa_t\c+Z_3(\j,\c)\g^{tt}(\pa_t\c)^2\Big)~,
\eal
where the functions $W(\j,\c)$, $Z_1(\j,\c)$, $Z_2(\j,\c)$ and $Z_3(\j,\c)$ are to be determined.
Inserting these general forms for $\cu_{(0)}$ and $\cu_{(2)}$ in the Hamilton-Jacobi equation \eqref{HJ-S} and matching terms of equal weight under $\d_\g$ leads to a system of equations for the functions $W(\j,\c)$, $Z_1(\j,\c)$, $Z_2(\j,\c)$ and $Z_3(\j,\c)$. We  
find that $W$ and $Z_3$ satisfy the system of equations
\bal\label{HJ-U-ansatz}
&\frac12 W\pa_\j W+\frac13(\pa_\c W)^2-\frac{Q^2R^2}{4}e^{\j+3\c}-\frac{1}{4R^2}e^{\j-3\c}+e^{-4\j}\Big(\frac{1}{R^2}e^{3\j}+\frac{6}{\ell_5^2}e^{\j+\c}\Big)=0~,\NO\\
%%%%%%%%%%%%%%%%%
&\rule{0cm}{0.8cm}	\frac43\pa_\c W\pa_\j\(\frac{Z_3}{W}\)+\pa_\c\[\frac{2e^{-4\j}}{W}+\(\frac{4\pa_\c W}{3W}\)^2Z_3\]=0~,
\eal
while the remaining functions $Z_1$ and $Z_2$ can be expressed in terms of $W$ and $Z_3$ as
\be\label{Z1Z2}
Z_1=\frac{2e^{-4\j}}{W}+\(\frac{4\pa_\c W}{3W}\)^2Z_3~,\qquad
Z_2=-\frac{8\pa_\c W}{3W}Z_3~.
\ee

In principle, the two coupled equations \eqref{HJ-U-ansatz}, together with \eqref{Z1Z2}, solve the dynamical problem completely up to second order in time 
derivatives, because the linear flow equations \eqref{flow-eqs-S} then determine the solutions of the equations of motion \eqref{eoms-extremal-gf}. An {\em exact} solution of the Hamilton-Jacobi equation \eqref{HJ-S}, i.e. valid to all orders in time derivatives and throughout the RG flow, is presented in  
appendix \ref{UV-analysis-flat} for the case $\ell_5^{-1}=0$ and $\c$ constant.
However, since the equations \eqref{HJ-U-ansatz} are nonlinear, in general we must resort to perturbation theory. Our primary interest is perturbation theory around the IR fixed point, developed in the following subsection. The solution of \eqref{HJ-U-ansatz} in the UV, i.e. far away from the IR fixed point, is discussed in appendix \ref{UV-analysis-AdS} where it is compared with the well known solution of the radial Hamilton-Jacobi equation for pure AdS$_5$ gravity. This comparison allows us to determine also the four-derivative term $\cu_{(4)}$ near the UV.

%%%%%%%%%%%%%%%%%%%%%%%%%%%%%%%%%%%%%%%%%%%%%%%%%%%%%%%%%%%%%%%%%%%%%%%%%
\subsection{Effective Superpotential for Near IR Solutions}\label{sec:effIR}

In this subsection we solve the two equations \eqref{HJ-U-ansatz} near the IR fixed point. We verify that the corresponding flow equations \eqref{flow-eqs-S} lead to the perturbative near IR solutions previously obtained in subsection \ref{sec:IRperturbations} using Lagrangian 
methods. Importantly, the covariant form of the asymptotic solution obtained here also determines the boundary counterterms necessary to holographically renormalize the theory. This application is the subject of the next section. 

A solution of the two equations \eqref{HJ-U-ansatz} near the IR fixed point can be sought in the form of a Taylor expansion around the constant scalar values $\j_0$ and $\c_0$ at the IR fixed point, exhibited in \eqref{eqn:psi0eqn} and \eqref{eqn:chi0eqn}. We denote the deviations of the scalar fields $\j$ and $\c$ away from their IR fixed point values by $\cy$ and $\cx$, respectively, as in 
\eqref{eqn:cycxdef}. This gives 
\be
\label{eqn:cycxpert}
e^{-2\j}=e^{-2\j_0}+\cy~,\qquad \pa_\j=-2(e^{-2\j_0}+\cy)\pa_\cy~,\qquad \c=\c_0+\cx~,\qquad \pa_\c=\pa_\cx~.
\ee
Using these identities and inserting the Taylor expansion 
\be\label{W-exp}
W\sbtx{pert}=w_{00}+w_{10}\cy+w_{01}\cx+w_{20}\cy^2+w_{11}\cy\cx+w_{02}\cx^2+\cdots~,
\ee
in the first equation in \eqref{HJ-U-ansatz} we determine
\bal
\label{W-sol}
	\begin{aligned}
		&w_{00}=w_{01}=0~,\quad w_{10}=\frac{1}{\ell_2}~,\quad w_{11}=\frac{3q}{1+2q}\frac{(\D_\c-2)}{\ell_2}~,\quad 
		w_{02}=\frac{3e^{-2\j_0}(1-\D_\c)}{4\ell_2}~,\\
		&w_{20}=-\frac{e^{2\j_0}}{\ell_2(1+2q)^2}\(\frac{(1+8q)(1+4q-16q^2)}{2(1+12q)}+3q^2\D_\c\)~,
	\end{aligned}
\eal
where $\ell_2$ and $q$ are defined in \eqref{eq:l2} and \eqref{eq:q}, respectively, and $\D_\c$ is given in \eqref{chi-dim}. The overall sign of $W\sbtx{pert}$ is not determined by the equations \eqref{HJ-U-ansatz}, but can be fixed through the first order flow equations by demanding that the leading asymptotic form of the solutions matches that of the near IR solutions in subsection \ref{sec:IRperturbations} (see \eqref{eqn:focycx} below). The perturbative solution $W\sbtx{pert}$ given in \eqref{W-exp} with coefficients \eqref{W-sol} is a particular solution for the function $W(\phi,\chi)$ near the IR fixed point. 

The equation for $W(\phi,\chi)$ in \eqref{HJ-U-ansatz} is first order in derivatives with respect to both fields $\phi$ and $\chi$, and so a complete integral of this equation must contain two integration constants. Of course, since $W$ satisfies a partial differential equation, the general solution for $W$ contains an arbitrary function. However, a complete integral, i.e. a special two-parameter family of solutions, suffices for obtaining the general solution of the equations of motion. An important caveat to this statement is that typically it holds only locally in configuration space. In particular, although a complete integral suffices to obtain the general solution of the equations of motion in a specific neighborhood of configuration space, a different complete integral may be necessary for another neighborhood.

The perturbative solution $W\sbtx{pert}$ given in \eqref{W-sol} does not contain any integration constants and so is uniquely determined. A complete integral in the neighborhood of configuration space defined by $W\sbtx{pert}$ can be obtained by finding a two-parameter family of small deformations around the perturbative solution \eqref{W-exp}. Inserting $W=W\sbtx{pert}+\D W$ with $\D W$ small relative to $W\sbtx{pert}$ into \eqref{HJ-U-ansatz} we find that $\D W$ satisfies the linear equation 
\be\label{W-def}
\pa_\j(W\sbtx{pert}\D W)+\frac43\pa_\c(W\sbtx{pert})\pa_\c \D W=0~.
\ee
However, using the solution $W\sbtx{pert}$ in \eqref{W-exp} we find the {\em three}-parameter family of small deformations
\be\label{delta-W}
\D W=c_0\frac{\ell_2}{2}\Big(\cy^{-1}+\co(1)\Big)+c_1\Big(\cx^{-\frac{1}{\D_\c-1}}+\co(1)\Big)+c_2\Big(\frac{\cx^2}{\cy^2}+\frac{4\ell_2e^{2\j_0}w_{11}}{3(\D_\c-1)}\frac{\cx}{\cy}+\co(1)\Big)~,
\ee
where $c_0$, $c_1$ and $c_2$ are arbitrary integration constants and the ellipses again denote terms subleading in the fluctuations around the IR fixed point. The first two terms in \eqref{delta-W} are similar in nature, as we see by recalling that the dilaton $\psi$, represented by the fluctuation $\cy$, has dimension $\D_\j=2$. We will see shortly that $c_0$ is the same constant that was introduced from a Lagrangian point of view in \eqref{eq:s2} when solving the equations of motion near the IR fixed point. $c_1$ is then an analogue for the fluctuation $\cx$. The role of the integration constant $c_2$ is less clear at this point, but we will see below that its value is uniquely determined by requiring that $W=W\sbtx{pert}+\D W$, with $\D W$ given in \eqref{delta-W}, is a complete integral for the near IR solutions obtained in subsection \ref{sec:IRperturbations}.

It is interesting that the family of small deformations \eqref{delta-W} is non perturbative in the field fluctuations $\cy$ and $\cx$, which is why it was not found using the Taylor expansion \eqref{W-exp}. As we will see shortly, through the first order flow equations, the perturbative terms $W\sbtx{pert}$ determine the sources for the system while the non perturbative terms $\D W$ are related to the vacuum expectation values, i.e. the one-point functions. 

In summary, in this subsection we have found that to quadratic order in fluctuations away from the IR fixed point the solution for 
the $\cu_{(0)}$ takes the form
\be\label{U0-sol}
	\cu_{(0)}=\frac{1}{\k_2^2}\int dt\sqrt{-\g}\;\Big(\D W+w_{10}\cy+w_{20}\cy^2+w_{11}\cy\cx+w_{02}\cx^2+\cdots\Big)~,
\ee
where $\D W$ is given in \eqref{delta-W}. Inserting $W$, the integrand of this solution for $\cu_{(0)}$, into the second equation in \eqref{HJ-U-ansatz} we find 
\be
Z_3=-\frac{3\ell_2e^{-2\j_0}}{4(2\D_\c-3)}+\cdots~,
\ee
where the ellipses denote terms that are higher order in fluctuations away from the IR fixed point. The expressions \eqref{Z1Z2} for $Z_1$ and $Z_2$ then determine that to quadratic order in the fluctuations around the IR fixed point
\bal\label{U2-sol}
	\cu_{(2)}=&\;\frac{1}{\k_2^2}\int dt\sqrt{-\g}\Bigg(-\Big(\frac{3\ell_2e^{-2\j_0}}{4(2\D_\c-3)}+\co(\cy,\cx)\Big)\g^{tt}(\pa_t\cx)^2\NO\\
	&\hskip-0.5cm+\frac{\ell_2}{(2\D_\c-3)}\Big(\frac32(\D_\c-1)e^{-2\j_0}\frac{\cx}{\cy}-\ell_2w_{11}+\co(\cy,\cx)\Big)\g^{tt}\pa_t\cx\pa_t\cy\\
	&\hskip-0.5cm-\frac{\ell_2}{2}\Bigg(\frac{3e^{-2\j_0}(\D_\c-1)}{4(2\D_\c-3)}\frac{\cx^2}{\cy^2}-\frac{\ell_2w_{11}}{(2\D_\c-3)}\frac{\cx}{\cy}+\ell_2\Big(w_{20}+\frac{2\ell_2e^{2\j_0}w_{11}^2}{3(2\D_\c-3)}\Big)-\frac{1}{\cy}+\co(\cy,\cx)\Bigg)\g^{tt}(\pa_t\cy)^2\Bigg)~.\NO
\eal
The expressions (\ref{U0-sol},\,\ref{U2-sol}) give the reduced effective superpotential $\cu$ in \eqref{U-exp} to second order in time derivatives and to quadratic order in the fluctuations 
near the IR fixed point. Hamilton's principal function $\cs[\g_{tt},\j,\c, A_t]$ then follows from \eqref{S-sol}, by adding the contribution from the gauge field.

Having determined Hamilton's principal function, we can now use the relations \eqref{flow-eqs-S} to obtain the corresponding first order flow equations for the fluctuations of the fields. For example, for the scalar fluctuations $\cy$ and $\cx$ we obtain 
\bal
\dot\cy=&\;\frac{1}{\ell_2}\cy-\frac{\ell_2}{2}\cy^{-1}\g^{tt}(\pa_t\cy)^2+\frac{\ell_2c_0}{2\cy}+c_1\cx^{-\frac{1}{\D_\c-1}}+\cdots,\NO\\
%%%%%%%%%%%%%%%%%%%%%%%%%%%%%%%%%%%%%%%%%%%%%%%%%%%%%%%%%%%%%%%%%%%%%%%%%%%%%%%%%%%%%%%%%%%%%%%
\rule{0.cm}{0.8cm}
-\frac32e^{-2\j_0}\dot\cx=&\;w_{11}\cy+2w_{02}\cx+c_2\Big(\frac{2\cx}{\cy^2}+\frac{4\ell_2e^{2\j_0}w_{11}}{3(\D_\c-1)}\frac{1}{\cy}\Big)-\frac{c_1}{(\D_\c-1)}\cx^{-\frac{1}{\D_\c-1}-1}\NO\\
&+\frac{\ell_2}{2}\Big(\frac{3e^{-2\j_0}(\D_\c-1)}{2(2\D_\c-3)}\frac{\cx}{\cy^2}+\frac{\ell_2w_{11}}{(2\D_\c-3)}\frac{1}{\cy}\Big)\g^{tt}(\pa_t\cy)^2\NO\\
&+\frac{3\ell_2e^{-2\j_0}}{2(2\D_\c-3)}\square_t\cx+\Big(-\frac{3\ell_2e^{-2\j_0}(\D_\c-1)}{2(2\D_\c-3)}\frac{\cx}{\cy}+\frac{\ell_2^2w_{11}}{(2\D_\c-3)}\Big)\square_t\cy+\cdots~.
\label{eqn:focycx}
\eal
Integrating this system of first order equations we find that the solution for $\cx$ is of the form $\cx=\cx\sbtx{hom}+\cx\sbtx{inhom}$. The inhomogeneous solution,  $\cx\sbtx{inhom}$, is given in \eqref{eq:x1}, and in this context it corresponds to a solution of \eqref{eqn:focycx}
provided the integration constant $c_2$ is related to $c_0$ as
\be\label{eqn:c2}
c_2=-\frac{3\ell_2 e^{-2\j_0}(\D_\c-1)}{8(2\D_\c-3)}c_0~.
\ee
In particular, for this value of $c_2$ and $c_1=0$, setting $\cx=\cx\sbtx{inhom}$ with $\cx\sbtx{inhom}$ given in \eqref{eq:x1} the first order equation for $\cx$ in \eqref{eqn:focycx} reduces to a multiple of the first order equation for $\cy$. Moreover, the homogeneous solution for $\cx$ takes the form\footnote{The normalizable mode in the homogeneous solution \eqref{eq:X-hom} is in fact not the most general allowed by the linearized equations of motion in  \eqref{eoms-near-extremal-gf}. In the general solution of the linearized equations of motion the normalizable mode of the homogeneous solution for $\cx$ is non local in time derivatives, leading to a non trivial two-point function for the operator dual to $\cx$. However, by writing \eqref{HJ-U} for the function $\cu$ we sought the solution in a derivative expansion, which is why we find the special homogeneous solution \eqref{eq:X-hom}. Of course, we will see in the next section that a derivative expansion for $\cu$ is sufficient for determining the boundary terms required to renormalize the theory.}
\be\label{eq:X-hom}
\cx\sbtx{hom}=\z(t)e^{(\D_\c-1)\r/\ell_2}(1+\cdots)-\frac{2\ell_2e^{2\j_0}}{3(\D_\c-1)(2\D_\c-1)}c_1\z(t)^{\frac{\D_\c}{1-\D_\c}}e^{-\D_\c\r/\ell_2}(1+\cdots)~.
\ee
In subsection \ref{sec:IRperturbations} we omitted the homogeneous solution for $\cx$, for brevity, so the integration ``constant'' $\z(t)$ did not appear previously. It will soon be identified with the independent source of the scalar operator dual to $\c$.

The solution of \eqref{eqn:focycx} for the dilaton fluctuation $\cy$ is the perturbative solution \eqref{eq:s1} written in terms of the dilaton source $\nu(t)$ and the function $\vth(t)$ that agrees with \eqref{eq:s2}, except that there is now a term for the independent source $\z(t)$:
\be\label{eq:s2-full}
\vth(t)=-\frac{\ell_2^2}{4\n}\(c_0+\frac{(\partial_t\n)^2}{\a^2}\)-\frac{\ell_2}{2}c_1\z^{\frac{1}{1-\D_\c}}~.
\ee
In particular, the integration constant $c_0$ in the solution \eqref{U0-sol} is the same constant that was introduced in subsection \ref{sec:IRperturbations}, as promised. Moreover, we should point out that although the full homogeneous solution \eqref{eq:X-hom} for $\cx$ can be obtained by solving the corresponding linearized equation in \eqref{eoms-near-extremal-gf}, the backreaction of $\cx$ on $\cy$, which corresponds to the term involving $\z(t)$ in \eqref{eq:s2-full}, goes beyond the linearized approximation of the equations of motion \eqref{eoms-extremal-gf}, which is why this term was not seen when solving the linearized equations.

Finally, the leading order metric was introduced in \eqref{eqn:leading} in terms of the source $\alpha(t)$ and the function $\beta(t)$. The expression for $\b(t)$ given in \eqref{eq:s2} does not get modified in the presence of $\z(t)$. The flow equation for the metric fluctuation can also be obtained from \eqref{flow-eqs-S}. It reproduces the solution \eqref{eq:s3}, except for additional terms related to the independent source $\z(t)$ turned on by the homogeneous solution $\cx\sbtx{hom}$, namely
\be\label{eq:metric-fluct-zeta}
\sqrt{-\g_1}=-\frac{(1+10q+8q^2)}{(1+2q)(1+12q)}e^{2\psi_0}\le[\sqrt{-\g_0}\;\cy+{2\ell^2_2}\partial_t\le(\frac{\partial_t\n}{\a}\ri)\ri]+\frac{3q(\D_\c-2)}{(1+2q)(\D_\c-1)}\sqrt{-\g_0}\;\z e^{\frac{(\D_\c-1)\r}{\ell_2}}+\cdots~.
\ee

%%%%%%%%%%%%%%%%%%%%%%%%%%%%%%%%%%%%%%%%%%%%%%%%%%%%%%%%%%%%
%%%%%%%%%%%%%%%%%%%%%%%%%%%%%%%%%%%%%%%%%%%%%%%%%%%%%%%%%%%%

\section{Holographic Renormalization}
\label{dictB}

Holographic renormalization is best understood as a canonical transformation on the space of fields, here $\g_{tt},\j,\c, A_t$, and their conjugate radial momenta \cite{Papadimitriou:2010as}. This canonical transformation is generated by a specific boundary  term that renders the variational problem well posed. A well defined variational principle automatically ensures that the corresponding on-shell action is finite \cite{Papadimitriou:2005ii}. In contrast, a boundary term that leads to a finite on-shell action is not necessarily compatible with the symplectic structure of the theory and may not lead to a well posed variational problem. Moreover, as we will see below, there are cases where certain boundary terms do not contribute to the on-shell action, but are nevertheless necessary for the renormalization of the canonical variables. 

The boundary terms required to render the variational problem well posed can be determined by solving the radial Hamilton-Jacobi equation \cite{Papadimitriou:2010as}. In the present context, the boundary terms can therefore be obtained from Hamilton's principal function $\cs$ in \eqref{S-sol}, where the reduced principal function $\cu$ is given in (\ref{U0-sol},\,\ref{U2-sol}). Specifically, only the perturbative solution for $\cs$ near the IR fixed point is required, since this solution for $\cs$ controls the asymptotic behavior of the fields near the IR fixed point. 

There are two important subtleties in the holographic renormalization of nAdS$_2$ backgrounds which we need to address before delving into the structure of the renormalized theory. The first concerns the special treatment required by gauge fields, which was also discussed extensively in \cite{Cvetic:2016eiv}; see also \cite{GonzalezGrumillerSalzer2018}. The second subtlety is related to the fact that we are interested in the effective action near an IR fixed point, without making reference to any possible UV completion.\footnote{For finite AdS$_5$ radius the UV completion is provided by pure 5D gravity, which is dual to a subsector of $\cn=4$ super Yang-Mills theory in four dimensions. Similarly, the UV completion of the 2D model considered in \cite{Cvetic:2016eiv} was provided by pure AdS$_3$ gravity and its dual CFT$_2$. However, our present analysis is intended to address both asymptotically flat and asymptotically AdS$_5$ Kerr black holes, which is why we focus exclusively on the effective theory near the IR, without reference to any UV completion.} 
This requires a UV cutoff and involves conformal perturbation theory in the presence of irrelevant couplings. The treatment of irrelevant deformations in the context of holographic renormalization was first discussed in \cite{vanRees:2011fr}.

%%%%%%%%%%%%%%%%%%%%%%%%%%%%%%%%%%%%%%%%%%%%%%%%%%%%%%%%%%%%%%%%%%%%%%%%%%%%%%%%%%%%%%%%	
\subsection{The Gauge Field in AdS$_2$}

The subtlety in the holographic renormalization of gauge fields in nAdS$_2$ amounts to the fact that the canonical transformation, and hence the boundary term, required to render the variational problem well posed is qualitatively different from those typically arising in higher dimensions. The reason for this is that the conserved electric charge dominates the asymptotic behavior of a gauge field in nAdS$_2$, as can be seen from the IR fixed point solution \eqref{eqn:leading}. This is in contrast to the more familiar asymptotic behavior of Maxwell fields in AdS$_{d+1}$ with $d\geq 3$, which is dominated by the chemical potential. The conserved charge also dominates the asymptotic behavior of Maxwell fields in AdS$_3$ \cite{Jensen:2010em}, and more generally of $p$-form fields in AdS$_{d+1}$ with $p\geq [d/2]$ \cite{thermo2}.

In order to render the variational problem well posed it is necessary to identify the canonical transformation that diagonalizes the symplectic map from the space of fields and momenta, here parameterized by $A_t$ and $\p^t$, to the space of asymptotic solutions, here parameterized by the conserved charge $Q$ and the chemical potential $\m(t)$ in \eqref{eqn:leading} \cite{Papadimitriou:2010as}. Since the canonical momentum $\p^t$ is proportional to the conserved charge $Q$, the required canonical transformation need only modify the gauge potential $A_t$. Taking also into account that the electric charge is the leading term in two dimensions, it takes the form
\be\label{can-trans}
\(\begin{matrix}
A_t \\ \p^t
\end{matrix}\)\to \(\begin{matrix}
 - \p^t \\ A_t^{\rm ren}
\end{matrix}\),
\ee
where $A_t^{\rm ren}$ is the canonically transformed gauge field. Moreover, in order for the canonical transformation to diagonalize the aforementioned symplectic map, $A_t^{\rm ren}$ must be asymptotically proportional to the chemical potential $\m(t)$. 

The canonical transformation \eqref{can-trans} is generated by a boundary term of the form     
\be\label{bt1}
I_b=-\int dt\;\p^t A_t+I_{c}[\g_{tt},\j,\c,\p^t]~,
\ee 
where $I_{c}[\g_{tt},\j,\c,\p^t]$ is a yet undetermined local functional of its arguments. Note that the first term in \eqref{bt1} implements a Legendre transform on the gauge field. Such a Legendre transform for gauge fields in AdS$_2$ has been considered before in various contexts, including the quantum entropy functional on AdS$_2$ \cite{Sen:2008vm}, as well as dilaton-gravity models in \cite{Grumiller:2014oha}. Adding the boundary term \eqref{bt1} and the Gibbons-Hawking term \eqref{GH} to the 2D bulk action \eqref{2Daction-extremal} results in the (on-shell) variational principle
\be\label{var1}
\d(I\sbtx{reg}+I_b)=\int dt\;\big(\p_{\rm ren}^{tt}\d\g_{tt}+\p^{\rm ren}_\j\d\j+\p^{\rm ren}_\c\d\c-A^{\rm ren}_t\d\p^t\big)~,
\ee
where the renormalized (i.e. canonically transformed) variables are given by 
\be\label{ren-vars}
\p_{\rm ren}^{tt}=\p^{tt}+\frac{\d I_c}{\d \g_{tt}}~,\qquad \p^{\rm ren}_\j=\p_\j+\frac{\d I_c}{\d\j}~,\qquad \p^{\rm ren}_\c=\p_\c+\frac{\d I_c}{\d\c}~,\qquad A_t^{\rm ren}=A_t-\frac{\d I_c}{\d\p^t}~,
\ee
while their canonical conjugates are not transformed.

Having established that the appropriate boundary term is of the form \eqref{bt1}, it remains to determine the functional $I_{c}[\g_{tt},\j,\c,\p^t]$ in a covariant expansion near the IR fixed point. The dependence of $I_{c}[\g_{tt},\j,\c,\p^t]$ on $\p^t$ in the vicinity of the IR fixed point can be deduced from the IR solution \eqref{eqn:leading} and its correction following from the linearized perturbations in subsection \ref{sec:IRperturbations}. Since the leading asymptotic behavior of the renormalized gauge field $A^{\rm ren}_t$ must be proportional to the chemical potential $\m(t)$, the term $-\d I_c/\d\p^t$ in the expression for $A_t^{\rm ren}$ in \eqref{ren-vars} must cancel the term proportional to the charge in the IR solution \eqref{eqn:leading}. Using the value of the momentum $\p^t$ given in terms of the electric charge in \eqref{Maxwell-momentum}, we can express the leading asymptotic behavior of the gauge potential $A_t$ in \eqref{eqn:leading} as
\be\label{At-asymptotic}
A_t\sim \frac{2\k_2^2\ell_2}{R^2}e^{3\j_0+3\c_0}\sqrt{-\g}\;\p^t~.
\ee
In order for $A_t^{\rm ren}\sim \m(t)$ near the IR fixed point, therefore, $I_c$ must satisfy 
\be
\frac{\d I_c}{\d \p^t}\sim \frac{2\k_2^2\ell_2}{R^2}e^{3\j_0+3\c_0}\sqrt{-\g}\;\p^t~.
\ee
Integrating this determines that to leading order asymptotically
\be\label{ct-vector-leading}
I_c\sim\int dt\sqrt{-\g}\;\frac{\k_2^2\ell_2}{R^2}e^{3\j_0+3\c_0}\Big((\p^t)^2-\frac{Q^2R^4}{4\k_2^4}\Big)+I'_c[\g_{tt},\j,\c]~,
\ee
for some functional $I'_c[\g_{tt},\j,\c]$ that does not depend on $\p^t$. The integration constant was determined so that on-shell $I_c$ and $I_c'$ coincide on-shell due to the identity \eqref{Maxwell-momentum}. Thus the term in the parenthesis in \eqref{ct-vector-leading} is an example of a boundary counterterm that vanishes on-shell but is nevertheless crucial for renormalizing the canonical variables and rendering the variational problem well posed. This illustrates the fact that it is the variational problem that dictates the correct boundary terms, and not the divergences of the on-shell action.

The expression \eqref{ct-vector-leading} for $I_c$ holds only to leading asymptotic order near the IR fixed point, since it was obtained through the leading asymptotic form of the gauge field in \eqref{At-asymptotic}. Perturbations away from the IR fixed point lead to additional terms in $I_c$, whose form can be parameterized as
\be\label{ct-vector}
I_c[\g_{tt},\j,\c,\p^t]=\int dt \sqrt{-\g}\; \cg[\g_{tt},\j,\c]\Big((\p^t)^2-\frac{Q^2R^4}{4\k_2^4}\Big)+I_c'[\g_{tt},\j,\c]~,
\ee
where $\cg[\g_{tt},\j,\c]$ is a local function of its arguments and their time derivatives. An equation for this function can be derived by demanding that asymptotically $A_t$ coincides with $\d I_c/\d\p^t$, i.e.  
\be
A_t=\frac{\d I_c}{\d\p^t}=2\p^t \sqrt{-\g}\;\cg[\g_{tt},\j,\c]~.
\ee
Since $\p^t$ is a constant, taking the derivative with respect to $\r$ on both sides of this relation gives 
\be
\dot A_t=2\p^t\Big(\sqrt{-\g}\;K\cg+\int dt\sqrt{-\g}\; \Big(2K_{tt}\frac{\d\cg}{\d\g_{tt}}+\dot\j\frac{\d\cg}{\d\j}+\dot\c\frac{\d\cg}{\d\c}\Big)\Big)~,
\ee
where the extrinsic curvature $K_{tt}$ was defined in \eqref{eq:K}. Using  \eqref{S-sol} and substituting the first order equation for $A_t$ in \eqref{H-eqs} and the first order equations for $\g_{tt}$, $\j$ and $\c$ in \eqref{flow-eqs-S} leads to the functional differential equation 
\be\label{eq:G-eq}
\frac12e^{2\j}\frac{\d\cu}{\d\j}\cg+\int dt\;e^{2\j}\Big(\g_{tt}\frac{\d\cu}{\d\j}\frac{\d\cg}{\d\g_{tt}}+\g_{tt}\frac{\d\cu}{\d\g_{tt}}\frac{\d\cg}{\d\j}+\frac23\frac{\d\cu}{\d\c}\frac{\d\cg}{\d\c}\Big)+\frac{1}{R^2}{\sqrt{-\g}}\;e^{3\j+3\c}=0~. 
\ee
Using the near IR solution for $\cu$ in (\ref{U0-sol},\ref{U2-sol}), this equation determines the near IR expansion of the function $\cg$ to any desired order. To first subleading order we find  
\be\label{G-sol}
\cg=\frac{\k_2^2\ell_2}{R^2}e^{3\j_0+3\c_0}\Big[1-\Big(\frac34 e^{2\j_0}+\ell_2 w_{20}+\frac{e^{2\j_0}\ell_2 w_{11}(\ell_2 w_{11}-3)}{3\D_\c}\Big)\cy+\frac{3-\ell_2 w_{11}}{\D_\c}\cx+\cdots\Big]~,
\ee
where the constants $w$ are given in \eqref{W-sol} and the ellipses denote asymptotically subleading terms near the IR fixed point. As we will see in the next subsection, the terms shown in \eqref{G-sol} suffice in order to renormalize the gauge field $A_t$ and so we need not determine any higher order terms. Note that the leading term in \eqref{G-sol} coincides with the leading asymptotic expression in \eqref{ct-vector-leading}, as required. 

%%%%%%%%%%%%%%%%%%%%%%%%%%%%%%%%%%%%%%%%%%%%%%%%%%%%%%%%%%%%%%%%%%%%%%%%%%%%%%%%%%%%%%%%

\subsection{Conformal Perturbation Theory}

Finally, we need to determine the form of the functional $I'_c[\g_{tt},\j,\c]$ in \eqref{ct-vector} near the IR fixed point. As we now show, this functional must agree asymptotically with $-\left.\cu\right|_{\r_c}$, where $\cu$ is the effective superpotential introduced in \eqref{S-sol}, and whose asymptotic form we determined in the previous section by solving the Hamilton-Jacobi equation. To see this we observe that adding the boundary term \eqref{bt1} to the regularized action \eqref{reg-action-Sa} and using  \eqref{S-sol} gives
\be
\label{eqn:iregib}
I\sbtx{reg}+I_b=\left.\cu\right|_{\r_c}+I_c - \int\limits_{\rho=\r_c}\hskip-0.15cm dt\; \Big(\pi^t + {QR^2\over 2\kappa^2_2}\Big) A_t + I\sbtx{global}~,
\ee
where 
\be\label{Sglobal}
I\sbtx{global}=\frac{1}{2\k^2_2}\int\limits_{\r=\r_h}\hskip-0.15cm dt\sqrt{-\g}\;e^{-2\j}2K-\left.\cu\right|_{\r_h}-\int\limits_{\rho=\r_h}\hskip-0.15cm dt\;\Big(-\frac{QR^2}{2\k^2_2}\Big)A_t~,
\ee
accounts for contributions from a possible horizon. The coefficient of the gauge field in \eqref{eqn:iregib} vanishes identically on-shell due to \eqref{Maxwell-momentum} and so it does not contribute to the divergences of the on-shell action. It follows that the counterterm $I_c[\g_{tt},\j,\c,\p^t]$, and hence $I'_c[\g_{tt},\j,\c]$ since the first term in \eqref{ct-vector} also vanishes identically on-shell, must asymptotically coincide with the effective superpotential $-\left.\cu\right|_{\r_c}$.

The asymptotic form of the effective superpotential $\cu$ near the IR fixed point was determined in the previous section and is given in \eqref{U0-sol} and \eqref{U2-sol}. Not all terms in the solution for $\cu$ should be included in the counterterms $I'_c[\g_{tt},\j,\c]$, however. In the more familiar situation where irrelevant deformations are absent, the divergent terms of the on-shell action are local, i.e. analytic in the fields and polynomial in boundary derivatives, and the divergences of the on-shell action are in one to one correspondence with the divergences of the one- and higher-point functions. In those cases only the local and divergent terms in the solution of the Hamilton-Jacobi equation should be included in the counterterms. However, in the presence of irrelevant deformations both of these properties cease to hold in general and identifying the terms that should be included in the counterterms is more subtle. 

First, the notion of ``divergent'' in this context must take into account that we deform away from the IR fixed point by two irrelevant operators with couplings $\sim e^{2\j_0}\n(t)$ and $\sim \z(t)$, respectively. On the dual conformal quantum mechanics side the appropriate formalism for dealing with this situation is conformal perturbation theory, which has a well defined analogue in the bulk (see e.g. \cite{vanRees:2011fr,Korovin:2013bua} for other examples of conformal perturbation theory in holography). Namely, we need to introduce a UV cutoff at $\r=\r_c$ and work with irrelevant couplings $\n(t)$ and $\z(t)$ that satisfy
\be\label{UV-cutoff}
e^{2\j_0}|\n(t)|e^{\r_c/\ell_2} \ll 1~,\qquad |\z(t)|e^{(\D_\c-1)\r_c/\ell_2} \ll 1~, 
\ee 
i.e. each of these numbers are kept small even for large UV cutoff $\rho_c$. Therefore, the divergent terms are those that grow faster, as $\r_c\to\infty$, than $e^{n\r_c/\ell_2}$ at $\co(\n^n)$ in perturbation theory and/or $e^{m(\D_\c-1)\r_c/\ell_2}$ at order ${\cal O}(\z^m)$. There are typically an infinite number of such terms, but only a finite number at each order in the irrelevant couplings. 

Second, in the presence of irrelevant couplings, removing the divergences of the on-shell action does not ensure that higher-point functions are finite. In particular, the more insertions of an irrelevant operator there are in a correlation function, the more terms need to be included in the boundary counterterms to cancel the divergences in the correlation function. This is another example of a situation where vanishing terms in the on-shell action are required in order to cancel divergences in higher-point functions, the divergent terms cannot be identified from the on-shell action alone. The divergences of correlation functions must be considered as well. In the subsequent analysis we are interested in renormalizing the on-shell action and all one-point functions, and so we will identify the terms in $\cu$ that contribute to divergences in these observables only.   

Starting from the on-shell action, the counting of divergences in conformal perturbation theory is precisely such that the fluctuations $\cy$, $\cx$ defined in \eqref{eqn:cycxpert}
are treated as finite and small, due to the restrictions \eqref{UV-cutoff}. The perturbative terms in the asymptotic solution that were collected in $W_{\rm pert}$ \eqref{W-exp} are therefore all finite. However, because the effective superpotential $\cu_{(0)}$ defined in \eqref{HJ-U} includes an overall factor of the volume measure $\sqrt{-\gamma}$, the corresponding terms in $\cu_{(0)}$ are all divergent. In contrast, the two-derivative terms in $\cu_{(2)}$, given in \eqref{U2-sol}, are all finite, because there the volume divergence is compensated by additional factors of $\gamma^{tt}$. This argument generalizes to higher derivative terms in $\cu$, i.e. $\cu_{(2k)}$ with $k>1$, none of which contain divergent terms in the sense of conformal perturbation theory. The non perturbative terms $\Delta W$ in \eqref{delta-W} also make a finite contribution to $\cu_{(0)}$  because their coefficients $c_0,\, c_1$ are respectively $\co(\n^2)$ and $\co(\z^{\D_\c/(\D_\c-1)})$ in perturbation theory, which is sufficient to compensate for the divergent volume measure. We conclude that the counterterm $I'_c$ must include all terms in $\cu_{(0)}$ that are contained in the perturbative expansion $W_{\rm pert}$ \eqref{W-exp}. That is, up to quadratic terms in the fluctuations 
\be\label{ct-prelim}
I'_c=-\frac{1}{\k_2^2}\int dt\sqrt{-\g}\;\big(w_{10}\cy+w_{20}\cy^2+w_{11}\cy\cx+w_{02}\cx^2\big)~.
\ee

The counterterms \eqref{ct-prelim} cancel the divergences of the on-shell action up to second order in conformal perturbation theory. As it turns out they also suffice in order to renormalize the one-point functions of the operators dual to the fields $\j$ and $\g_{tt}$. However, additional boundary terms are required to cancel the divergences of the one-point function of the operator dual to $\c$. The full set of boundary counterterms necessary to renormalize the on-shell action and all one-point functions is obtained by including those subleading terms in the solution for $\cu$ given in (\ref{U0-sol}, \ref{U2-sol})  that depend on $\mathcal{X}$:
\bal\label{ct}
	I'_c=&\;-\frac{1}{\k_2^2}\int dt\sqrt{-\g}\;\big(w_{10}\cy+w_{20}\cy^2+w_{11}\cy\cx+w_{02}\cx^2\big)\NO\\
	&\hskip-.in+\frac{3\ell_2 e^{-2\j_0}(\D_\c-1)}{8(2\D_\c-3)\k_2^2}\int dt\sqrt{-\g}\Bigg(c_0\Big(\frac{\cx^2}{\cy^2}+\frac{4\ell_2e^{2\j_0}w_{11}}{3(\D_\c-1)}\frac{\cx}{\cy}\Big)+\frac{2}{\D_\c-1}\g^{tt}(\pa_t\cx)^2\NO\\
	&\hskip.5in-4\Big(\frac{\cx}{\cy}-\frac{2\ell_2 e^{2\j_0}w_{11}}{3(\D_\c-1)}\Big)\g^{tt}\pa_t\cx\pa_t\cy+\Bigg(\frac{\cx^2}{\cy^2}-\frac{4\ell_2 e^{2\j_0}w_{11}}{3(\D_\c-1)}\frac{\cx}{\cy}\Bigg)\g^{tt}(\pa_t\cy)^2\Bigg)~.
\eal
Notice that these counterterms depend explicitly on the integration constant $c_0$, which will shortly be associated with the vacuum expectation value of operators in the dual conformal quantum mechanics. This is a well known property of conformal perturbation theory in the presence of irrelevant operators \cite{vanRees:2011fr}.

%%%%%%%%%%%%%%%%%%%%%%%%%%%%%%%%%%%%%%%%%%%%%%%%%%%%%%%%%%%%%%%%%%%%%%%%%%%%%%%%%%%%%%%%
\subsection{The Renormalized Theory}
At this point we can finally collect the results of holographic renormalization. The action $I\sbtx{reg}+I_b$ obtained by adding the boundary term \eqref{bt1} coincides with the renormalized on-shell action obtained by integrating out the Maxwell field directly in \eqref{2Daction-extremal}. As can be seen from \eqref{var1}, the variational problem for this action is well posed provided Neumann boundary conditions are imposed on $A_t$, i.e. the charge $Q$ is kept fixed by Dirichlet conditions on the gauge momentum $\pi^t$. 
In that case the dual 1D theory does not possess a $U(1)$ current operator. However, 
we can consider different boundary conditions for the gauge field by adding the extra {\em finite} boundary term
\be\label{bt2}
I_b'=\int dt\; \p^t A_t^{\rm ren}~,
\ee
so that the renormalized action becomes 
\be\label{Sren-def}
I\sbtx{ren}\equiv I\sbtx{reg}+I_b+I'_b~.
\ee
It follows from \eqref{var1} that the variational principle for this action takes the form
\be\label{var12}
\d I\sbtx{ren}=\int dt\;\big(\p_{\rm ren}^{tt}\d\g_{tt}+\p^{\rm ren}_\j\d\j+\p^{\rm ren}_\c\d\c+\p^t\d A^{\rm ren}_t\big)~,
\ee
and is therefore well posed provided Dirichlet boundary conditions are imposed on all fields. The dual theory in this case contains a $U(1)$ current and is the theory we will focus on in our analysis. 

The conformal perturbation theory developed in this subsection is equivalent to the procedure for cutting-off AdS$_2$ space described in the context of the Jackiw-Teitelboim model in \cite{Maldacena:2016upp}. A more general effective action can be obtained if the 
irrelevant couplings are not treated perturbatively. In the bulk, this is possible if the corresponding UV solutions can be found. For 
the 2D model \eqref{2Daction-extremal} we address this problem in appendix \ref{UV-analysis}. This analysis illustrates that the dependence 
of the counterterm \eqref{ct} on the charge $Q$ (through the AdS$_2$ radius and other variables) is a feature of renormalization around the IR fixed point; the UV boundary counterterms do not depend on $Q$. The non-perturbative renormalization 
for the Jackiw-Teitelboim model with a Maxwell field obtained from the circle reduction of pure AdS$_3$ gravity was done in \cite{Cvetic:2016eiv}. A special feature of that model is that the perturbative renormalization near the IR fixed point and the non-perturbative one result in the 
same effective actions.

\paragraph{Renormalized Canonical Variables} The asymptotic expansions of the fields and of their conjugate momenta implement a symplectic map $\ci$ from the space of fields and momenta to the space of modes that can be identified with sources and one-point functions in the dual theory. The goal of holographic renormalization is to diagonalize this map, thus rendering the variational problem well posed. This means that the pullback of the symplectic potential (i.e. the variational principle) \eqref{var12} by the map $\ci$ is diagonal in the modes parameterizing the asymptotic expansions and independent of the radial cutoff $\r_c$ \cite{Papadimitriou:2010as}. 

Defining $\Hat\p^t\equiv \p^t/\a(t)$ and using the leading asymptotic expansions of the variables that are not affected by the canonical transformation \eqref{bt1}, namely  
\be\label{sources}
\d\g_{tt}\sim \d(-\a^2) e^{2\r_c/\ell_2}~, \quad \d\j\sim \d(-e^{2\j_0}\d\n/2) e^{\r_c/\ell_2}~,\quad \d\c\sim \d\z e^{(\D_\c-1)\r_c/\ell_2}~,\quad \Hat\p^t= -\frac{QR^2}{2\k_2^2\a}~,
\ee
the pullback of the symplectic potential \eqref{var12} takes the form 
\bal
\label{var13}
\d I\sbtx{ren}= &\;\int dt\;\a\Big(\Hat\p^{tt}\d(-\a^2)+\Hat\p_\j(-e^{2\j_0}\d\n/2)+\Hat\p_\c\d\z+\Hat \p^t\d \Hat A_t\Big)~,
\eal
where the hatted variables (other than $\Hat\p^t$) are {\em defined} as the symplectic conjugates of $-\a^2$, $-e^{2\j_0}\n/2$, $\z$ and $\Hat\p^t$, respectively. Inserting the leading asymptotic expansions \eqref{sources} in the symplectic potential \eqref{var12} and comparing with \eqref{var13} we determine that 
\be\label{renmap}
\Hat\p^{tt} = \frac{1}{\a} e^{2\r_c/\ell_2}\p^{tt}_{\rm ren}~, \quad \Hat\p_\j=\frac{1}{\a} e^{\r_c/\ell_2}\p_\j^{\rm ren}~,\quad \Hat\p_\c=\frac{1}{\a} e^{(\D_\c-1)\r_c/\ell_2}\p_\c^{\rm ren}~,\quad  \Hat A_t=A_t^{\rm ren}~.
\ee

These quantities can be evaluated explicitly using the expressions \eqref{ren-vars} for the renormalized momenta, the defining relations \eqref{momenta} for the canonical momenta, as well as the boundary counterterms \eqref{ct}. The resulting values for the variables are finite in the sense of conformal perturbation theory near the IR, as they must be: 
\bal\label{ren-vars-limit}
\Hat\p^{t}_t=&\;e^{\r_c/\ell_2}\Big(-\frac{1}{\k^2_2}e^{-2\j}\dot\j-\frac{1}{2\k_2^2}\big(w_{10}\cy+w_{20}\cy^2+w_{11}\cy\cx+w_{02}\cx^2\big)\Big)
=-\frac{\vth(t)}{\k_2^2\ell_2}~,\NO\\
%%%%%%%%%%%%%%%%%%%%%%%%%%%%%%%%%%%%%%%%%%%%%%%%%%%%%%%%%%%%%%%%%%%%%%%%%%%%%%%%%%%%%%%%
\Hat\p_\j=&\;e^{2\r_c/\ell_2}\Big(-\frac{1}{\k^2_2}e^{-2\j}2K+\frac{2(e^{-2\j_0}+\cy)}{\k_2^2}\big(w_{10}+2w_{20}\cy+w_{11}\cx\big)\Big)=\frac{4e^{-2\j_0}\b(t)}{\k_2^2\ell_2\a(t)}~,\NO\\
%%%%%%%%%%%%%%%%%%%%%%%%%%%%%%%%%%%%%%%%%%%%%%%%%%%%%%%%%%%%%%%%%%%%%%%%%%%%%%%%%%%%%%%%
\Hat\p_\c=&\;e^{\D_\c\r_c/\ell_2}\Big(-\frac{3}{2\k_2^2}e^{-2\j}\dot\c-\frac{1}{\k_2^2}\big(w_{11}\cy+2w_{02}\cx\big)+\cdots\Big)=\frac{c_1\z(t)^{\frac{\D_\c}{1-\D_\c}}}{(1-\D_\c)\k_2^2}~,
\eal
where the ellipses in the last line stand for the remaining terms obtained from the counterterm action \eqref{ct} and the functions $\b(t)$ and $\vth(t)$ are given in \eqref{eq:s2} and \eqref{eq:s2-full}, respectively. The final expression on the right hand side of each equation in \eqref{ren-vars-limit} is finite and valid to leading order in conformal perturbation theory. Similarly, using \eqref{renmap}, \eqref{ren-vars} and the asymptotic solution \eqref{G-sol} for the function $\cg$ we can evaluate the variable $\Hat A_t$ conjugate to $\Hat\p^t$. Up to terms subleading in conformal perturbation theory we obtain  
\begin{align}\label{ren-At-limit}
\Hat A_t&=\;A_t-\frac{2\k_2^2\ell_2}{R^2}e^{3\j_0+3\c_0}\sqrt{-\g}\;\p^t\Big[1-\Big(\frac34 e^{2\j_0}+\ell_2 w_{20}+\frac{e^{2\j_0}\ell_2 w_{11}(\ell_2 w_{11}-3)}{3\D_\c}\Big)\cy+\frac{3-\ell_2 w_{11}}{\D_\c}\cx\Big]\nonumber\\
&=\m(t)~,
\end{align}
which is finite, as required.  

The variational principle \eqref{var13} allows us to identify the arbitrary functions $\a(t)$, $\n(t)$, $\z(t)$ and $\m(t)$ with sources of local operators in the dual quantum mechanics, and the variables $\Hat\p^{tt}$, $\Hat\p_\j$, $\Hat\p_\c$ and $\Hat\p^t$ with the corresponding operators. More precisely we define the one-point functions
\bal\label{1pt-fns}
\<\ct\>\equiv&\;2\Hat\p^t_t=\frac{\d I\sbtx{ren}}{\d\a}=\frac{\ell_2}{2\k_2^2\n}\(c_0+\frac{(\partial_t\n)^2}{\a^2}\)+\frac{1}{\k_2^2}c_1\z(t)^{\frac{1}{1-\D_\c}}~,\NO\\
%%%%%%%%%%%%%%%%%%%%%%%%%%%%%%%%%%%%%%%%%%%%%%%%%%%%%%%%%%%%%%%%%%%%%%%%%%%%%%%%%%%%%%%%%%%%%%
\<\cj^t\>\equiv&\;-\Hat\p^t=-\frac{1}{\a}\frac{\d I\sbtx{ren}}{\d\m}=\frac{QR^2}{2\k_2^2\a}~,\NO\\ 
%%%%%%%%%%%%%%%%%%%%%%%%%%%%%%%%%%%%%%%%%%%%%%%%%%%%%%%%%%%%%%%%%%%%%%%%%%%%%%%%%%%%%%%%%%%%%%
\<\co_\j\>\equiv&\;-\Hat\p_\j=2e^{-2\j_0}\frac{1}{\a}\frac{\d I\sbtx{ren}}{\d\n}=\frac{\ell_2e^{-2\j_0}}{\k_2^2\partial_t\n}\pa_t\(\frac{1}{\n}\( c_0+\frac{(\partial_t\n)^2}{\a^2}\)\)~,\NO\\ 
%%%%%%%%%%%%%%%%%%%%%%%%%%%%%%%%%%%%%%%%%%%%%%%%%%%%%%%%%%%%%%%%%%%%%%%%%%%%%%%%%%%%%%%%%%%%%%
\<\co_\c\>\equiv&\;\Hat\p_\c=\frac{1}{\a}\frac{\d I\sbtx{ren}}{\d\z}=\frac{c_1}{(1-\D_\c)\k_2^2}\z^{\frac{\D_\c}{1-\D_\c}}~,
\eal
where we have used the expression for $\b(t)$ in \eqref{eq:s2} and for $\vth(t)$ in \eqref{eq:s2-full}. As in (\ref{ren-vars-limit},\,\ref{ren-At-limit}), the final expressions on the right hand side hold up to terms subleading in conformal perturbation theory in the irrelevant couplings.
	
\paragraph{Renormalized On-shell Action} 
We previously wrote the regularized action including boundary counterterms in \eqref{eqn:iregib}. Taking the gauge field momentum $\pi^t$ 
on-shell it is  
\be
I\sbtx{reg}+I_b=\left.\cu\right|_{\r_c}+I_c  + I\sbtx{global}~,
\ee
where the terms from a possible horizon $I\sbtx{global}$ were given in \eqref{Sglobal}. We also saw that the counterterm $I_c$, given in \eqref{ct-vector}, on-shell coincides with $I_c'$ given in \eqref{ct}. This counterterm was constructed such that it cancels all divergent terms in the effective superpotential $\left.\cu\right|_{\r_c}$. Therefore, the sum $\left.\cu\right|_{\r_c}+I_c$ is given by the finite contributions to $\left.\cu\right|_{\r_c}$. 

As we discussed after \eqref{Sglobal}, most of the terms in the leading superpotential $\cu_{(0)}$ \eqref{U0-sol} are divergent and so they do
not contribute to the sum $\left.\cu\right|_{\r_c}+I_c$. The only exception is
the non perturbative contribution $\Delta W$ identified in \eqref{delta-W}. The higher derivative terms \eqref{U2-sol} in the superpotential contain no divergences and so they all contribute to the renormalized action, in principle, but the leading order of conformal perturbation theory retains only the last term in \eqref{U2-sol}. We therefore find 
\be
\left.\cu\right|_{\r_c}+I_c=\frac{1}{\k_2^2}\int dt\sqrt{-\g}\;\Big(\frac{\ell_2}{2}\cy^{-1}\Big(c_0+\g^{tt}(\pa_t\cy)^2\Big)+c_1\cx^{-\frac{1}{\D_\c-1}}+\cdots\Big)~.
\ee	
The leading asymptotic behavior of these fields, namely $\sqrt{-\gamma}\sim\alpha e^{\rho_c/\ell_2}$, $\cy\sim\nu e^{\rho_c/\ell_2}$, $\cx\sim\zeta e^{(\Delta_\cx-1)\rho_c/\ell_2}$
are sufficient for evaluating this expression. Incorporating also the finite boundary term \eqref{bt2}, we then obtain the full renormalized on-shell action %%%%%%
\be\label{Sren}
I\sbtx{ren}=\frac{\ell_2}{2\k_2^2}\int dt\;\Big(\frac{\a c_0}{\n}-\frac{(\pa_t\n)^2}{\n\a}+\frac{2c_1}{\ell_2}\a\z^{\frac{1}{1-\D_\c}}-\frac{QR^2}{\ell_2}\m\Big)+I\sbtx{global}~,
\ee
up to terms subleading in conformal perturbation theory.
With the exception of $I\sbtx{global}$, the renormalized action \eqref{Sren} can also be obtained by integrating the one-point functions \eqref{1pt-fns} 
with respect to the corresponding source. 

The renormalized action \eqref{Sren} is identical to the one found in \cite{Cvetic:2016eiv} for the 2D dilaton model obtained from a circle reduction of pure AdS$_3$ gravity, except that \eqref{Sren} contains an additional term due to the fluctuations of the second scalar $\c$. However, while in the present context this effective action holds only in the vicinity of the IR fixed point, for the model considered in \cite{Cvetic:2016eiv} it was valid all the way to the UV, since the analysis there was non perturbative in the dilaton coupling. 

\paragraph{Ward Identities and Trace Anomaly} By inspection, the one-point functions \eqref{1pt-fns} satisfy the Ward identities
\bal\label{WIDs}
&\pa_t\<\ct\>-\frac12e^{2\j_0}\pa_t\n\;\<\co_\j\>-\pa_t\z\;\<\co_\c\>=0~,\NO\\
%%%%%%%%%%%%%%%%%%%%%%%%%%%%%%%%%%%%%%%%%%%%%%%%%%%%%%%%%%%%%%%%%%%%%%%%%%%%%%%%%%%%%%%%%%%%%%%%%%
&\<\ct\>+\frac12 e^{2\j_0}\n\;\<\co_\j\>+(\D_\c-1)\z\;\<\co_\c\>=\frac{\ell_2}{\k_2^2\a}\pa_t\(\frac{\pa_t\n}{\a}\)=\ca~,\NO\\
%%%%%%%%%%%%%%%%%%%%%%%%%%%%%%%%%%%%%%%%%%%%%%%%%%%%%%%%%%%%%%%%%%%%%%%%%%%%%%%%%%%%%%%%%%%%%%%%%%
&\cd_t\<\cj^t\>=0~,
\eal	
where $\cd_t$ stands for the covariant derivative with respect to the boundary metric $-\a^2$. These identities reflect the global symmetries of the dual theory and can alternatively be derived by renormalizing the first class constraints \eqref{constraints0}, without using the explicit form of the one-point functions. 

The Ward identity for the trace of the energy momentum tensor contains, in addition to the beta functions for the two scalar operators of dimension $\D_\j=2$ and $\D_\c$, a conformal anomaly $\ca$ that is local in the sources $\a(t)$ and $\n(t)$. This trace anomaly was previously derived in \cite{Cvetic:2016eiv}, and it was identified as the origin of the Schwarzian effective action. In particular, the term $-(\pa_t\n)^2/(\n\a)$ in the renormalized on-shell action \eqref{Sren} corresponds to the effective action of the conformal anomaly, i.e. the analogue of the non-local Polyakov action $\sim c\;R\square^{-1}R$ for a 2d CFT with central charge $c$ \cite{Polyakov:1981rd}. Indeed, it was shown in \cite{Cvetic:2016eiv} that this term can be obtained by a circle reduction from the Polyakov action of the 2d CFT at the UV. It is well known that in conformal gauge the Polyakov action reduces to the 2d Liouville action with zero Liouville coupling. We will see in subsection \ref{sec:schwarzian} that in conformal gauge the term $-(\pa_t\n)^2/(\n\a)$ similarly reduces to the 1d Liouville action with zero Liouville coupling, which can be mapped to the Schwarzian effective action. We revisit this argument in the following two subsections, where now we incorporate the coupling $\z(t)$ for the second scalar operator $\co_\c$ that was not present in the analysis of \cite{Cvetic:2016eiv}.

%%%%%%%%%%%%%%%%%%%%%%%%%%%%%%%%%%%%%%%%%%%%%%%%%%%%%%%%%%%%%%%%%%%%%%%%%%%%%%%%%%%%%%%%%%%%%%%%%%
	
\subsection{Residual Gauge Symmetries}\label{sec:pbh}

The Fefferman-Graham gauge \eqref{FG-gauge} for the geometry and the radial gauge $A_\r=0$ for the Maxwell field do not completely fix the local symmetries in the bulk. The residual bulk diffeomorphisms, known as Penrose-Brown-Henneaux (PBH) diffeomorphisms \cite{Penrose,Brown:1986nw,Imbimbo:1999bj}, and residual $U(1)$ gauge transformations are interesting because they are directly related to 
the asymptotic symmetry algebra and to the Ward identities \eqref{WIDs}. 

In order to determine these residual local symmetries and how they act on the sources and operators of the dual theory, we consider a generic infinitesimal bulk diffeomorphism generated by the vector field $\x^a(\r,t)$ and a generic infinitesimal gauge transformation corresponding to the gauge parameter $\L(\r,t)$. Under such a combined transformation the bulk fields transform as
\bal\label{local-transformations}
&\d_{\x+\L} g_{\r\r} =\cl_\x g_{\r\r}=\dot\x^\r~,\quad
\d_{\x+\L} g_{tt} = \cl_\x g_{\r t}=\g_{tt}(\dot\x^t+\g^{tt}\pa_t\x^\r)~,\quad
\d_{\x+\L} g_{tt} = \cl_\x g_{tt}=L_\x\g_{tt}+2K_{tt}\x^\r~,\NO\\
%%%%%%%%%%%%%%%%%%%%%%%%%%%%%%%%%%%%%%%%%%%%%%%%%%%%%%%%%%%%%%%%%
&\d_{\x+\L} \j= \cl_\x\j =L_\x\j+\x^\r\dot\j~,\qquad \d_{\x+\L}\c= \cl_\x\c =L_\x\c+\x^\r\dot\c~,\NO\\
%%%%%%%%%%%%%%%%%%%%%%%%%%%%%%%%%%%%%%%%%%%%%%%%%%%%%%%%%%%%%%%%%
&\d_{\x+\L} A_{\r} =\cl_\x A^\L_{\r}+\d_\L A_\r=\dot\x^{t}A_{t}+\dot \L~,\qquad
\d_{\x+\L} A_{t} =\cl_\x A_{t}+\d_\L A_{t}=L_\x A_{t}+\x^\r\dot  A_{t}+\pa_t\L~,
\eal
where $\cl_\x$ is the bulk Lie derivative generated by the vector field $\xi^\mu$ and $L_\x$ denotes the Lie derivative with respect to the time component $\x^t$. As in previous sections, a dot $\dot{}$ denotes a derivative with respect to the radial coordinate 
$\r$ and $K_{tt}$ is the extrinsic curvature \eqref{eq:K}. The conditions that the Fefferman-Graham gauge \eqref{FG-gauge} and the radial gauge $A_\r=0$ are preserved by the transformations \eqref{local-transformations} are
\be
\cl_\x g_{\r\r}=\cl_\x g_{\r t}=0~,\qquad (\cl_\x+\d_\L)A_{\r}=0~,
\ee
which amounts to a set of differential equations for the gauge parameters $\x^\mu(\r,t)$ and $\L(\r,t)$ for the residual gauge symmetries. 
The general solution of these equations is \cite{Papadimitriou:2005ii}
\begin{align}\label{gen-PBH}
\x^\r=\s(t)~,\quad
\x^t=\h(t)+\pa_t\s(t)\int_\r^\infty d\r'\g^{tt}(\r',t)~,\cr
\L=\vf(t)-\pa_{t}\s(t)\int_\r^\infty d\r'\g^{tt}(\r',t)A_{t}(\r',t)~, 
\end{align}
where $\h(t)$, $\s(t)$, and $\vf(t)$ are arbitrary functions of time. The boundary diffeomorphism generated by $\eta(t)$, the boundary Weyl transformation generated by $\s(t)$, and the
boundary gauge transformation generated by $\vf(t)$ all act independently of one another. 

The residual gauge symmetries transform the field components that were not fixed by the gauge choices, such as
$\gamma_{tt}$ and $A_t$. The transformations \eqref{local-transformations} are such that the 
solutions (\ref{eqn:leading}, \ref{eq:s1}, \ref{eq:X-hom})
retain their form, but with integration ``constants" modified according to 
\bal\label{PBH-sources}
\d\sbtx{PBH}\a =&\;\pa_t(\h\a)+\a\s/\ell_2~,\NO\\
\d\sbtx{PBH}\n =&\;\h\pa_t\n+\n\s/\ell_2~,\NO\\
\d\sbtx{PBH}\z =&\;\h\pa_t\z+(\D_\c-1)\z\s/\ell_2~,\NO\\
\d\sbtx{PBH}\m =&\;\pa_t(\h\m+\vf)~.
\eal
Since we know the one-point functions in terms of these sources explicitly from \eqref{1pt-fns}, the transformations \eqref{PBH-sources} allow us to determine the transformations of the one-point functions as well. We find
\bal\label{PBH-1pt-fns}
\d\sbtx{PBH}\<\ct\> =&\;\h\pa_t\<\ct\>-\frac{\s}{\ell_2}\<\ct\>+\frac{\pa_t\n\pa_t\s}{\k_2^2\a^2}~,\NO\\
%%%%%%%%%%%%%%%%%%%%%%%%%%%%%%%%%%%%%%%%%%%%%%%%%%%%%%%%%%%%%%%%%%%%%%%%%%%%%%%%%%%%%%%%%%
\d\sbtx{PBH}\<\co_\j\> =&\;\h\pa_t\<\co_\j\>-2\frac{\s}{\ell_2}\<\co_\j\>+\frac{ e^{-2\j_0}}{\k_2^2\a}\pa_t\Big(\frac{\pa_t\s}{\a}\Big)~,\NO\\
%%%%%%%%%%%%%%%%%%%%%%%%%%%%%%%%%%%%%%%%%%%%%%%%%%%%%%%%%%%%%%%%%%%%%%%%%%%%%%%%%%%%%%%%%%
\d\sbtx{PBH}\<\co_\c\> =&\;\h\pa_t\<\co_\c\>-\D_\c\frac{\s}{\ell_2}\<\co_\c\>~,\NO\\
%%%%%%%%%%%%%%%%%%%%%%%%%%%%%%%%%%%%%%%%%%%%%%%%%%%%%%%%%%%%%%%%%%%%%%%%%%%%%%%%%%%%%%%%%%
\d\sbtx{PBH}\<\cj^t\> =&\;-\(\frac{\pa_t(\h\a)}{\a}+\frac{\s}{\ell_2}\)\<\cj^t\>~.
\eal
It is also instructive to consider the transformation of the renormalized on-shell action  
under the residual gauge symmetries \eqref{PBH-sources}. Either using the explicit form of the action in terms of the 
sources \eqref{Sren}, or using the one-point functions \eqref{1pt-fns} together with the Ward identities \eqref{WIDs}, we find  
\be
\label{eqn:PGHsigma}
\d\sbtx{PBH}I\sbtx{ren}=\frac{1}{\k_2^2}\int dt\; \s\pa_t\(\frac{\pa_t\n}{\a}\)=\int dt\;\a \frac{\s}{\ell_2}\ca~,
\ee
where $\ca$ is the trace anomaly that appears in the trace Ward identity in \eqref{WIDs}. It follows that the renormalized action is invariant under time reparameterizations and $U(1)$ gauge transformations, but generally not under the boundary Weyl transformations.

%%%%%%%%%%%%%%%%%%%%%%%%%%%%%%%%%%%%%%%%%%%%%%%%%%%%%%%%%%%%%%%%%%%%%%%%%%%%%%%%%%%%%%%%%%%%%%%%%%
	
\subsection{The Schwarzian Effective Action}
\label{sec:schwarzian}
The residual gauge transformations are parameterized by three arbitrary functions of time: $\eta(t)$, $\s(t)$, and $\vf(t)$. The sources 
$\a(t)$, $\n(t)$, and $\m(t)$ for the local operators in the dual 1D theory are scalar functions of time, so these sources can be generated entirely through residual symmetries, at least locally. Therefore, all these sources can be formally interpreted as pure gauge and it is illuminating to 
do so. However, the source $\z(t)$ for the squashing mode is not pure gauge and cannot be traded for any local gauge symmetry. 

The fact that the sources $\a(t)$, $\n(t)$, and $\m(t)$ can be traded for local symmetries implies that the number of independent functions of time on which the renormalized on-shell action \eqref{Sren} depends on can be reduced. We first recall that, as discussed after \eqref{eqn:PGHsigma}, the renormalized action is invariant under infinitesimal time reparameterizations and $U(1)$ gauge transformations. From the explicit form of the renormalized action in \eqref{Sren} it is clear that this invariance applies to the corresponding finite transformations as well, namely
\be
\label{eqn:symm}
\a(t)\to \a(\h(t))/\pa_t\h(t)~, \quad \n(t)\to \n(\h(t))~,\quad \m(t)\to\m(\h(t))/\pa_t\h(t)+\pa_t\vf(t),\quad \z(t)\to\z(\h(t))~,
\ee
for any finite functions $\h(t)$ and $\vf(t)$. However, although a boundary Weyl transformation generated by $\s(t)$ corresponds to a bulk diffeomorphism, it is not a symmetry of the boundary theory: the renormalized action does in fact depend on $\s(t)$, as exhibited in \eqref{eqn:PGHsigma}. This dependence is a manifestation of the conformal anomaly. 

The fact that the sources $\a(t)$, $\n(t)$, and $\m(t)$ are pure gauge and can be traded for the three independent functions $\eta(t)$, $\s(t)$, and $\vf(t)$ that generate residual gauge symmetries, together with the observation that the renormalized on-shell action is independent of $\eta(t)$, $\vf(t)$ but it does depend on $\s(t)$, imply that the total dependence of the renormalized on-shell action on the sources can be parameterized by the single function $\s(t)$, as well as the source $\z(t)$, which is not pure gauge. Without loss of generality, therefore, we can parameterize the sources as a finite Weyl transformation starting from the reference point with constant $\a=1$, $\n=\n_0$, $\m=\m_0$ and an arbitrary $\z_0(t)$, namely
\be
\label{Weyl-background}
\a(t)=e^{\s(t)/\ell_2}~,\qquad
\n(t)=\n_0 e^{\s(t)/\ell_2}~,\qquad \m=\m_0~, \qquad \z(t)=e^{(\D_\c-1)\s(t)/\ell_2}\z_0(t)~.
\ee
Inserting this form of the sources in \eqref{Sren} gives
\be
\label{Liouville}
I\sbtx{ren}=\frac{\ell_2}{2\k_2^2}\int dt\Big(\frac{c_0}{\n_0}-\frac{\n_0}{\ell_2^2}(\pa_t\s)^2+\frac{2c_1}{\ell_2}\z_0(t)^{\frac{1}{1-\D_\c}}-\frac{QR^2}{\ell_2}\m_0\Big)+I\sbtx{global}~.
\ee
Thus the entire time dependence of the renormalized action \eqref{Sren} is due to the conformal factor $\sigma(t)$ and the dynamical scalar source $\z_0(t)$. This form of the renormalized on-shell action makes is manifest that the Weyl mode $\s(t)$ can interpreted as the Goldstone boson of spontaneous conformal symmetry breaking, i.e. the boundary dilaton, through the matching of the conformal anomaly effective action in the spontaneously broken and unbroken phases \cite{Schwimmer:2010za}. 

The effective action for the Weyl factor $\s(t)$ in \eqref{Liouville} is directly related to the Schwarzian effective action. In particular, parameterizing $\s(t)$ in terms of an arbitrary auxiliary function $\t(t)$ as
\be
\label{d-time}
\s=\ell_2\log\pa_t\t(t)~,
\ee
and adding a suitable total derivative term the effective action \eqref{Liouville} becomes
\be
\label{Schwarzian}%\boxed{
I\sbtx{ren}=\frac{\ell_2}{2\k_2^2}\int dt\Big(\frac{c_0}{\n_0}+2\n_0\{\t(t),t\}+\frac{2c_1}{\ell_2}\z_0(t)^{\frac{1}{1-\D_\c}}-\frac{QR^2}{\ell_2}\m_0\Big)+I\sbtx{global}~,%}
\ee
where $\{\t(t),t\}$ denotes the Schwarzian derivative
\be
\label{Schwarzian-def}
\{\t(t),t\}=-\frac12\left(\frac{\pa_t^2\t}{\pa_t\t}\right)^2 +\left(\frac{\pa_t^2\t}{\pa_t\t}\right)^\prime = \frac{\pa_t^3\t}{\pa_t\t}-\frac32\left(\frac{\pa_t^2\t}{\pa_t\t}\right)^2~.
\ee
%

%%%%%%%%%%%%%%%%%%%%%%%%%%%%%%%%%%%%%%%%%%%%%%%%%%%%%%%%%%%%
%%%%%%%%%%%%%%%%%%%%%%%%%%%%%%%%%%%%%%%%%%%%%%%%%%%%%%%%%%%%

\section{Thermodynamics of 2D Black Holes}\label{sec:2Dthermo}

In this section we quantify the thermodynamic properties of the 2D backgrounds for our theory with special emphasis on black holes. We use our results from section \ref{dictB} to evaluate the conserved charges and corresponding chemical potentials. We compare and contrast this analysis with the near extremal results obtained from the 5D point of view in subsection \ref{sec:nIR5Dthermo}.

%%%%%%%%%%%%%%%%%%%%%%%%%%%%%%%%%%%%%%%%
\subsection{Killing Symmetries and Conserved Charges}

In subsection \ref{sec:IRperturbations}, complemented with the Hamiltonian analysis in subsection \ref{sec:effIR}, we presented the most general linearized solutions around the IR fixed point for our 2D theory. In the following we identify which of those background solutions possess a Killing symmetry and compute the corresponding conserved charges. 

It is useful to start from the PBH transformations in subsection \ref{sec:pbh}. From this point of view, a rigid symmetry is a transformation that leaves $\alpha(t)$, $\nu(t)$, $\z(t)$ and $\mu(t)$ unchanged, i.e. setting the left hand side of  \eqref{PBH-sources} to zero. This simultaneously restricts the background, such that a solution to $\delta_{\rm PBH}(\ldots)=0$ exists, and determines the Killing symmetry. There are two background configurations that have a non-trivial symmetry. The first background amounts to setting $\z(t)=0$, i.e. removing the source for $\chi$, and keeping all other parameters arbitrary. In this case the background is unchanged for the PBH transformations with parameters\footnote{The parameters $\x^a$ and $\L$ as defined in \eqref{local-transformations}, or $\s$, $\h$ and $\vf$ defined in \eqref{gen-PBH},  have dimensions of length. However, the Killing symmetry parameters should be dimensionless in order for the corresponding conserved charges to be correctly normalized. The Killing symmetry parameters are therefore only proportional to the corresponding local symmetry parameters. To avoid introducing additional notation, however, in this section we will use $\x^a$ and $\L$, as well as $\s$, $\h$ and $\vf$, to refer to the dimensionless Killing parameters, rather than the dimensionful local symmetry parameters. }
\be\label{CKVs}
\h= k\frac{\n}{\a}~,\qquad \s=-k\ell_2 \frac{\pa_t\n}{\a}~,\qquad \vf=- k\frac{\n\m}{\a}~,
\ee
where $k>0$ is an arbitrary dimensionless constant that accounts for an inherent ambiguity in the normalization of the timelike Killing vector, as we will discuss shortly. From \eqref{gen-PBH}, the generators of the corresponding residual symmetry transformation are
\be\label{gen-PBH1}
\x^\r=-k\ell_2\frac{\pa_t\n}{\a}~,\quad
\x^t= k \frac{\n}{\a}+O(e^{-2\rho/\ell_2})~,\quad
\L=- k \frac{\n\m}{\a}+O(e^{-\rho/\ell_2})~.
\ee
This Killing transformation reduces to the bulk Killing vector $\x^a=-\ell_2 k\e^{ab}\nabla_be^{-2\j}$ \cite{Mann:1992yv} in the special case of vanishing chemical potential $\m$. 
 
The second background with a rigid symmetry corresponds to setting the sources for $\j$ and $\chi$ to be time independent and non-zero, i.e. $\nu(t)=\nu_0$ and $\z(t)=\z_0$, while $\alpha$ and $\mu$ are arbitrary functions. Solving for $\delta_{\rm PBH}(\ldots)=0$ in \eqref{PBH-sources} for this situation gives
\be\label{eq:CKVs1}
\h= k\frac{\n_0}{\a}~,\qquad \s=0~,\qquad \vf=- k\frac{\n_0\m}{\a}~.
\ee
We have chosen the constants here such that the two backgrounds with Killing symmetry can be discussed as one: \eqref{eq:CKVs1} is simply obtained by setting $\nu(t)=\nu_0$ in \eqref{CKVs}.

Having identified \eqref{CKVs} and \eqref{eq:CKVs1} as the relevant Killing symmetries, we turn to the corresponding conserved charges, which can be derived from the Ward identities \eqref{WIDs}. Starting with the conservation of the U(1) current in \eqref{WIDs},  
\be
\cd_t\<\cj^t\>=\frac{1}{\a}\pa_t\big(\a\<\cj^t\>\big)=0~,
\ee 
which leads to the conserved electric charge
\be\label{e-charge}
\cq\equiv-\a\<\cj^t\>=\p^t=-\frac{QR^2}{2\k^2_2}~.
\ee  

To obtain the conserved charge associated with the conformal Killing symmetry \eqref{CKVs} (and  \eqref{eq:CKVs1} as a special case) we multiply the first Ward identity in \eqref{WIDs} with $\h$ in \eqref{CKVs} to get
\be
\h\pa_t\<\ct\>-\frac12e^{2\j_0}\h\pa_t\n\;\<\co_\j\>-\h\pa_t\z\;\<\co_\c\>=0~.
\ee
Note that this is the correct Ward identity that generates the combined diffeomorphism and U(1) gauge transformation that is compatible with the Killing symmetry \eqref{CKVs}.
Using the fact that $\h$ is a conformal Killing symmetry and hence satisfies \eqref{PBH-sources} with zero on the left hand side, this identity becomes
\be
\h\pa_t\<\ct\>+\Big(\frac12e^{2\j_0}\n\;\<\co_\j\>+(\D_\c-1)\z\;\<\co_\c\>\Big)\frac{\s}{\ell_2}=0~.
\ee
The trace Ward identity in \eqref{WIDs} and the explicit form of the conformal Killing parameters in \eqref{CKVs} then allow us to rewrite this identity as 
\be
k\frac{\n}{\a}\pa_t\<\ct\>+\big(\<\ct\>-\ca\big)k\frac{\pa_t\n}{\a}=0 \Leftrightarrow \frac{1}{\a}\pa_t\big(k\n\<\ct\>\big)=k\frac{\pa_t\n}{\a}\ca=\frac{k\ell_2}{2\k_2^2\a}\pa_t\(\frac{\pa_t\n}{\a}\)^2~.
\ee

This identity implies that the quantity
\be\label{mass2D}
M\sbtx{2D}= -k\n \(\<\ct\>-\frac{\ell_2}{2\k_2^2\n}\Big(\frac{\pa_t\n}{\a}\Big)^2\)=-\a\h\(\<\ct\>-\frac{\ell_2}{2\k_2^2\n}\Big(\frac{\pa_t\n}{\a}\Big)^2\)~,
\ee
is independent of time for arbitrary sources, i.e. it is a conserved charge. This is indeed the Noether charge that generates the conformal Killing symmetry \eqref{CKVs} (see e.g. (5.12)--(5.16) in \cite{Cvetic:2016eiv}). The charge is simply the time component of the corresponding current since a spatial slice of the boundary is just a point. Moreover, notice that it is only because the conformal anomaly is a total derivative that we can always define a conserved charge associated with the conformal Killing symmetry \eqref{CKVs}, even when the anomaly does not vanish numerically. In more generic situations conformal Killing symmetries lead to conserved charges only in the absence of a conformal anomaly \cite{Papadimitriou:2005ii}. 

Using the one point function of the stress tensor in \eqref{1pt-fns} and the conformal Killing symmetry \eqref{CKVs} we can evaluate the mass \eqref{mass2D}, namely   
\be
M\sbtx{2D}=-\frac{k\ell_2}{2\k_2^2}c_0~,
\ee
which applies for backgrounds with $\z(t)=0$ and arbitrary $\a(t)$, $\n(t)$ and $\m(t)$. For our second configuration, $\nu(t)=\nu_0$ and $\z(t)=\z_0$, the rigid Killing symmetry \eqref{eq:CKVs1} in \eqref{mass2D} leads  to 
\be
\label{eqn:M2D}
M\sbtx{2D}=-\frac{k\ell_2}{2\k_2^2}\(c_0+\frac{2\n_0}{\ell_2}c_1\z_0^{\frac{1}{1-\D_\c}}\)~.
\ee
In black hole applications explicit expressions such as \eqref{c0} have $c_0<0$. The equations for $M\sbtx{2D}$ given here establish $|c_0|$ as a measure of the excitation energy, as expected. 

%%%%%%%%%%%%%%%%%%%%%%%%%%%%%%%%%%%%%%%%
\subsection{Thermodynamics of 2D Black Holes}\label{sec:2dthermo}
In this subsection we study the thermodynamics of {\em static} configurations in the 2D theory and in particular focus on black holes. To simplify the analysis, and since the 5D black holes we are interested in have $c_1=0$, we will set $c_1=0$ throughout our discussion of the thermodynamics.

The static limit of the IR fixed point solution introduced in subsection \ref{sec:nIRfix}, 
\begin{align}\label{eq:mm1}
ds^2= d\rho^2- (\alpha_0 e^{\rho/\ell_2} + \beta_0 e^{-\rho/\ell_2})^2 dt^2 ~,
\end{align}
is the canonical example of a 2D black hole geometry. The outer horizon is located at $\r_h$ given by
\be
\label{horizon}
e^{\r_h/\ell_2}=\sqrt{-\b_0/\a_0}\;.
\ee	
The black hole is generally not extremal: the extremal limit corresponds to $\b_0\to0$ or $\r_h\to-\infty$. In these formulae (and others below) the subscript ``$0$'' emphasizes the point that the coefficients of the solution are time independent. 

We study the black hole {\it near} the IR fixed point so the time dependent dilaton is engaged. The resulting backreaction on the geometry 
modifies the metric so that $\sqrt{-\gamma_0}\to \sqrt{-\gamma_0}+\sqrt{-\gamma_1}$, but for static backgrounds the linear fluctuation $\sqrt{-\gamma_1}$ given in \eqref{eq:s3} (or more generally in \eqref{eq:metric-fluct-zeta} when the source $\z(t)$ is turned on) is proportional to the zero order solution $\sqrt{-\gamma_0}$. The position of the horizon \eqref{horizon} determined from the leading order solution therefore applies also to linear order in the fluctuations around the IR fixed point.
	
In the following we evaluate the thermodynamic variables of our 2D black hole. We start with the temperature and the electric potential, the potentials conjugate to the conserved charges $M_{\rm 2D}$ and $ \cq$, respectively; and then proceed to evaluate the entropy and the Gibbs free energy around the IR fixed point. 
	
%%%
\paragraph{Temperature:} Expanding the metric around the horizon radius \eqref{horizon} and demanding that the Euclidean section has no conical singularity gives the periodicity condition  
\be\label{conical}
\wt t_E\equiv t_E\left.\pa_\r\sqrt{-\g}\right|_{\r_h}~,\qquad \wt t_E\sim \wt t_E+2\p~,
\ee
where $t_E$ is the analytic continuation of the time coordinate $t$. This periodicity condition is unambiguous, but the relation between the time coordinate ``$t$" and the appropriate physical time may not be the same from the IR and UV viewpoints. The physical time in the IR defined by the timelike Killing vector $\h$ in \eqref{CKVs} is $t\h^{-1}$ and so the physical temperature is determined by the periodicity condition $t_E\h^{-1}\sim t_E\h^{-1}+T\sbtx{2D}^{-1}$. Therefore \eqref{conical} determines the 2D temperature 
\be\label{2Dtemperature}
T\sbtx{2D}=\frac{\h}{2\p}\left.\pa_\r\sqrt{-\g}\right|_{\r_h}~.
\ee
In appendix \ref{Tpot} we show that this expression for the temperature is proportional to the value of the scalar potential at the horizon. Evaluating the expression \eqref{2Dtemperature} at the horizon radius \eqref{horizon} we find 
\be\label{2Dtemperature-explicit}
T\sbtx{2D}={k\over 2\pi}\sqrt{-c_0} +\cdots \,,
\ee
where the ellipses denote higher order terms in the expansion near the IR fixed point and we have used the linearized solution \eqref{eq:s1}, as well as the static limit of the expression \eqref{eq:s2} for $\b(t)$. 

%%%
\paragraph{Electric Potential:} From the solution for the gauge field in \eqref{eqn:leading}	and the horizon radius in \eqref{horizon} follows that the value of the gauge potential on the horizon is	
\be
A_t(\r_h)=\mu_0-2 Q \ell_2 e^{3\chi_0+3\psi_0}\sqrt{-\a_0\b_0}~.
\ee
The electric potential is therefore given by	
\be
\label{eqn:elpot}
\F_e=\h\(\mu_0-A_t(\r_h)\)=2\p Q \ell_2^2 e^{3\chi_0+3\psi_0}T\sbtx{2D}\;.
\ee

\paragraph{Entropy:} The entropy of the 2D black hole is given by the value of the dilaton on the horizon \cite{Myers:1994sg,Gegenberg:1994pv,Cadoni:1999ja}
\be\label{2Dentropy}
S=\frac{2\p}{\k_2^2}e^{-2\j(\r_h)}=\frac{2\p}{\k_2^2}\Big(e^{-2\j_0}+\ell_2\sqrt{-c_0}+\cdots\Big)~,
\ee
where we used the perturbative solution for $\cy$ in \eqref{eq:s1}, as well as the expression for $\b$ and $\vth$ in \eqref{eq:s2}.

%%%
\paragraph{Gibbs Free Energy:} Our expression \eqref{Sren} for the renormalized on-shell action comprises a conventional term that controls the boundary dynamics and a ``global" term that is due to the black hole horizon. The boundary term receives contributions from the 2D mass \eqref{eqn:M2D} and the renormalized gauge potential $\m_0$. The global term was presented in \eqref{Sglobal} with a contribution from the extrinsic curvature at the horizon
\be
\frac{1}{\k^2_2}\left.\sqrt{-\g}\;e^{-2\j}K\right|_{\r_h}=\frac{1}{\k^2_2}\left.e^{-2\j}\pa_\r\sqrt{-\g}\right|_{\r_h}=ST\sbtx{2D}\h^{-1}~,
\ee
a contribution from the reduced Hamilton's principal function at the horizon that happens to vanish $\left.\cu\right|_{\r_h}=0$ (see footnote \ref{zero-sup} in appendix \ref{Tpot}), and a term proportional to value of the gauge potential at the horizon. After continuation to Euclidean signature we can recast the three non vanishing contributions to \eqref{Sren} as
\bal
\label{eqn:gibbsI}
I\sbtx{ren}^E=&\;\int_0^{T\sbtx{2D}^{-1}\h}\hskip-0.3cm dt_E\h^{-1}M\sbtx{2D}-\int_0^{T\sbtx{2D}^{-1}\h}\hskip-0.3cm dt_E\h^{-1}ST\sbtx{2D}-\cq\int_0^{T\sbtx{2D}^{-1}\h}\hskip-0.3cm dt_E\h^{-1}\F_e~,
\eal
or 
\be
I\sbtx{ren}^E=T\sbtx{2D}^{-1}(M\sbtx{2D}-T\sbtx{2D}S-\F_e\cq)~.
\ee
This is the expected relation between the Euclidean renormalized on-shell action and the Gibbs free energy. It is interesting that the mass term is due to the dynamical term in the renormalized action \eqref{Sren} while the entropy is entirely due to the global contribution. The term involving the electric charge gets contributions from both the dynamical and global parts, since the gauge invariant electric potential $\F_e$ is the difference between the (renormalized) gauge potential at the boundary and the horizon. 

%%%
\paragraph{Mass Gap:} 
The entropy \eqref{2Dentropy} can be expressed in the form 
\be\label{eq:entropy2d}
S=S\sbtx{ext}+\frac{2}{M\sbtx{gap}}T\sbtx{2D}~,
\ee
where
\be
S\sbtx{ext}= \frac{2\p}{\k_2^2} e^{-2\j_0}~,
\ee
and  the ``mass gap'' is given by
\be\label{eqn:gap}
M\sbtx{gap}=\frac{k\k_2^2}{2\p^2\ell_2}~.
\ee
The temperature independent term $S\sbtx{ext}$ corresponds to the entropy at the IR fixed point which is interpreted as the entropy at extremality, i.e. ground state entropy. As we move away from the fixed point, the entropy depends linearly on $T_{\rm 2D}$ with a strength controlled by $M\sbtx{gap}$ \cite{Preskill:1991tb}. If we pick the normalization constant $k$ as a numerical constant of ${\cal O}(1)$ that is independent of black hole parameters, our equation for the mass gap \eqref{eqn:gap} agrees with most other studies of nAdS$_2$/CFT$_1$ holography: it depends only on the AdS$_2$ scale $\ell_2$ and is proportional to the gravitational coupling $\kappa_2^2$. This is also the coefficient in front of the Schwarzian effective action \eqref{Schwarzian}  .

The black hole mass \eqref{eqn:M2D} can also be expressed in terms of the mass gap in the near extremal limit and takes the form
\be\label{eq:mass2d}
M_{\rm 2D}=\frac{2\p^2\ell_2}{k\k_2^2}T\sbtx{2D}^2=\frac{1}{M\sbtx{gap}}T\sbtx{2D}^2~.
\ee
As we will see momentarily, this is consistent with the first law of black hole thermodynamics. 

\paragraph{First Law:} Since the AdS$_2$ radius $\ell_2$, given in \eqref{eq:l2}, depends on the electric charge $\cq$, this charge must be kept fixed in order to have a well defined first law near the IR fixed point.\footnote{It is interesting to note that the AdS$_2$ radius for the near extremal BTZ black hole is half of the AdS$_3$ radius and hence independent of the AdS$_2$ electric charge \cite{Strominger:1998yg,Cvetic:2016eiv}. As a consequence, the thermodynamic ensemble, as well as the boundary conditions on the AdS$_2$ gauge field, is not fixed in that case.} 
Using the thermodynamic variables computed above it is straightforward to show that the first law  
\be\label{2Dfirstlaw}
\d M\sbtx{2D}=T\sbtx{2D}\d S~,
\ee
holds provided the sources are kept fixed, along with the charge $\cq$ and the AdS$_2$ radius $\ell_2$. As promised, the near extremal behavior of the entropy  
\eqref{eq:entropy2d} and the 2D mass \eqref{eq:mass2d} are compatible with the first law \eqref{2Dfirstlaw}, where $S\sbtx{ext}$ and $M\sbtx{gap}$ are kept fixed.

%%%%%%%%%%%%%%%%%%%%%%%%%%%%%%%%%%%%%%%%
\subsection{5D versus 2D thermodynamics}\label{sec:5d2dthermo} 

In this subsection we confront our findings in 2D with the derivation of black hole thermodynamics near extremality from a 5D point of view presented in subsection  \ref{sec:nIR5Dthermo}. When needed for comparison, we will denote with a subscript `(5D)'  or `(2D)' to the quantities in section  \ref{sec:5d} and  section \ref{sec:2dthermo}, respectively. The relation between these two sections relies on the dictionary in subsection \ref{sec:2D5DBH}.\footnote{Note that we have set $\a_0=1$ in this and the subsequent relations since the 5D thermodynamic expressions in subsection \ref{sec:2D5DBH} are given for $\a_0=1$ only.}

We recall that the {\it nearly} extreme black hole has small temperature $T_{5D}\ll M$ while keeping the angular momentum $J$ fixed. Fixing the angular momentum $J$ in 5D corresponds to keeping ${\cal Q}$ fixed in 2D, as we have done in this section. In this limit, the 5D Hawking temperature \eqref{eqn:TOmega} becomes
\begin{align}
T_{\rm 5D}&={x^2(2x^2-1)\over \pi a_0^2({1+x^2})} {\varepsilon \lambda} + O(\lambda^2)\cr
%&= {\varepsilon \lambda\over \pi \ell_2\lambda_0} 
&=  {\sqrt{-c_0}\over 2\pi |\nu_0|}{ \lambda\over  \lambda_0} + O(\lambda^2)~,
\end{align}
where in the first line we used the result from the near extreme limit in 5D \eqref{eq:extlimit} and in the second line we translated the expression to 2D variables via \eqref{eq:mapl2}, \eqref{eq:l1}, and \eqref{c0}. Comparison with the 2D result \eqref{2Dtemperature} leads to
\be
{T_{\rm 5D}\over T_{\rm 2D}}={1\over k\nu_0} { \lambda\over\lambda_0} ~.
\ee
Similarly, if we compare the mass gap given by \eqref{eqn:CJT=0} to the one deduced in \eqref{eqn:gap}, we have
\be
{M_{\rm gap}^{(5D)}\over M_{\rm gap}^{(2D)}}={1\over k\nu_0} { \lambda\over\lambda_0} ~.
\ee
We can trace these apparent discrepancies to our normalization \eqref{gen-PBH1} of the 2D Killing vector in the IR as $\xi^t\partial_t = \eta\partial_t = {k\nu_0}\partial_t$, while in the UV we normalize it as $\partial_t$. Moreover, in taking the near horizon/near extreme limit of the 5D solution \eqref{eq:extlimit2} we rescale the temporal coordinate $t\to {\lambda_0\over\lambda} t$.

A possible remedy for the discrepancy between the 5D thermodynamic variables and their 2D counterparts is to fix the normalization $k$ as 
\be
k= { \lambda_0\over  \lambda} \nu_0~. 
\ee  
After all, the variable $k$ was introduced in \eqref{CKVs} precisely to parameterize the ambiguity in the normalization of the timelike Killing vector. 
However, this choice is awkward from a 2D perspective: it amounts to boundary conditions on nAdS$_2$ that depend on the strength of the dilaton source and other parameters that are not specified at the IR fixed point. Instead, the deviations away from AdS$_2$ depend on the asymptotic data at the UV fixed point, and hence $M_{\rm gap}$ is not dictated by considerations intrinsic to nAdS$_2$/nCFT$_1$.

It is the interplay between the AdS$_5$ radius and angular momentum that causes this discrepancy. More concretely, for finite values of $\ell_5$ ($x\neq1$) the rescaling parameter $\lambda_0$  in \eqref{eq:extlimit2} reflects that there is a different notion of time in the UV versus the IR. The variable $x$ measures the strength of the AdS$_5$ curvature relative to the black hole rotation and a proper understanding of the near extreme black hole entropy must account for this dependence physically rather than inserting it as an input. The dependence of nAdS$_2$ on UV data and the resulting lack of universality was also discussed in \cite{Larsen:2018iou}. 

The effect we encounter here does not arise in the coupling between the AdS$_5$ radius and a electric/magnetic charge studied in \cite{Almheiri:2016fws,NayakShuklaSoniEtAl2018}. Thus the Kerr-AdS$_5$ black holes are not in the same universality class as their charged counterparts. However, in the absence of a 5D cosmological constant, there is only one relevant scale $\ell_2$ so then $k\sim\ell_2$ and we recover the notion of universality advocated in \cite{Maldacena:2016upp}.

It is worth stressing that the entropy does not suffer from the ambiguity discussed here:
\begin{align}
S_{\rm 2D}&= S_{\rm ext} + {2\over M_{\rm gap}^{(2D)}} T_{2D} \cr
&= S_{\rm ext} + {2\over M_{\rm gap}^{(5D)}} T_{5D} \cr
&= S_{\rm 5D}~.
\end{align}
This agreement is expected since the entropy in \eqref{eq:sbh} and \eqref{2Dentropy} are just the area of the horizon which is not sensitive to the normalization of a timelike Killing vector.

%%%%%%%%%%%%%%%%%%%%%%%%%%%%%%%%%%%%%%%%%%%%%%%%%%%%%%%%%%%%%%%%%%%%%%%%%%%%%%%%
\section{Summary and Future Directions}\label{sec:discussion}

We have used the nAdS$_2$/nCFT$_1$ correspondence to study aspects of Kerr-AdS$_5$ black holes with their two rotation parameters equal. We derived a 2D effective theory that is a consistent truncation of 5D Einstein gravity with and without cosmological constant, for which this black hole is one solution. This truncation contains the ingredients needed to discuss physics near the IR fixed point (i.e. the nAdS$_2$ region), the dynamics near the UV theory (i.e. either the asymptotically flat or the AdS$_5$ region), and the flow between these limits. 

Our 2D model contains two scalar fields $\{\psi, \chi\}$ in addition to the 2D metric and a $U(1)$ gauge field. We identify $\psi$ as the dilaton field and view $\chi$ as additional matter. A central aspect of our analysis is to develop the AdS$_2$/CFT$_1$ holographic dictionary for this theory near its IR fixed point, which we carry out in detail. A novel feature of our setup is the coupling between $\psi$ and $\chi$ in the nAdS$_2$ region that forces us to keep track of $\chi$ as $\psi$ breaks the conformal symmetry of the  AdS$_2$ background. From a five dimensional perspective, the persistence of $\chi$ is due to the coupling between the angular momentum of the black hole and the five dimensional cosmological constant.

Some aspects of our 2D model are conventional and expected: after appropriately diagonalizing the linearized fluctuations around the IR, we find a Goldstone mode in nAdS$_2$ whose effective action is a Schwarzian. The coefficient of the Schwarzian in \eqref{Schwarzian} depends on the AdS$_2$ radius in Planck units and the source of the dilaton. It leads to an entropy that is a linear function of the temperature with mass gap \eqref{eqn:gap} given by the AdS$_2$ radius in Planck units. These features fall into the universality class captured by the 2D Jackiw-Teitelboim (and related) models studied recently in, for example, 
\cite{Maldacena:2016upp,Almheiri:2016fws,Cvetic:2016eiv,Jensen2016,Engelsoy:2016xyb,GrumillerSalzerVassilevich2017,GrumillerMcNeesSalzerEtAl2017,ForsteGolla2017,CadoniCiuluTuveri2018,GaikwadJoshiMandalEtAl2018,KolekarNarayan2018,NayakShuklaSoniEtAl2018,GonzalezGrumillerSalzer2018}.

However, our discussion in subsection \ref{sec:5d2dthermo} exhibits this situation as unsatisfactory for the Kerr-AdS$_5$ black hole: the 2D theory does not capture the value of the mass gap (and hence the heat capacity) derived from the five dimensional black hole thermodynamics. The reason is a mismatch between the natural unit for time in the UV and in the IR. The black hole thermodynamics depends on the former and the nAdS$_2$ theory on the latter. This disagreement on units depends on physical quantities and is a consequence of the coupling between angular momentum and the AdS$_5$ radius.  From the perspective of the dual nCFT, the 5D thermodynamics indicate that there are two scales --$\nu_0$ and $\lambda_0$--  in the near IR, while the nAdS$_2$ determines the value of only one intrinsic scale --$\nu_0$. Finding a way how to predict the value of $\lambda_0$ in the nCFT is an open question that we leave for future work. 

There are several interesting open directions that we leave for future research, including: 
\begin{enumerate}
\item {\it Explore the dual nCFT$_1$:} The SYK model stands as the preeminent example of a nCFT$_1$. Various variants of the model include global symmetries, supersymmetry, tensor models, among other properties  (see \cite{Sarosi2018} for a partial list of references). It would be interesting to find within all these models the ones that can capture the couplings between $\psi$ and $\chi$. A step in this direction would be to evaluate holographic correlation functions of these fields and study their characteristic features. 
\item {\it Explore the RG flow:} Our 2D model is embedded in AdS$_5$ so we have significant control over the dual theory in the UV. It would be interesting to characterize the sector of ${\cal N}=4$ SYM  that accommodates our consistent truncation and cast it in terms of a suitable nCFT$_1$, as was done for different truncations in \cite{Taylor:2017dly}.
\item {\it Explore the black hole zoo:} Our analysis highlights a striking difference between Kerr-AdS$_5$ black holes and their charged cousins, the RN-AdS black holes. It would be interesting to generalize our study to other black holes (or black rings) and classify which of them display similar, or more general, features as those uncovered here. 
\item {\it Explore the phase diagram:} Black holes present a rich arena to study phase transitions and critical phenomena. It is an important goal to account for the renowned phase diagram of  AdS black holes in terms of the nCFT$_1$. An interesting direction was pursued in \cite{Anninos2017} for the four dimensional Kerr-Newman solution. 
\item {\it Explore quantum corrections:} The IR modes of the effective theory of quantum gravity control not just the area law, but also the logarithmic corrections to the Bekenstein-Hawking entropy \cite{Sen2013}. Recent developments \cite{CharlesLarsen2015,CharlesLarsenMayerson2017,CastroGodetLarsenEtAl2018} show non-trivial patterns in the coefficients that accompany these log-terms.  It would be interesting to account for such patterns through the nAdS$_2$/nCFT$_1$ correspondence.
\end{enumerate}

\section*{Acknowledgements}
We thank Mirjam Cveti\v c, Daniel Grumiller, Kostas Skenderis, Wei Song, Marika Taylor, David Turton and Boyang Yu for useful discussions. We also thank TSIMF, Sanya  for their hospitality during this collaboration. A.C. would like to thank the Leinweber Center for Theoretical Physics at the University of Michigan for hospitality and partial support during the completion of this work. I.P. thanks the Yau Mathematical Sciences Center at Tsinghua University for hospitality and partial financial support during the completion of this work. This work was supported in part by the U.S. Department of Energy under grant DE-FG02-95ER40899. A.C. is supported by the Nederlandse Organisatie voor Wetenschappelijk Onderzoek (NWO) via a Vidi grant and through the Delta ITP consortium, a program of the NWO that is funded by the Dutch Ministry of Education, Culture and Science (OCW).

\appendix 

%\section{3D to 2D reduction}
%If needed

\section{Asymptotic Solutions and Superpotentials in the UV}
\label{UV-analysis}

In this appendix we solve the Hamilton-Jacobi equation \eqref{HJ-S} away from the IR fixed point. The first order equations \eqref{flow-eqs-S} can then be integrated to obtain the corresponding solutions of the second order equations of motion. We will consider two cases corresponding to two opposite limits. Firstly, we set $\ell_5=\infty$ with constant $\c$, in which case the exact solution can be obtained throughout the RG flow. Secondly, we solve the Hamilton-Jacobi equation for $\ell_5/R\ll 1$ and show that the solution coincides with the Kaluza-Klein reduction of the known boundary counterterms for AdS$_5$.       

%%%%%%%%%%%%%%%%%%%%%%%%%%%%%%%%%%%%%%%%%%%%%%%%%%%%%
\subsection{Asymptotically Flat Solutions with Constant $\c$}
\label{UV-analysis-flat}

Setting $\ell_5=\infty$ and $\c$ constant, equations (\ref{HJ-U-ansatz}-\ref{Z1Z2}) can be solved exactly to obtain 
\be
W=\pm\sqrt{\frac{4}{R^2}e^{-\j}+2Q^2R^2e^{3\c_0+\j}-4m}\,,\qquad Z_1=\frac{2e^{-4\j}}{W},\qquad Z_2=Z_3=0~,
\ee
where $m$ is an integration constant. This determines the Hamilton-Jacobi functional $\cu$ up to two derivatives in time. However, in this case we can do much better: an exact solution of the Hamilton-Jacobi equation \eqref{HJ-S} to all orders in time derivatives can be obtained. Up to a choice of sign in front of the square roots the solution takes the form
\bal\label{Uexact}
\cu=&\;\frac{1}{\k_2^2}\int dt\sqrt{-\g}\Bigg[\sqrt{\frac{4}{R^2}e^{-\j}-\g^{tt}(\pa_te^{-2\j})^2+2Q^2R^2e^{3\c_0+\j}-4m}\NO\\
&-\frac{\pa_t e^{-2\j}}{\sqrt{-\g}}\log\Bigg(\frac{\frac{\pa_t e^{-2\j}}{\sqrt{-\g}}+\sqrt{\frac{4}{R^2}e^{-\j}-\g^{tt}(\pa_te^{-2\j})^2+2Q^2R^2e^{3\c_0+\j}-4m}}{\sqrt{\frac{4}{R^2}e^{-\j}+2Q^2R^2e^{3\c_0+\j}-4m}}\Bigg)\Bigg]~.
\eal

In order to confirm that \eqref{Uexact} is an exact solution of the Hamilton-Jacobi equation \eqref{HJ-S} it is instructive to first evaluate 
the functional derivatives of $\cu$ with respect to the fields $\j$ and $\g_{tt}$, namely
\bal\label{U-der}
\frac{\d\cu}{\d\j}=&\;-\frac{\sqrt{-\g}\; e^{-2\j}\Big(\frac{1}{R^2}e^{3\j}-\frac{1}{2}Q^2R^2e^{3\c_0+5\j}-\square_t\Big)e^{-2\j}}{\k_2^2\sqrt{\frac{1}{R^2}e^{-\j}-\frac14\g^{tt}(\pa_te^{-2\j})^2+\frac{1}{2}Q^2R^2e^{3\c_0+\j}-m}}\,,\NO\\
%%%%%%%%%%%%%%%%%%%%%%%%%%%%%%%%%%%%%%%%%%%%%%%%%%%%%%%%%%%%%%%%%%%%%%%%%%%%%%%%%%%%
\rule{0cm}{0.7cm}\g_{tt}\frac{\d\cu}{\d\g_{tt}}=&\;\frac{\sqrt{-\g}}{\k_2^2}\sqrt{\frac{1}{R^2}e^{-\j}-\frac14\g^{tt}(\pa_te^{-2\j})^2+\frac{1}{2}Q^2R^2e^{3\c_0+\j}-m}\,.
\eal
Using these expressions for the functional derivatives of $\cu$, together with the relation between the Hamilton-Jacobi functionals $\cs$ and $\cu$ in \eqref{S-sol}, it is straightforward to show that \eqref{Uexact} solves \eqref{HJ-S}. 

The exact solution \eqref{Uexact} for $\cu$ allows us to obtain the general solution of the equations of motion by solving the first order equations \eqref{flow-eqs-S}, which can be expressed in the form
\bal
\label{flow-eqs-S-simple}
\pa_\r\sqrt{-\g}&=-\frac{\k^2_2}{2}e^{2\j}\frac{\d\cu}{\d\j}~,\NO\\
%%%%%%%%%%%%%%%%%%%%%%%%%%%%%%%%%%%%%%%%%%%%%%%%%%%%%%%%%%%%%%%%%%%%%%%%%%%%%%%%%%%%
%%%%%%%%%%%%%%%%%%%%%%%%%%%%%%%%%%%%%%%%%%%%%%%%%%%%%%%%%%%%%%%%%%%%%%%%%%%%%%%%%%%%
\pa_\r e^{-2\j}&=\frac{2\k_2^2}{\sqrt{-\g}}\g_{tt}\frac{\d\cu}{\d\g_{tt}}~,\NO\\
%%%%%%%%%%%%%%%%%%%%%%%%%%%%%%%%%%%%%%%%%%%%%%%%%%%%%%%%%%%%%%%%%%%%%%%%%%%%%%%%%%%%
\pa_\r A_t&=-Qe^{3\j+3\c_0}\sqrt{-\g}~.
\eal
Using the functional derivatives \eqref{U-der} these first order equations imply that the combination $\pa_t e^{-2\j}/\sqrt{-\g}$ is independent of the radial coordinate $\r$. As a result, the general solution for the dilaton $\j(\r,t)$ can be expressed in the form
\be
\int^{\j(\r,t)}\hskip-0.4cm \frac{-2e^{-2\wt\j} d\wt \j}{\sqrt{\frac{4}{R^2}e^{-\wt\j}+2Q^2R^2e^{3\c_0+\wt\j}+4(\pa_t\vf)^2-4m}}=\r+\o(t)~,
\ee
where $\vf(t)$ and $\o(t)$ are arbitrary functions of time only. The solution for the metric then is 
\be
\sqrt{-\g}=\left\{\begin{matrix}
\hskip-2.6cm\frac{1}{2\pa_t\vf}\pa_t e^{-2\j}, && \hskip0.5cm\pa_t e^{-2\j}\neq 0~,\\
\rule{0cm}{.8cm}\a_0\sqrt{\frac{4}{R^2}e^{-\j}+2Q^2R^2e^{3\c_0+\j}-4m}\,\,, && \hskip0.5cm\pa_t e^{-2\j}=0~,
\end{matrix}\right.
\ee
where $\a_0$ is a positive constant. Finally, the gauge field is determined from the last equation in \eqref{flow-eqs-S-simple}. The 5D uplift of the static solution is an asymptotically Taub-NUT geometry with a four dimensional Reissner-Nordstr\"{o}m black hole base, which becomes extremal 
when 
\be
m=\sqrt{2}e^{3\c_0/2}Q~.
\ee

In addition to determining the general solution of the equations of motion, the exact solution \eqref{Uexact} of the Hamilton-Jacobi equation allows one to determine the boundary terms that render the variational problem well posed for asymptotically Taub-NUT solutions in 5D, or asymptotically flat solutions in 4D, far away from the AdS$_2$ IR region that was the focus of this paper. Moreover, the residual asymptotic local symmetries in this case are likely related to the BMS group in four and five dimensions and would be interesting to study how this is encoded in the effective action \eqref{Uexact}. We hope to address these questions in future work.

%%%%%%%%%%%%%%%%%%%%%%%%%%%%%%%%%%%%%%%%%%%%%%%%%%%%%%
\subsection{Asymptotically AdS$_5$ solutions}
\label{UV-analysis-AdS}

In the limit $\ell_5/R\ll 1$ the solution of the system of equations  \eqref{HJ-U-ansatz}-\eqref{Z1Z2} takes the form
\bal\label{HJ-sol-5D}
W=&\;\frac{3}{\ell_5}e^{-3\j/2+\c/2}+\frac{\ell_5}{8R^2}\Big(4e^{\j/2-\c/2}-e^{5\j/2-7\c/2}\Big)-\frac{\ell_5^3}{18R^4}e^{5\j/2-3\c/2}(e^{2\j-3\c}-1)^2\j+\cdots~,\NO\\
Z_1=&\;\frac{\ell_5}{2}e^{-5\j/2-\c/2}-\frac{5\ell_5^3}{18R^2}e^{3\j/2-9\c/2}\j+\cdots~,\NO\\
Z_2=&\;\frac{\ell_5}{2}e^{-5\j/2-\c/2}+\frac{5\ell_5^3}{6R^2}e^{3\j/2-9\c/2}\j+\cdots~,\NO\\
Z_3=&\;-\frac{3\ell_5}{8}e^{-5\j/2-\c/2}-\frac{5\ell_5^3}{8R^2}e^{3\j/2-9\c/2}\j+\cdots~,
\eal
where the ellipses indicate subleading terms. The first order equations \eqref{flow-eqs-S} then imply that
\be
e^{-2\j}\sim e^{-3\c}\sim \r^{12/5},\qquad e^{(\j+\c)/2}=\frac45(\r/\ell_5)^{-1}+\cdots~,
\ee
to leading order as $\r\to\infty$. 

However, the asymptotic solution of the Hamilton-Jacobi equation for the 5D pure gravity action \eqref{eq:action5d} is well known \cite{deHaro:2000vlm}
\be\label{5DHJ}
\cs^{(5)}=\frac{1}{\k_5^2}\int d^4x\sqrt{-\g^{(4)}}\;\Bigg(
	\frac{3}{\ell_5} 
	+ \frac{\ell_5}{4} \car[\g^{(4)}] 
	-\frac{\ell_5^3}{16}\Big(\car[\g^{(4)}]_{ij}\car[\g^{(4)}]^{ij}-\frac13\car[\g^{(4)}]^2\Big) \log(e^{-2\r_5/\ell_5})\Bigg)~,
\ee
where $\r_5$ is the canonical radial coordinate in five dimensions and $\g^{(4)}_{ij}$ denotes the induced metric on the radial slice, i.e.
\be
ds_5^2=d\r_5^2+\g^{(4)}_{ij}dx^idx^j\,.
\ee 
Reducing this solution to two dimensions using the Kaluza-Klein ansatz \eqref{eqn:mansatz} we reproduce exactly the solution \eqref{HJ-sol-5D} of the Hamilton-Jacobi equation in two dimensions, up to second order in time derivatives. However, the reduction also provides the four-derivatives terms
\bal
\cu_{(4)}&\;=-\frac{\ell_5^3}{16\k_2^2}\int dt\sqrt{-\g}\;e^{-7\j/2-3\c/2}\Bigg(\frac16(2\square_t\j-3\square_t\c)^2+\frac32\Big(\pa\j\cdot\pa\j-3\pa\j\cdot\pa\c\Big)\square_t\c-\frac{11}{18}(\pa\j\cdot\pa\j)^2\NO\\
&\hskip-0.5cm+\frac{9}{8}(\pa\c\cdot\pa\c)^2+\frac{69}{8}(\pa\j\cdot\pa\c)^2-\frac{19}{12}(\pa\j\cdot\pa\j)(\pa\j\cdot\pa\c)-3(\pa\c\cdot\pa\c)(\pa\j\cdot\pa\c)\Bigg)\log(e^{-2\r_5/\ell_5}),
\eal
where we have introduced the shorthand notation $\pa\j\cdot\pa\c\equiv\g^{tt}\pa_t\j\pa_t\c$. The Kaluza-Klein ansatz \eqref{eqn:mansatz} relates the 5D and 2D radial coordinates as $d\r_5=e^{(\j+\c)/2}d\r$, which implies that as $\r_5\to\infty$
\be
\r\sim e^{\frac{5\r_5}{4\ell_5}}~,
\ee
and so
\be
e^{-2\j}\sim e^{-3\c}\sim e^{3\r_5/\ell_5}~.
\ee
It is straightforward to verify that this asymptotic behavior of the 2D scalars renders the 5D metric \eqref{eqn:mansatz} asymptotically locally AdS$_5$.

%%%%%%%%%%%%%%%%%%%%%%%%%%%%%%%%%%%%%%%%%%%%%%%%%%%%%
\section{Black Hole Temperature from Scalar Potential}
\label{Tpot}

In this appendix we demonstrate that the temperature of any black hole solution of the 2D model \eqref{2Daction-extremal} is given by the value of the scalar potential on the horizon. We expect this result to hold more generally for 2D dilaton gravity theories.

It is instructive to first consider the case of constant $\c$, and so necessarily $\ell_5=\infty$, since in that case the global timelike Killing vector can be expressed covariantly as $\x^a=-\ell_2 k\e^{ab}\nabla_be^{-2\j}$. The Hawking temperature can then be easily obtained from the surface gravity $\Hat\k$, namely 
\be
T\sbtx{2D}=\frac{\Hat\k}{2\p}~,
\ee
where
\be
\Hat\k^2\equiv-\frac12\left.(\nabla^a\x^b)(\nabla_a\x_b)\right|_{\r_h}~.
\ee
Using the expression for the Killing vector and the first equation in \eqref{eoms-extremal} 
\be\label{2Dtemperature-constant-chi}
T\sbtx{2D}=\frac{k\ell_2}{2\p R^2}\Big(e^{\psi(\r_h)}-\frac{|Q|R^2}{2}e^{3\psi(\r_h)}\Big)~.
\ee

The above result can be generalized to non-constant $\c$ and $\ell_5<\infty$. To this end let us consider the general static, near-horizon solution of the equations of motion in an expansion in the deviation $v$ of the radial coordinate away from the horizon, i.e. $\r=\r_h+v$. The horizon at $\r=\r_h$ is not assumed to be extremal. A straightforward calculation determines that in the vicinity of the horizon\footnote{\label{zero-sup} From \eqref{flow-eqs-S}, \eqref{S-sol} and \eqref{HJ-U} follows that for static solutions $\pa_\r e^{-2\j}=W$. The near-horizon solution \eqref{near-horizon}, therefore, implies that $\left.W\right|_{\r_h}$=0.} 
\be\label{near-horizon}
\j=\j(\r_h)+\co(v^2),\qquad \c=\c(\r_h)+\co(v^4),\qquad \pa_\r e^{-2\j}=h_0 \sqrt{-\g}+\co(v^8)~,
\ee
where $h_0\neq 0$ is a dimensionful integration constant, and by the definition of the horizon, $\sqrt{-\g}=\co(v)$. The constant $h_0$ can be determined by inserting the near IR solutions \eqref{eqn:leading} and \eqref{eq:s1}, giving $h_0=\n_0/(\a_0\ell_2)$, where the subscripts ``0'' denote that we are considering static solutions in two dimensions. From the expression \eqref{2Dtemperature} for the black hole temperature and \eqref{CKVs} then follows that
\be\label{2Dtemperature-psi}
T\sbtx{2D}=\frac{\h}{2\p}\left.\pa_\r\sqrt{-\g}\right|_{\r_h}=\frac{\h}{2\p h_0}\left.\pa^2_\r e^{-2\j}\right|_{\r_h}=\frac{k\ell_2}{2\p}\left.\pa^2_\r e^{-2\j}\right|_{\r_h}~.
\ee

The expression \eqref{2Dtemperature-psi} for the temperature not only implies that $\x^a=-\ell_2k\e^{ab}\nabla_be^{-2\j}$ remains a Killing vector in the vicinity of the horizon even when $\c$ is not constant, but also it allows us to express the temperature in terms of the values of the scalars on the horizon. Namely, for static solutions the third equation in \eqref{eoms-extremal-gf} can be rewritten in the form 
\be
\frac{1}{\sqrt{-\g}}\pa_\r(\sqrt{-\g}\;\pa_\r e^{-2\j})=-{1\over 2R^2}e^{-3\chi+3\psi}(1+{R^4Q^2}e^{6\chi})+{2\over R^2}e^{\psi}+\frac{12}{\ell_5^2}e^{-\j+\c}.
\ee
Notice that the expression on the right hand side is the scalar potential in the 2D action \eqref{2Daction-extremal}. Evaluating the same expression using the near horizon solution \eqref{near-horizon} gives
\begin{align}
\frac{1}{\sqrt{-\g}}\pa_\r(\sqrt{-\g}\;\pa_\r e^{-2\j})&=\frac{h_0}{\sqrt{-\g}}\pa_v((\sqrt{-\g})^2+\co(v^9))\cr &=2h_0\pa_v\sqrt{-\g}+\co(v^7)=2\pa_v^2e^{-2\j}+\co(v^7)~,
\end{align}
from which we conclude that 
\be\label{2Dtemperature-potential}
T\sbtx{2D}=\frac{k\ell_2}{4\p}\Big(-{1\over 2R^2}e^{-3\chi(\r_h)+3\psi(\r_h)}(1+{R^4Q^2}e^{6\chi(\r_h)})+{2\over R^2}e^{\psi(\r_h)}+\frac{12}{\ell_5^2}e^{-\j(\r_h)+\c(\r_h)}\Big).
\ee
Therefore, even when $\c$ is not constant, the temperature is given by the value of the scalar potential at the horizon. Moreover, extremizing the scalar potential with respect to $\c(\r_h)$ leads to the relation 
\be\label{chi-constraint}
Q^2R^4e^{6\c(\r_h)}-\frac{8R^2}{\ell_5^2}e^{-4\j(\r_h)+4\c(\r_h)}-1=0~.
\ee
Using this relation and sending $\ell_5\to\infty$ we recover the expression \eqref{2Dtemperature-constant-chi} for constant $\c$.

\section{A Neutral Solution}\label{app:neutral}

One exact solution to the equations of motion \eqref{eoms-extremal} with $Q=0$ is
\be\label{eq:qq1}
ds_2^2 = -(1+{r^2\over \ell^2}) dt^2 +{dr^2\over 1+{r^2\over \ell^2}}~,
\ee
and 
\begin{align}
e^{\chi} &=R\, \le(f(t)\sqrt{r^2+\ell^2} + \kappa\, r\ri)^{-1}~, \cr
e^{2\psi} &= R^3\, \le(f(t)\sqrt{r^2+\ell^2} + \kappa\, r\ri)^{-3}~, 
\end{align}
where
\be
f(t)= f_1 \cos(t/L) +f_2\sin(t/L)~, \qquad \kappa^2= f_1^2+f_2^2 +{1\over 4}~.
\ee
Here $f_{1,2}$ and $\kappa$ are constants. The case of global AdS$_5$ corresponds to $f_1=f_2=0$.  This is a rather simple representative of a neutral solution to our effective 2D theory in section \ref{sec:2Deff}, but by no means general nor exhaustive.

\bibliographystyle{utphys}
\bibliography{Kerr_nAdSbib} 

\end{document}